\documentclass[aps,prb,twocolumn,showpacs,superscriptaddress]{revtex4-2}
\usepackage{amsmath,amssymb,color,graphicx,hyperref,amsthm,bm,epstopdf,mathrsfs}
\usepackage{hyperref}
\usepackage{graphicx,grffile}
\usepackage{amsmath}
\usepackage{amsfonts}
\usepackage{array}
\usepackage{url}
\usepackage{comment}
\usepackage[normalem]{ulem}
\usepackage{float}
\usepackage[toc,page]{appendix}
\usepackage{subcaption}

\renewcommand{\AA}{\mathcal{A}} 
\newcommand{\BB}{\mathcal{B}} 
\newcommand{\CC}{\mathcal{C}} 
\newcommand{\Q}{\mathbb{Q}} 
\newcommand{\C}{\mathbb{C}} 
\newcommand{\R}{\mathbb{R}} 
\newcommand{\eps}{\varepsilon}
\newcommand{\reK}{\R}
\newcommand{\imK}{\mathbb I}

\newcommand{\wt}{\widetilde}
\renewcommand{\ol}{\overline}
\newcommand{\diag}{{\rm diag}}
\newcommand{\I}{{\mathbb{I}}}

\newcommand{\nvec}{\mbox{\boldmath{$n$}}}

\newcommand{\zvec}{\mbox{\boldmath{$z$}}}
\newcommand{\im}{\mbox{Im}}

\newcommand{\vvec}{v}

\newcommand{\area}{\mathcal{A}}

\renewcommand{\eqref}[1]{eq. \ref{#1}}
\newcommand{\figref}[1]{fig. \ref{#1}}

\graphicspath{{figures/}}

\makeatletter
\let\cat@comma@active\@empty
\makeatother

\begin{document}

\title{Symmetries, Length Scales, Magnetic Response and Skyrmion Chains in Nematic Superconductors}
\author{Martin Speight}
\affiliation{School of Mathematics, University of Leeds, Leeds LS2 9JT, United Kingdom}
\author{Thomas Winyard}
\affiliation{Maxwell Institute of Mathematical Sciences and School of Mathematics, University of Edinburgh, Edinburgh, EH9 3FD, United Kingdom}
\author{Egor Babaev}
\affiliation{Department of Physics, KTH-Royal Institute of Technology, Stockholm, SE-10691 Sweden}

\begin{abstract}
Nematic systems are two component superconductors that break rotational symmetry, but exhibit a mixed symmetry that couples spatial rotations and phase difference rotations. We show that a consequence of this induced spatial anisotropy is mixed normal modes, that is the linear response to a small perturbation of the system about its ground state, generally couples magnetic and condensate degrees of freedom. 
We will study the effect of mode mixing on the magnetic response of a nematic system as the strength of applied field is increased. In general we show that the coupled modes generate magnetic field perpendicular to the applied field, causing the magnetic response to spontaneously twist direction. We will study this for the Meissner effect with weak fields and also for stronger applied fields, which produce a mixture of Skyrmions and composite vortices, forming orientation dependent bound states. We will also calculate the anisotropies of the resulting first and second critical fields $H_{c_1}$ and $H_{c_2}$.
The Skyrmion lattices for $H_{c_1} \leq H \leq H_{c_2}$ in nematic superconductors are shown to be structurally complicated, in contrast to the triangular or square vortex lattices in conventional superconductors. For low fields the magnetic response of the system involves a loosely bound collection of parallel Skyrmion chains. As the external field is increased the chains attract one another, causing a transition where the unit cell becomes triangular for high applied fields. This unique Skyrmion lattice and the magnetic twisting are clear indicators that could be used experimentally to identify materials that exhibit nematic superconductivity. To obtain these results we develop and present a novel method to find the unit cell of a vortex lattice that can be applied to other kinds of superconducting systems.
\end{abstract}

\maketitle
\section{Introduction}\label{sec:intro}
Nematic superconductors have been of increasing interest in recent years and a number of their properties have been demonstrated to be  unconventional \cite{fu2010odd,fu2014odd,venderbos2016odd,zyuzin2017nematic,wu2017majorana,chirolli2018chiral,barkman2019antichiral,uematsu2019chiral,dentelski2020effect,yiphalf,chubukovTBG}. One material that has been intensively studied experimentally is the doped topological insulator $Bi_2 Se_3$
\cite{hor2010superconductivity,wray2010observation,kriener2011bulk}. 
Superconductivity has also been reported in $Cu_x Bi_2 Se_3$, $\mathrm{Sr_xBi_2Se_3}$ 
\cite{Shruti_SrBiSe,Liu_SrBiSe,Pan_Hc2} and
$\mathrm{Nb_xBi_2Se_3}$ \cite{Asaba_NbBiSe} which also exhibits rotational symmetry breaking.

The possibility of nematic superconductivity in $Cu_x Bi_2 Se_3$ and other materials, motivates this study of their macroscopic properties. Two experimental signatures have previously been predicted: that the upper critical field of a magnetic field exhibits anisotropy \cite{venderbos2016identification,yonezawa2017thermodynamic}; and that topological excitations exhibit fractional vortex splitting, forming Skyrmions \cite{zyuzin2017nematic}.

Many questions concerning the properties of this state still remain. These include the form of the normal modes, coherence and magnetic field penetration lengths and collective properties of topological excitations. By normal modes, we mean those that govern the response of the system to small perturbations about the homogeneous superconducting ground state. In this paper, we demonstrate that the modes of nematic superconductors are non-trivially mixed. This is shown to result in a non-trivial magnetic response for nematic superconductors. We will demonstrate the effect of the mode mixing by considering the effect of an applied magnetic field $\boldsymbol{H}$ on the superconductor.
    
We will first consider $|\boldsymbol{H}| < H_{c_1}$ and magnetic field screening. We will show that the mixed modes cause the local magnetic field $B$ to locally twist direction, as has been demonstrated for $s+is$ and $s+id$ superconductors \cite{speight2021magnetic}. We will then consider vortex solutions in the bulk, which for many orientations exhibit fractional vortex splitting, forming Skyrmions. Orientation here refers to the direction of the centre of the vortex line (or flux tube), which is modelled by taking a cross-section and assuming the fields are homogeneous in the direction of the vortex line. Hence, if we refer to the basal plane, we mean the cross-section is in the basal plane and the vortex line is perpendicular to this plane. We will demonstrate that in the basal plane, the mixed modes do not excite the magnetic field components orthogonal to the vortex line, hence the magnetic field is always parallel to the vortex line. This leads to no magnetic field twisting and hence the results of \cite{zyuzin2017nematic} are valid. However, once the vortex plane is not the basal plane, the mixed modes excite the components of the magnetic field orthogonal to the vortex lines, causing the magnetic field to twist direction. This leads to non-trivial bound states, that are dependent on the orientation of the vortex line.

We then calculate $H_{c_1}$ and $H_{c_2}$, demonstrating that they are anisotropic, confirming the results for $H_{c_2}$ in \cite{venderbos2016identification}. We then consider the vortex lattice solutions in the bulk (when $H_{c_1} \leq H \leq H_{c_2})$, namely the periodic solution that minimises the Gibbs free energy per unit area $G/\mathcal{A}$. We present a new method to find such solutions, that minimises  $G/\mathcal{A}$ with respect to the unit cell geometry and the field configuration over the cell. This demonstrates that in general the lattices initially form Skyrmion chains with rectangular unit cells and winding $N=2$. These chains attract as the external field strength $|\boldsymbol{H}|$ is increased, until the cell eventually becomes triangular.

\section{The model}
Our starting point is a general anisotropic effective model. We will describe the methods used for this general model and restrict to the specific case of a nematic superconductor when we present the results. The most general 3-dimensional anisotropic Ginzburg-Landau free energy is given as,
\begin{equation}
F = \int_{\mathbb{R}^3}\left\{\frac{1}{2}Q^{\alpha\beta}_{ij} \overline{D_i \psi_\alpha} D_j \psi_\beta + \frac{1}{2} |B|^2 + F_P(|\psi_\alpha|, \theta_{12})\right\} d^3 x,
\label{eq:F}
\end{equation}
where $D_i = \partial_i - iA_i$ is the covariant derivative associated with the $U(1)$ gauge field $A_i$, leading to the magnetic field $B_i = \eps_{ijk}\partial_j A_k$. We will consider a 2-component model, where the two complex fields $\psi_\alpha =  \rho_\alpha e^{i\varphi_\alpha}$ represent the different superconducting components. Note that Greek indices $\alpha = 1,2$ will always enumerate components of the order parameter and Latin indices $i = 1,2,3$ indicate spatial components, while summation over repeated indices is implied for both. $F_p$  collects together the potential terms, which due to gauge invariance, depend only on the condensate magnitudes $\rho_\alpha$ and the phase difference between the condensates $\theta_{12} := \theta_1 - \theta_2$.

The anisotropy of the model is given by the anisotropy matrices $Q^{\alpha\beta}$, which must satisfy the minimal condition $Q^{\alpha\beta}_{ij} = \overline{Q}^{\beta\alpha}_{ji}$ so that the energy is real. Note that our methods will be presented for general values of $Q^{\alpha\beta}_{ij}$ and hence are applicable to any two-component anisotropic Ginzburg-Landau theory.

Nematic superconductors can be modelled by making the following restriction (for a  microscopic derivation, see \cite{zyuzin2017nematic}),
\begin{align}
Q^{11} = Q^{22} = \left( \begin{array}{ccc} 1 & 0 & 0 \\ 0 & 1 & 0 \\ 0 & 0 & \beta_z\end{array}\right), \quad Q^{12} = \beta_\perp\left( \begin{array}{ccc} 1 & i & 0 \\ i & -1 & 0 \\ 0 & 0 & 0 \end{array}\right),
\label{eq:nematic}
\end{align}
where $\beta_\perp$ and $\beta_z$ are positive parameters to be fixed. Other papers have extended this choice for the anisotropy, adding terms designed to break rotational symmetry in the basal plane. We will neglect such terms,
considering the model where rotational symmetry breaking in the basal plane is weak. The potential terms can be represented as follows  
\begin{equation}
F_p = \frac{\eta^2}{2}\left(-|\psi_1|^2 - |\psi_2|^2 + \frac{1}{2}(|\psi_1|^4 + |\psi_2|^4) + \gamma|\psi_1|^2 |\psi_2|^2\right).
\label{eq:Fp}
\end{equation}
The derivation of the above parameters is briefly discussed in the appendix, along with sensible values. It is worth noting that this is a chiral representation of the order parameter $\psi_{1,2} = \Delta_x \pm i \Delta_y$. Where $\Delta$ represents an interorbital spin-triplet pairing of the form $\Delta = (\Delta_x, \Delta_y, 0)$.

For the class of models we consider, the ground state (minimal energy degree $0$) solution is constant and hence determined purely by the values $\rho_\alpha$ that minimize the potential term $F_p$. We write the ground state solution as $(|\psi_\alpha|, \theta_{12}, A_i) = (u_\alpha, \delta_{12}, 0)$, where the values of $u_\alpha$ are dependent on the value of $\gamma$, and  $\delta_{12}$ is a free choice as $F_p$ is independent of $\theta_{12}$. $\gamma = 1$ is a critical parameter choice, as the ground state solutions are any point on the circle $u_1^2 + u_2^2 = 1$, leading to an enhanced $S^1\times U(1) \times U(1)$ symmetry (note that the phase difference is strictly ill-defined if either condensate is zero). If $\gamma < 1$ then the symmetry of the model becomes $U(1)\times U(1)$ with a single ground state,
\begin{equation}
u_1=u_2 = \frac{1}{\sqrt{1+\gamma}}.
\end{equation}
If $\gamma > 1$ then the symmetry becomes $Z_2\times U(1)$ and the two vacua,
\begin{equation}
(u_1, u_2) = (1,0), \qquad (u_1,u_2) = (0,1),
\end{equation}
clearly break $Z_2$ symmetry. Note that as one of the condensates is zero (not superconducting), the phase difference is ill-defined.

We will assume the parameter values $\beta_\perp = \frac{1}{3}$, $\beta_z = \frac{4}{3}$ and $\gamma = \frac{1}{3} < 1$ for the rest of this paper.
For the microscopic discussion of the parameters see \cite{zyuzin2017nematic}. Note that we have considered some alternate parameters to the ones above, to ensure that our results are robust.

Stationary configurations, such as vortices, are local minima of $F$. These satisfy the Ginzburg-Landau equations, obtained by variation of \eqref{eq:F} with respect to the constituent fields $\psi$ and $A$,
\begin{align}
Q^{\alpha\beta}_{ij} D_i D_j \psi_\beta &= 2 \frac{\partial F_p}{\partial \overline{\psi}_\alpha},\\
\partial_i ( \partial_j A_i - \partial_i A_j) &= J_i,
\label{eq:eom}
\end{align}
where the total supercurrent is defined as,
\begin{equation}
J_i := \im (Q^{\alpha\beta}_{ij} \overline{\psi}_\alpha D_j \psi_\beta). 
\end{equation}

\section{Symmetries}

While the potential in \eqref{eq:Fp} for our chosen parameters has a $U(1)\times U(1)$ symmetry, this, along with rotational symmetry, is broken to a mixed symmetry by the gradient terms if $\beta_\perp \neq  0$. To be precise, we consider a mapping of fields $(\psi,A)\mapsto (\wt\psi,\wt{A})$ to be a symmetry of the model if $F(\psi,A)=F(\wt\psi,\wt{A})$ for all configurations $(\psi,A)$. It is straightforward to verify that the following mappings are all symmetries of the nematic model:
\begin{enumerate}
\item Rotations in the basal plane:
\begin{equation}\label{eq:ariel}
\wt\psi(x)=S\psi(R^{-1}x),\qquad \wt{A}(x)=RA(R^{-1}x)
\end{equation}
where
\begin{equation}\label{eq:rotsym}
R=\left(\begin{array}{ccc}\cos\alpha & -\sin\alpha & 0 \\
\sin\alpha & \cos\alpha & 0 \\
0 & 0 & 1\end{array}\right),\qquad
S=\left(\begin{array}{cc} e^{i\alpha} & 0 \\ 0 & e^{-i\alpha}\end{array}\right).
\end{equation}
Note that this couples spatial rotations about the $x_3$ axis with rotations of the phase difference.
\item Reflexion in the basal plane: mapping \eqref{eq:ariel} with
\begin{equation}
R=\diag(1,1,-1),\qquad S=\I_2.
\end{equation}

\item Reflexion in a vertical plane: mapping \eqref{eq:ariel} with
\begin{equation}\label{eq:refsym}
R=\diag(1,-1,1),\qquad S=\left(\begin{array}{cc}0&1\\1&0\end{array}\right).
\end{equation}
Note that this swaps the condensates.

\item Complex conjugation:
\begin{equation}\label{eq:conjsym}
\wt\psi_1(x)=\ol{\psi_2(x)},\,
\wt\psi_2(x)=\ol{\psi_1(x)},\,
\wt{A}(x)=-A(x).
\end{equation}
Note that this also swaps the condensate's components, and coincides with the time reversal operation $\psi_{\pm}\mapsto \ol{\psi}_\pm$, $A\mapsto-A$, for the fields $\psi_\pm=\psi_1\pm i\psi_2$.
\end{enumerate}

Any composition of symmetries is also a symmetry, so by composing \eqref{eq:rotsym} and \eqref{eq:refsym}, for example, we can obtain a reflexion symmetry in any vertical plane. It is important to note that the rotation symmetry in \eqref{eq:rotsym} is orientation preserving (all the others are orientation reversing), and acts nontrivially on the phase difference $\theta_{12}$. Hence, non-homogeneous solutions of this model (such as vortices or Skyrmions), arise in one-parameter families, parametrized by the value of $\theta_{12}$ at spatial infinity. These distinct solutions coincide up to the rotation symmetry in \eqref{eq:rotsym}. Hence, the mixed symmetry \eqref{eq:rotsym} allows us to construct additional solutions from a given solution, by spatially rotating by an angle $\alpha$ about $x_3$ and simultaneously rotating the phase difference $\theta_{12}$ by $2\alpha$. Note, for vortices, if the $3$-dim vortex string is not oriented orthogonal to the basal plane, this mapping will change the spatial orientation of the string.

\section{Fundamental length scales:
 absence of conventional coherence and magnetic field penetration lengths}
 
We now consider a non-uniform superconducting state, where the fields locally deviate from their ground state values. Generally this is governed by the nonlinear GL equations \eqref{eq:eom}, which must be solved numerically. However, it is instructive to consider the fundamental length scales that govern how the fields decay to their ground state values.

In the isotropic case this is normally achieved by assuming that the fields are small when far from a defect. The magnetic penetration length is then found by fixing the matter fields to one of their ground state values and solving the resulting linear e.o.m., which is the famous London model for $B$. The coherence length is then found by fixing the gauge field to zero and solving the resulting linearised e.o.m. Hence, the crucial characteristics are the coherence and magnetic field penetration lengths, which determine the exponential law according to which the fields recover their ground state values.
Whether the superconductor is of type I, type II or type 1.5 can then be determined by comparing the fundamental matter and magnetic length scales, or coherence and magnetic field penetration lengths.

To understand why this will \emph{not work} in the anisotropic case, it is instructive to understand why it does work in the isotropic case. The correct way to find fundamental length scales is to linearise all of the e.o.m.\ simultaneously (the method for which we present below). In the isotropic case, the resulting linear e.o.m.\ decouple, leading to a single linear equation for the magnetic field $B$ (the London model) and some  linear matter equations in terms of $|\psi_\alpha|$, which match those from above. If this process is repeated for the anisotropic case however, the equations of motion will \emph{not} decouple into separate magnetic and matter equations and are in general mixed. It has been shown that this leads to mixed modes \cite{speight2021magnetic}, meaning the familiar London penetration depth and coherence lengths \emph{do not exist}. Hence in this paper, the London model will not describe the magnetic response of the system.

To find the linearised model we will first write our energy functional in terms of gauge invariant quantities. To achieve this we introduce a new gauge invariant vector field,  
\begin{equation}
p_i := A_i - \partial_i\theta_\Sigma,\qquad \theta_\Sigma:=\frac{1}{2}(\theta_1 + \theta_2),
\end{equation}
which is well defined wherever $\rho_1$ and $\rho_2$ are both nonzero. Since the aim is to describe the system in regions where the condensates are close to their (nonzero) ground state values, this restriction is not problematic. Note that in the isotropic model $p_i$ becomes proportional to the supercurrent. Since $p$ differs from $A$ by a gradient, its curl is still the magnetic field,  $B_k = \varepsilon_{ijk}\partial_i p_j$. This gives us the minimal set of gauge invariant quantities $(\rho_\alpha, \theta_\Delta, p_i)$ where
\begin{equation}
\theta_\Delta:=\frac12(\theta_1-\theta_2).
\end{equation}
The condensates may then be conveniently expressed as,
\begin{equation}
\psi_\alpha = \rho_\alpha e^{i(\theta_\Sigma+d_\alpha\theta_\Delta)},
\end{equation}
at the cost of defining the coefficients $d_\alpha=(-1)^{\alpha+1}$.

We assume that, far from any defect, the gauge invariant quantities decay to one of the possible ground state values $(\rho_\alpha, \theta_{\Delta},p_i) \to (u_\alpha, \theta_0, 0)$.  Note that $\theta_0 =0$ or $\pi/2$ in the phase (anti)locked case and $\theta_0 = \pm \pi/4$ for $s+is$, $s+id$ and $p+ip$ materials, which breaks time reversal symmetry. This is because we have defined $\theta_\Delta$ to be {\em half} the phase difference $\theta_{12}$. However, in the nematic case that we consider in this paper, $F_p$ is independent of $\theta_{12}$ and hence $\theta_0$ is a parameter of the model, related to the chosen orientation through the symmetry in \ref{eq:ariel}. Defining the quantities,
\begin{equation}
\varepsilon_\alpha:=\rho_\alpha-u_\alpha,\qquad \vartheta:=\theta_\Delta-\theta_0,
\end{equation}
the system is close to the chosen ground state precisely when $\varepsilon_\alpha$, $\vartheta$ and $p_i$ are small. As these are small, we then assume that only linear terms contribute to the field equations, which we may derive by expanding the free energy up to quadratic terms in $(\varepsilon_\alpha,\vartheta,p_i)$ and considering its variation. It will be convenient to define the matrices,
\begin{equation}
\Q^{\alpha\beta}_{ij} := Q^{\alpha\beta}_{ij} \exp i \left( d_\beta - d_\alpha \right) \theta_0,
\label{eq:Kmat}
\end{equation}
which enjoy the same symmetry as the anisotropy matrices: $\Q^{\alpha \beta}_{ij} = \overline{\Q^{\beta\alpha}_{ji}}$.  Note that $\Q^{11}=Q^{11}$, $\Q^{22}=Q^{22}$, $\Q^{12}=e^{-2i\theta_0}Q^{12}$ and $\Q^{21}=e^{2i\theta_0}Q^{21}$, so passing from $Q$ to $\Q$ amounts to twisting the off-diagonal matrices by the ground state value of the phase difference. With this notation, the free energy density to quadratic order is
\begin{widetext}
\begin{eqnarray}
\nonumber \mathcal{E}_{lin} &=& \frac{1}{2} \Q^{\alpha\beta}_{ij}(\partial_i \varepsilon_\alpha + i u_\alpha (p_i - d_\alpha \partial_i \vartheta) )(\partial_j \varepsilon_\beta - i u_\beta (p_j - d_\beta \partial_j \vartheta))\\
& & + \frac{1}{4} (\partial_i p_j - \partial_j p_i) (\partial_i p_j - \partial_j p_i) + \frac{1}{2} \mathcal{H}_{\alpha\beta} \varepsilon_\alpha \varepsilon_\beta + \mathcal{H}_{\alpha 3} \varepsilon_\alpha \vartheta + \frac{1}{2} \mathcal{H}_{33} \vartheta^2,
\end{eqnarray}
\end{widetext}
where $\mathcal{H}_{ab}$ is the $3\times 3$ Hessian matrix of second partial derivatives of $F_P$ with respect to the variables $(\rho_1,\rho_2,\theta_\Delta)$ evaluated at the chosen ground state, $(u_1,u_2,\theta_0)$.
This leads to the linear equations of motion,
\begin{eqnarray}
\nonumber -\reK^{\alpha\beta}_{ij} \partial_i \partial_j \varepsilon_\beta - \imK^{\alpha\beta}_{ij} u_\beta (\partial_i p_j - d_\beta \partial_i \partial_j \vartheta )& &\\ + \mathcal{H}_{\alpha\beta} \varepsilon_\beta + \mathcal{H}_{\alpha 3} \vartheta =0 \label{eq:eom_eps}\\
\nonumber - \reK^{\alpha\beta}_{ij} u_\alpha u_\beta d_\alpha (d_\beta \partial_i \partial_j \vartheta - \partial_i p_j) \\+ \imK^{\alpha\beta}_{ij} u_\beta d_\beta \partial_i \partial_j \varepsilon_\alpha + \mathcal{H}_{3\alpha} \varepsilon_\alpha + \mathcal{H}_{33} \vartheta = 0 \label{eq:eom_phase}\\
\nonumber -\partial_j^2 p_i + \partial_i \partial_j p_j - \imK^{\alpha\beta}_{ij} u_\alpha \partial_j \varepsilon_\beta \\+ \reK^{\alpha\beta}_{ij} u_\alpha u_\beta (p_j - d_\beta \partial_j \vartheta) =0, \label{eq:eom_p}
\end{eqnarray}
where $\reK$ and $\imK$ denote the real and imaginary parts of $\Q$.
From \eqref{eq:eom_p}, or by direct calculation, we may deduce that the total supercurrent, to linear order in small quantities, is
\begin{equation}
J_i = \imK^{\alpha\beta}_{ij} u_\alpha \partial_j \varepsilon_\beta - \reK^{\alpha\beta}_{ij} u_\alpha u_\beta (p_j - d_\beta \partial_j \vartheta ).
\end{equation}
We note that the coupling of the equations depends critically on whether $\imK$ is nonzero, and that this may happen even if the original $Q$ matrices are purely real if the ground state has complex phase difference (meaning $\theta_{12}\neq 0,\pi$).  

The linearized field equations are, in general, anisotropic, so the length scales describing decay from a localized defect to the ground state depend on the spatial direction along which decay occurs. To analyze this, we choose and fix a direction $\nvec$ in physical space and then impose on \eqref{eq:eom_eps}, \eqref{eq:eom_phase}, \eqref{eq:eom_p} the ansatz that $\eps_\alpha$, $\vartheta$ and $p_i$ are translation invariant orthogonal to $\nvec$. 

In practice, the most convenient way to implement this ansatz is to rotate to a new coordinate system $(x_1,x_2,x_3)$, such that the $x_1$ axis is aligned with our chosen direction $\nvec$. We then seek solutions which are independent of $(x_2,x_3)$. 
This amounts to choosing an $SO(3)$ matrix $R$ whose columns are the chosen orthonormal basis $(\hat{x}_1, \hat{x}_2, \hat{x}_3)$ and then transforming the matrices $Q^{\alpha\beta}$ according to the rule
\begin{equation}\label{eq:trid}
Q^{\alpha\beta}\mapsto R^TQ^{\alpha\beta}R.
\end{equation}
Note that the phase-twisted anisotropy matrices $\Q^{\alpha\beta}$ and their real and imaginary parts $\reK^{\alpha\beta},\imK^{\alpha\beta}$ also transform in the same way.

Having rotated our coordinate system and imposed the ansatz that $\eps_\alpha$, $\vartheta$ and $p_i$ depend only on $x_1$, the linearized field equations \eqref{eq:eom_eps}, \eqref{eq:eom_phase}, \eqref{eq:eom_p} reduce to a coupled linear system of ordinary differential equations for 
\begin{equation}
\vec{w}(x_1)=(\eps_1(x_1),\eps_2(x_1),\theta_\Delta(x_1),p_1(x_1),p_2(x_1),p_3(x_1))
\end{equation}
which may be economically written
\begin{equation}
\AA \frac{d^2\vec{w}}{dx_1^2} + \BB \frac{d\vec{w}}{dx_1} + \CC\vec{w} = 0,
\label{Eq:polynomial}
\end{equation}
where $\AA, \BB, \CC$ are the real $6 \times 6$ matrices
\begin{eqnarray}
\AA &=& \left( \begin{array}{cc} a & 0\\ 0 & a' \end{array} \right),\\
a &:=& \left( \begin{array}{ccc} -\reK^{11}_{11} & -\reK^{12}_{11} & \imK^{1\beta}_{11} u_\beta d_\beta \\ -\reK^{21}_{11} & -\reK^{22}_{11} & \imK^{2\beta}_{11}u_\beta d_\beta \\ \imK^{1\beta}_{11} u_\beta d_\beta & \imK^{2\beta}_{11} u_\beta d_\beta & -\reK^{\alpha\beta}_{11}u_\alpha u_\beta d_\alpha d_\beta \end{array}\right),\\
a' &:=& {\rm diag}(0,-1,-1),\\
\BB &=& \left( \begin{array}{cc} 0 & b \\ -b^T & 0 \end{array} \right), \\
b &:=& \left( \begin{array}{ccc} -\imK^{1\beta}_{11} u_\beta & -\imK^{1\beta}_{12} u_\beta & -\imK^{1\beta}_{13} u_\beta \\ -\imK^{2\beta}_{11}u_\beta & - \imK^{2\beta}_{12} u_\beta & - \imK^{2\beta}_{13} u_\beta \\ \reK^{\alpha\beta}_{11} u_\alpha u_\beta d_\alpha & \reK^{\alpha\beta}_{12} u_\alpha u_\beta d_\alpha & \reK^{\alpha\beta}_{13} u_\alpha u_\beta d_\alpha \end{array} \right),\quad \\\CC &=& \left( \begin{array}{cc} \mathcal{H} & 0 \\ 0 & \left< \reK \right> \end{array} \right)\\
\left< \reK \right>_{ij} &:=& u_\alpha \reK^{\alpha\beta}_{ij} u_\beta.
\end{eqnarray}
Note that $\AA$ and $\CC$ are symmetric while $\BB$ is skew, and that all these matrices depend implicitly on the chosen direction $\nvec$ through the transformation \eqref{eq:trid}.
 
The linearised system of field equations \eqref{Eq:polynomial} describes how a system recovers from a perturbation in the $x_1$-direction, under the assumption of translation invariance orthogonal to $\hat{x}_1$, for example, how the system behaves near the boundary of a superconductor with normal $\hat{x_1}$, subject to an external magnetic field.
We seek solutions of the form
\begin{equation}\label{klepto}
\vec{w}(x_1)=\vec{v}e^{-\mu x_1}
\end{equation}
where $\vec{v} \in \mathbb{C}^6$ is a constant vector and ${\rm re}\mu>0$, so that all fields decay to their ground state values as $x_1\to\infty$. We interpret $\vec{v}$ as a normal mode of the system about the chosen ground state, $\mu$ as the associated field mass, and $\lambda=1/\mu$ as the associated length scale. Given such a solution, let $\vec{z}=-\mu\vec{v}$. Then $(\vec{v},\vec{z})$ is a solution of the linear equation,
\begin{equation}
\Omega \left(\begin{array}{c} \vvec \\ \zvec \end{array}\right) = \mu  \Xi\left(\begin{array}{c} \vvec \\ \zvec \end{array}\right),
\label{eq:geneigen}
\end{equation}
where $\Omega$ and $\Xi$ are $12\times 12$ matrices,
\begin{equation}
    \Omega := \left( \begin{array}{cc} \mathcal{B} & \mathcal{A} \\ -I_6 & 0 \end{array}\right), \quad \Xi := \left( \begin{array}{cc} \mathcal{C} & 0 \\ 0 & I_6 \end{array}\right).
\end{equation}
If $\mathcal{C}$ is invertible (as assumed in \cite{speight2021magnetic}), then $\Xi$ is invertible allowing \eqref{eq:geneigen} to be written as an eigenvalue problem. However, if the potential $F_p$ is independent of $\theta_{12}$ ($\mathcal{H}_{3 \alpha} = \mathcal{H}_{33} = 0$), as for the nematic potential in \eqref{eq:Fp}, then $\mathcal{C}$ is singular. Hence, we must solve the generalised eigenvalue problem in \eqref{eq:geneigen}.

Given an eigenvector $(\vec{v},\vec{z})$ of \eqref{eq:geneigen} corresponding to a nonzero eigenvalue $1/\mu$, $\vec{z}=-\mu\vec{z}$ and \eqref{klepto} is a solution of \eqref{Eq:polynomial}. We conclude, therefore, that the length scales associated will decay to the ground state in the fixed direction $x_1$ are those eigenvalues with positive real part. Such eigenvalues are solutions of the degree 12 polynomial equation
\begin{equation}
\det\left(\AA - \lambda \BB + \lambda^2\CC\right) = 0.
\label{Eq:kernel}
\end{equation}
It follows from the symmetry properties of $\AA,\BB,\CC$
that \eqref{Eq:kernel} is actually a real degree 6 polynomial equation in $\lambda^2$, so if $\lambda$ is a solution, so are
$-\lambda,\ol\lambda$ and $-\ol\lambda$. Note that $0$ is an eigenvalue of $\Omega$ of algebraic multiplicity $2$ with eigenvector $(0,\ldots,0,1,0,0)$. This should be discarded as it does not correspond to a solution of \eqref{Eq:polynomial}. Of the remaining 10 eigenvalues, in general 5 have positive real part: these are the 5 length scales we seek. Let us order them by decreasing real part $\lambda_1,\lambda_2,\ldots,\lambda_5$. We call $\vec{v}_1$, the mode corresponding to the longest length scale $\lambda_1$, the {\em dominant mode} since, generically, at large $x_1$, this will dominate the solution of \eqref{Eq:polynomial}. Depending on the details of the defect being studied, it may be, however, that the dominant mode is unexcited, so subleading modes $\vec{v}_2,\vec{v_3},\ldots$ may still be phenomenologically important. 

It is important to note that we have retained all three components for $p_i$, as it was shown in \cite{speight2021magnetic} that reducing to a single component does not, in general, yield a solution of the equations of motion.

Note that in the case of a more conventional multicomponent superconductor, where $Q^{12} = 0$ and $Q^{11} = Q^{22}$ is real, the linear equations decouple into the London model for $p_i$, governing the magnetic response, and three matter equations for $(\eps_1, \eps_2, \theta_\Delta)$. This is handled by the usual approach of taking the London limit (see e.g.\ \cite{svistunov2015superfluid}), leading to a single magnetic field penetration length and multiple coherence lengths, each associated with different linear combinations of density fields \cite{babaev2010type,carlstrom2011type}.

The less restrictive case, where $Q^{12} = 0$ and $Q^{11}$, $Q^{22}$ are real was considered in \cite{silaev2018non,winyard2019hierarchies,winyard2019skyrmion}, in general leading to a modified London model, a pair for $(p_i, \theta_\Delta)$, and two matter equations, a pair for $(\eps_1, \eps_2)$. This gives the familiar multiple coherence lengths, but also multiple magnetic field penetration lengths, due to the hybridization of the matter modes and magnetic modes.

In general (and for nematic systems), the linear equations are all coupled, leading to modes that are linear combinations of all physical quantities. The implication of this is that the system can no longer be characterised by a distinct magnetic field (London) penetration depth and coherence length. Instead, the magnetic and density modes are mixed and one 
should construct a linear combination of the density and gauge fields to find the normal modes. In other words, the magnetic field decay will have several length scales that are shared with the matter fields.

Restricting to the specific case of nematic superconductors, $F_p$ is independent of $\theta_{12}$ leading to an additional two zero modes. Hence, for nematic superconductors we always have exactly four decaying modes, four equivalent growing modes and four massless (or zero) modes. The two additional zero modes point purely in the phase difference $\theta_\Delta$ direction, this mode is massless due to the absence of $\theta_{12}$ in $F_p$ in \eqref{eq:Fp}. Hence for the linearisation to be valid for a given excitation, one must be careful to match the value of $\delta_{12}$ to the value the phase difference $\theta_{12}$ decays to as $x_1 \rightarrow \infty$, where changing $\delta_{12}$ is equivalent to rotating the system around the $z$-axis, due to the symmetry in \eqref{eq:conjsym}.

Hence, in the nematic case we have four decaying modes and can rewrite the key linear quantities as,
\begin{align}
\nonumber \eps_\alpha &= \sum_{i=1}^4 c_i\vec{v}^\alpha_ie^{-\mu_i x_1},\\ B_{lin} &= \sum_{i=1}^4 -\mu_i c_i (0, \vec{v}_i^4, -\vec{v}_i^5) e^{-\mu_i x_1},
\end{align}
where $c_i$ are constants and $B_{lin}$ is the linear magnetic field. We have plotted the modes and length scales for $\delta_{12} = 0$ and various values of $\eta$ in the basal plane ($\hat{x}_1 = (\cos \theta, \sin \theta,0)$) in \figref{Fig:baselLengthScales} and in a tilted plane ($\hat{x}_1 = (\cos \theta,\sin \theta,1)/\sqrt{2}$) in \figref{Fig:halfLengthScales}. These plots have regions where a subset of the length scales have non-zero imaginary part (dashed lines), as well as mixed modes (discussed in the next section).

The non-zero imaginary part is due to the masses $\mu_i=1/\lambda_i$ associated with the mixed modes $\vec{v}_i$ being complex. This leads to the fields oscillating as they decay. Note that the period of such oscillations is large in comparison to the decay rate of the modes.

\subsection{Mixed modes}
In an isotropic multi-component superconductor, the normal modes $v_i$ are separated into matter modes: those associated with the coherence length or the modulus of the order parameters, and magnetic modes: those associated with the magnetic penetration depth or the massive vector field leading to the magnetic field. Our analysis reproduces these separate real length scales (coherence length and magnetic penetration depth) in the isotropic limit $Q^{\alpha\beta}_{ij}=\delta_{\alpha\beta}\delta_{ij}$.

Away from the isotropic limit, and in particular for the case of nematic superconductors, the normal modes are associated with linear combinations of magnetic and matter degrees of freedom. Hence, we should consider all excitations of our system in terms of these \emph{mixed modes} $\vec{v}_i$ and their corresponding length scales $\lambda_i$. This leads to important physical consequences, e.g.\ a density excitation can only be excited through coupled modes and hence induces magnetic field fluctuations. This leads to excitations that in the isotropic model would feature purely excited matter fields, such as domain walls and defects, spontaneously inducing localised magnetic field, as seen in \cite{Andrea,speight2021magnetic} and noted in \cite{comment}.

In addition, if we apply an external field $H$, such as for the Meissner state or vortices, if the magnetic component of the excited coupled modes is not parallel with $H$, the induced magnetic field will exhibit local twisting.  For example, take the Meissner state where the boundary conditions ensure $B=H$, but the excited linear modes (which dominate far from the boundary) have the magnetic field pointing in a different direction.  Hence, as $B$ decays it will twist direction, away from $H$ on the boundary,  to align with the excited mode with longest length scale.  A similar effect will happen for vortices (by a similar argument), where $B$ twists as it decays spatially from the centre of the vortex line.  We will refer to excited magnetic field orthogonal to the vortex line direction or applied field $H$ as spontaneous magnetic field, as without mixed modes these would not be excited.

It is useful to have a measure of how mixed a given mode is. By ignoring the $p_1$ contribution, we can achieve this by considering a general mode as a vector in a 5-dimensional space $(\eps_1, \eps_2, \vartheta, p_2, p_3)$, where we define the quantity $\theta^i_m$ as the mixing angle of the $i$th mode,
\begin{equation}
\cos \theta^i_m = \underbrace{\sqrt{|v^1_i|^2 + |v^2_i|^2 + |v^3_i|^2}}_{\text{matter modes}}, \quad \sin \theta^i_m = \underbrace{\sqrt{|v^4_i|^2 + |v^5_i|^2}}_{\text{magnetic modes}}.
\label{Eq:mixing}
\end{equation}
Conceptually, the mixing angle is then the angle that the 5-dimensional vector makes with the two regions in this space representing purely matter and purely magnetic modes.

This allows us to classify each mode as either purely matter ($\theta^i_m = n\pi$), purely magnetic ($\theta^i_m = \pi/2 + n\pi$) or mixed ($\theta^i_m \neq n\pi/2$), where $n\in \mathbb{Z}$. The angle can be used as a numerical value that determines the strength of the mixing. 

\begin{figure*}
\begin{subfigure}[b]{\textwidth}
\includegraphics[width=0.3\linewidth]{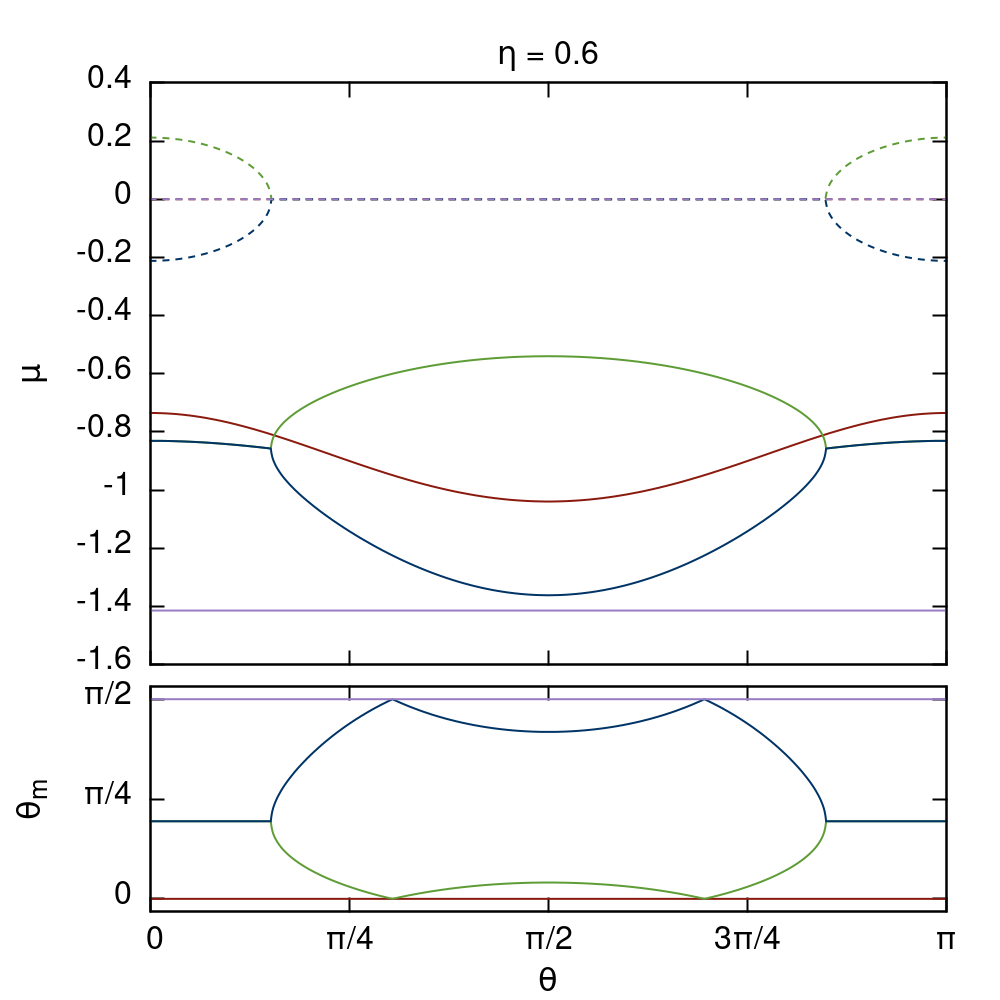}\includegraphics[width=0.3\linewidth]{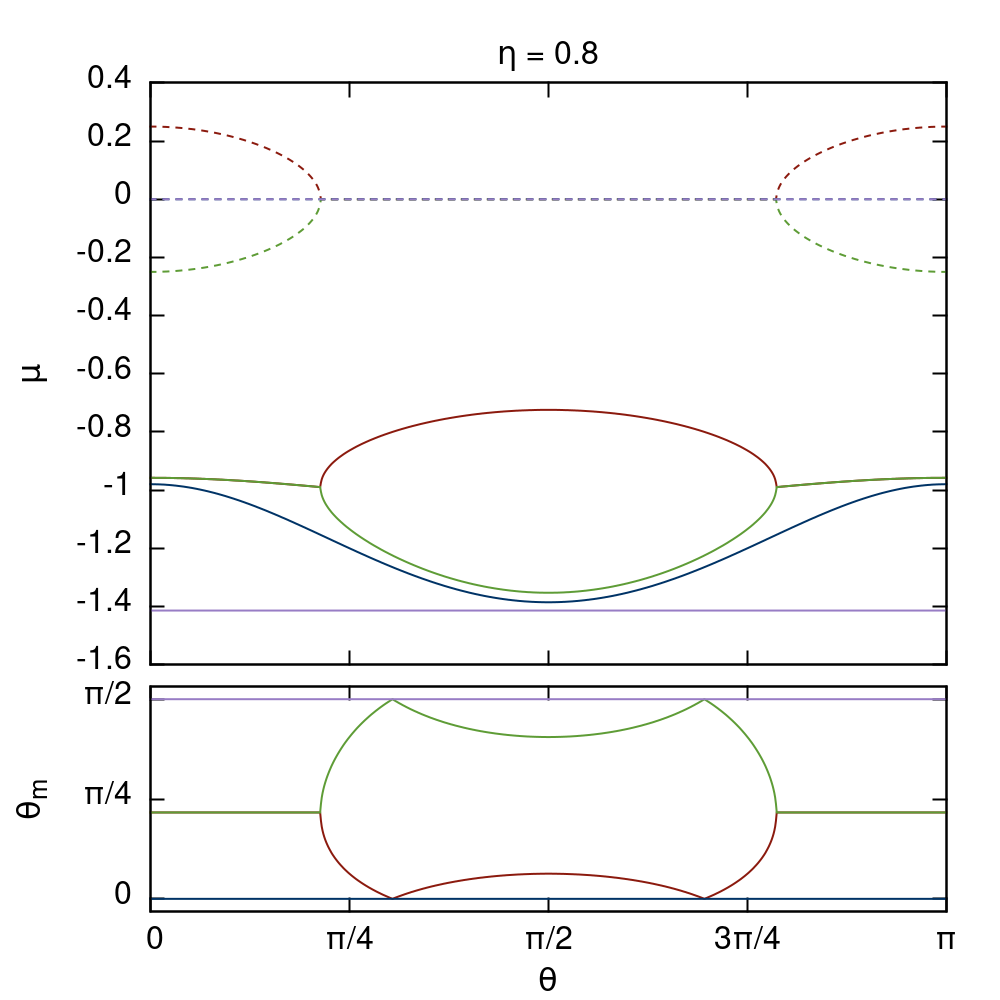}\includegraphics[width=0.3\linewidth]{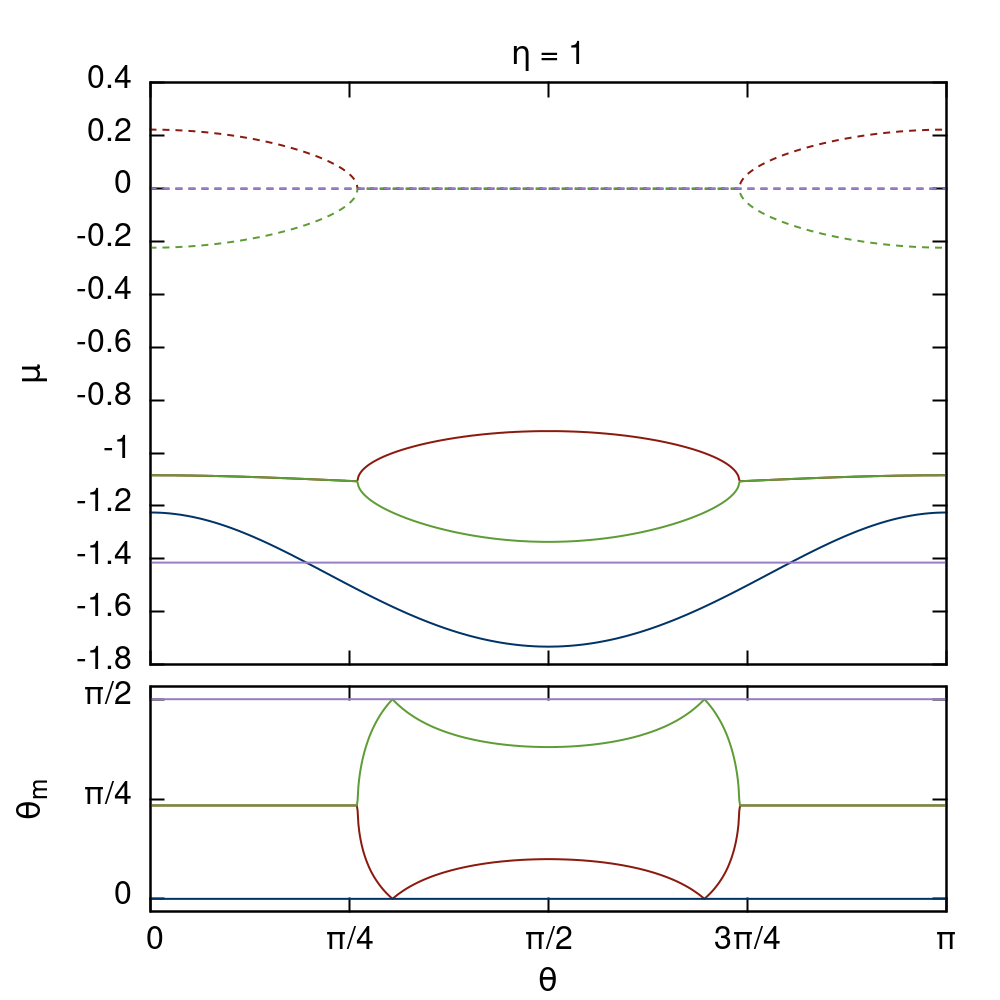}
\end{subfigure}
\begin{subfigure}{0.6\textwidth}
\includegraphics[width=0.5\linewidth]{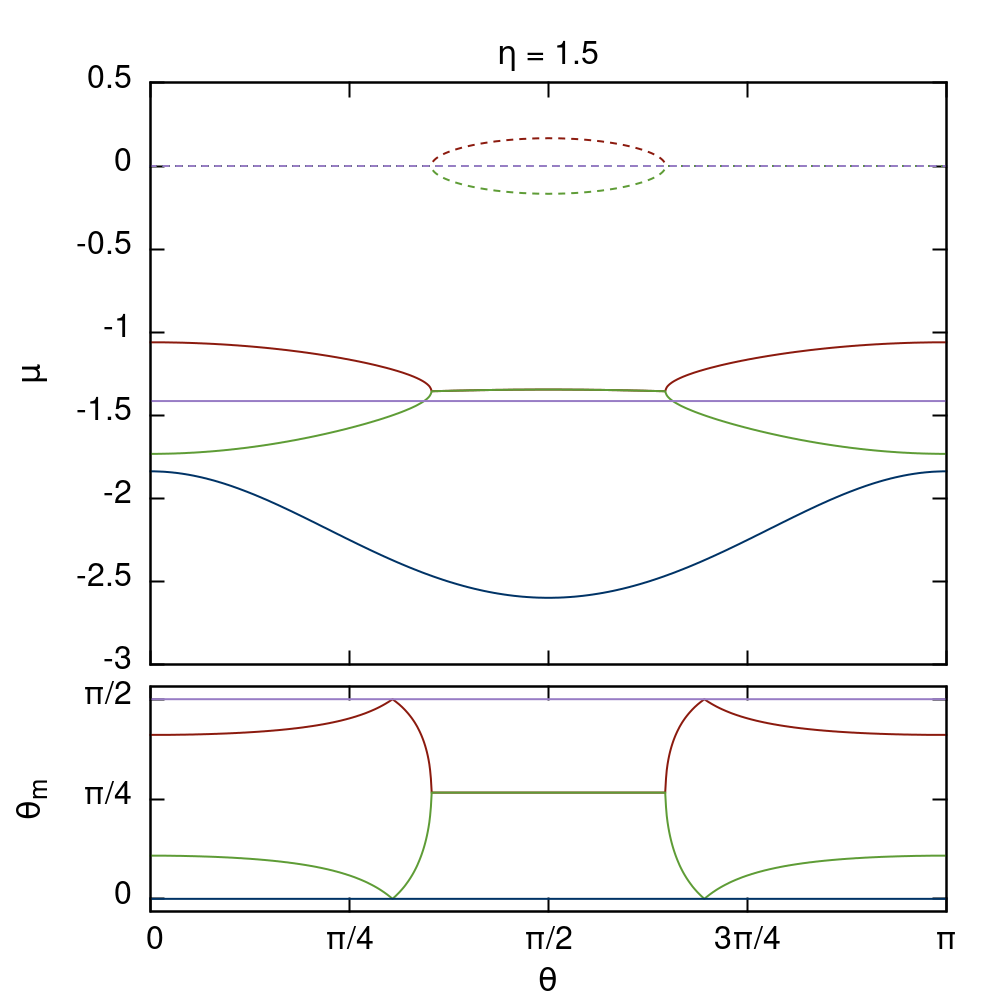}\includegraphics[width=0.5\linewidth]{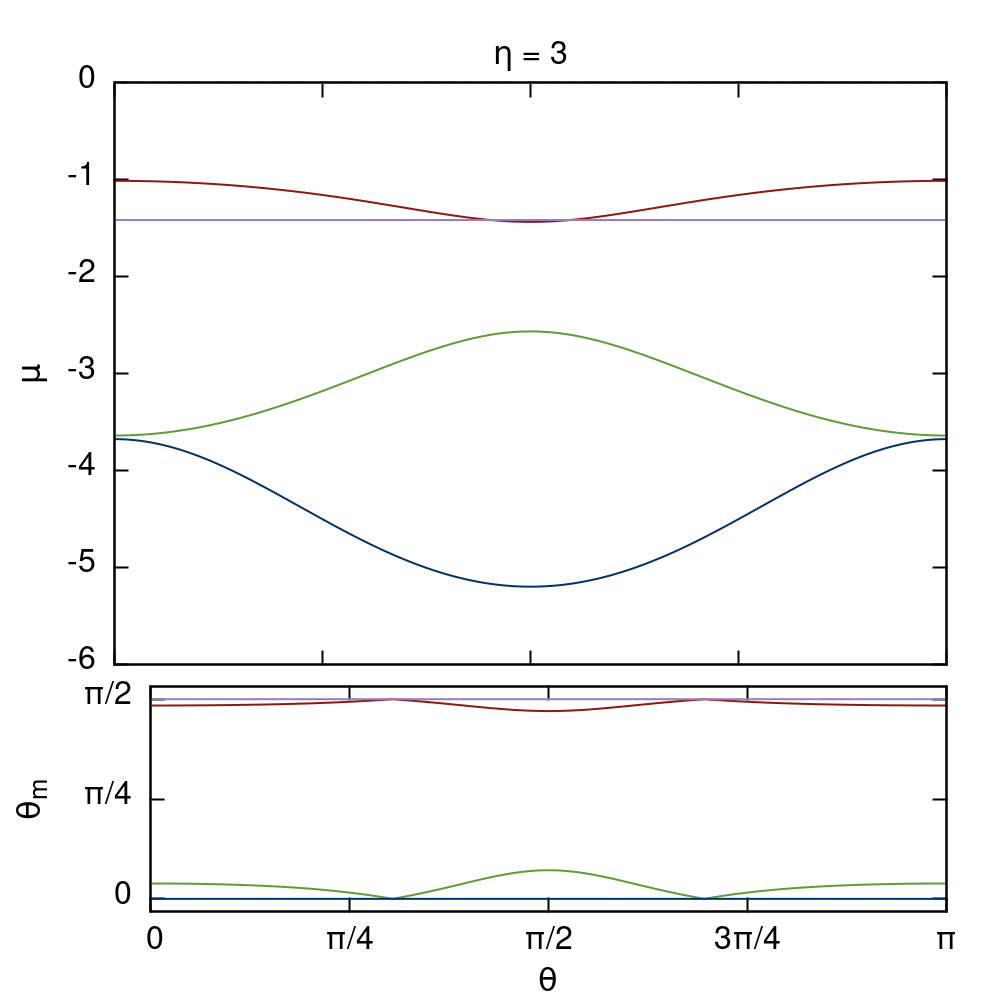}
\end{subfigure}
\caption{\label{Fig:baselLengthScales}
Plot of the linear modes and length scales in the basal plane with spatial dependence $\hat{x}_1 = (\cos\theta,\sin\theta,0)$, for successive values of $\eta$ and $\delta_{12} = 0$. Note that rotations $\theta$ are equivalent to rotating $\delta_{12}$ by $2\theta$ instead, due to \eqref{eq:conjsym}. The top panels plot the masses (inverse length scales) $\mu_i = \frac{1}{\lambda_i}$. Each $\mu_i$ is a different colour, with the real part plotted as a solid line and the imaginary part as a dashed line of the same colour. The bottom panel plots the mixing angle $\theta^i_m$ of each mode, where the colours of the modes match the colours of the corresponding mass above. The dominant mode is the highest solid line in the top plot and maximum coupling occurs at $\pi/4$ for the bottom plot. So for example for small $\theta$, $\eta=0.8$ the dominant mode exhibits strong mixing.}
\end{figure*}

\begin{figure*}
\begin{subfigure}[b]{\textwidth}
\includegraphics[width=0.3\linewidth]{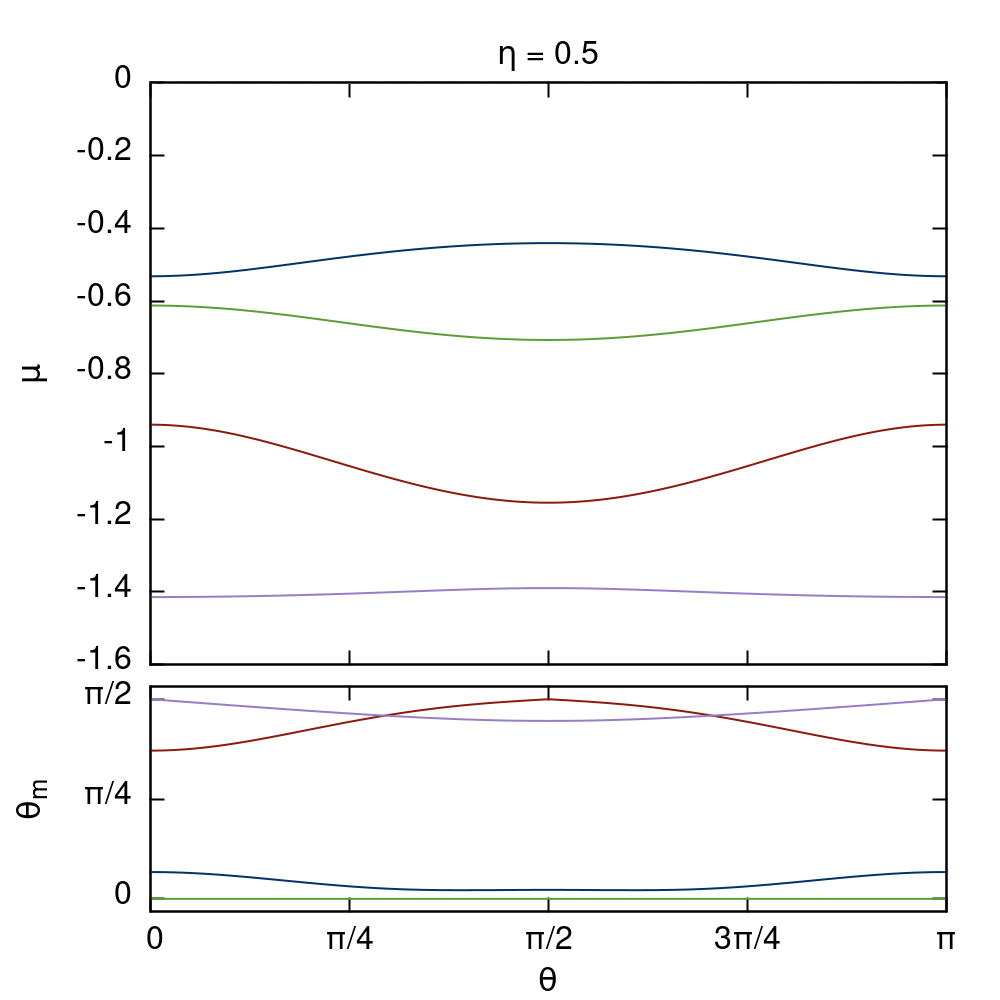}\includegraphics[width=0.3\linewidth]{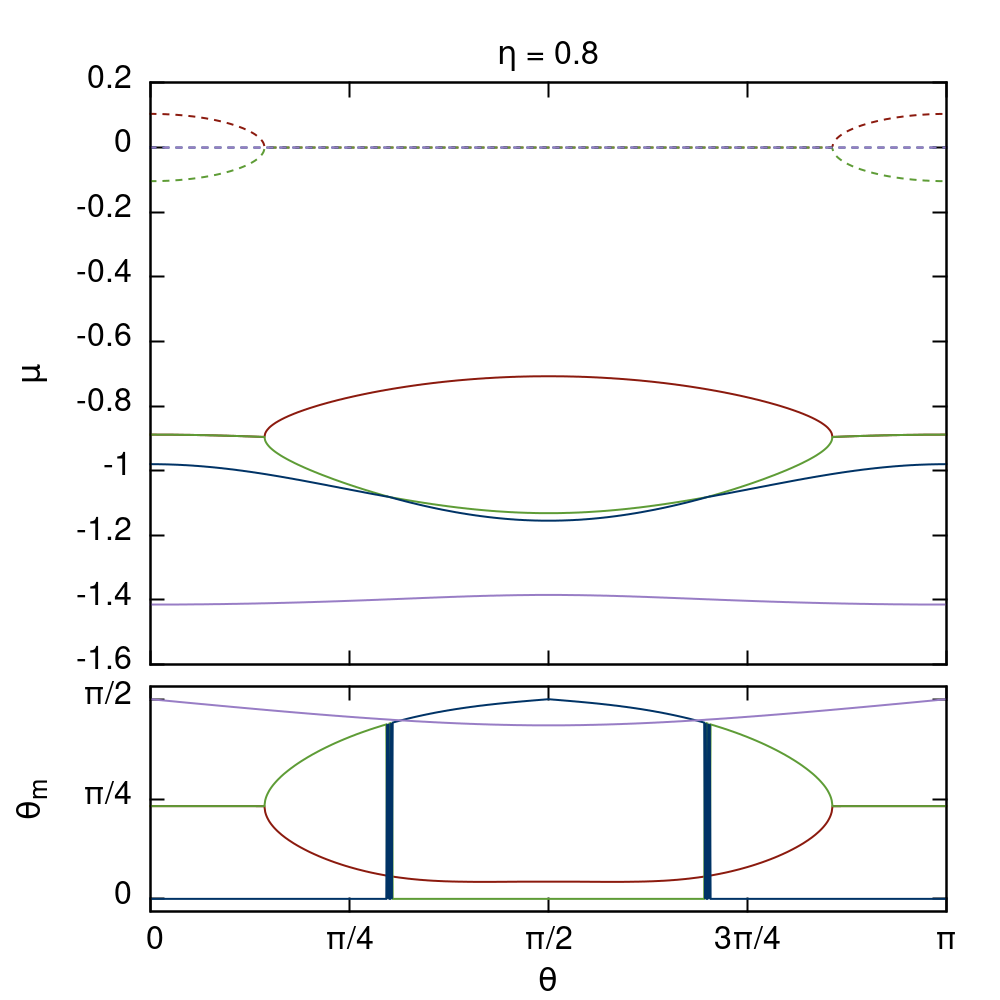}\includegraphics[width=0.3\linewidth]{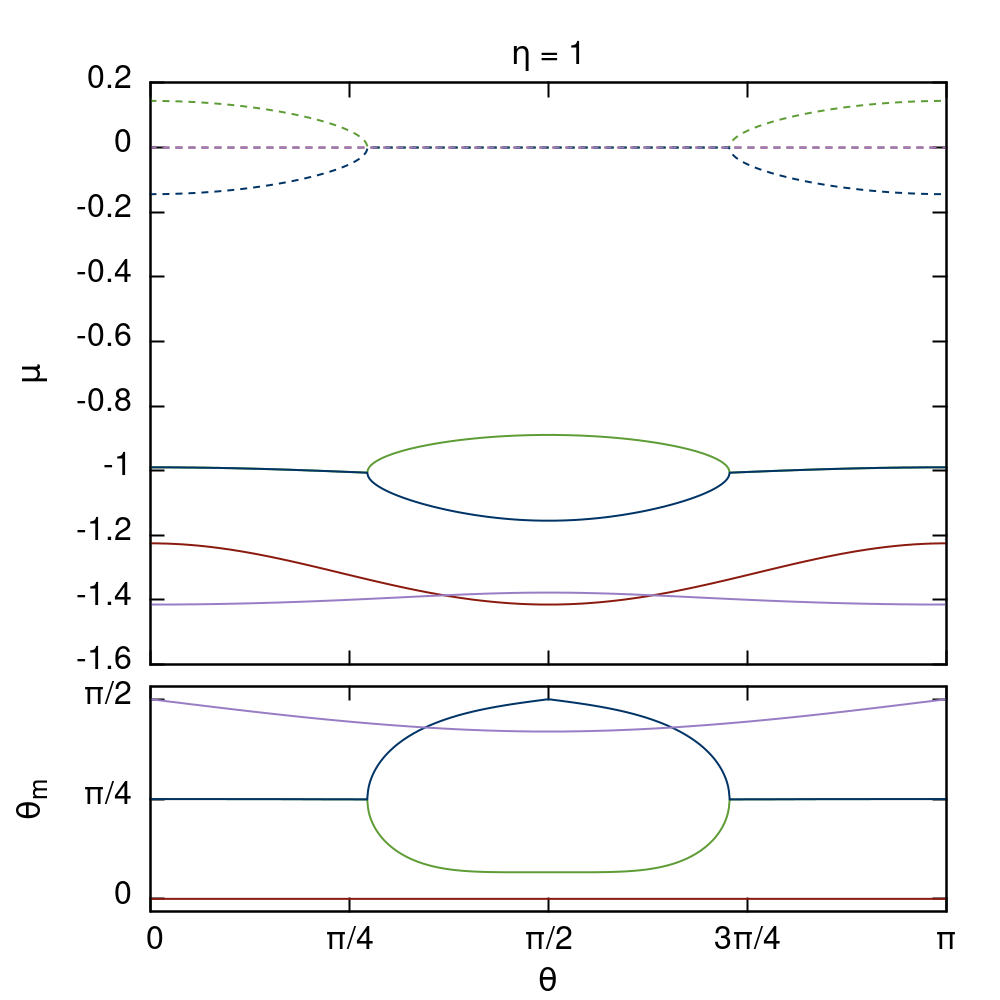}
\end{subfigure}
\begin{subfigure}[b]{0.6\textwidth}
\includegraphics[width=0.5\linewidth]{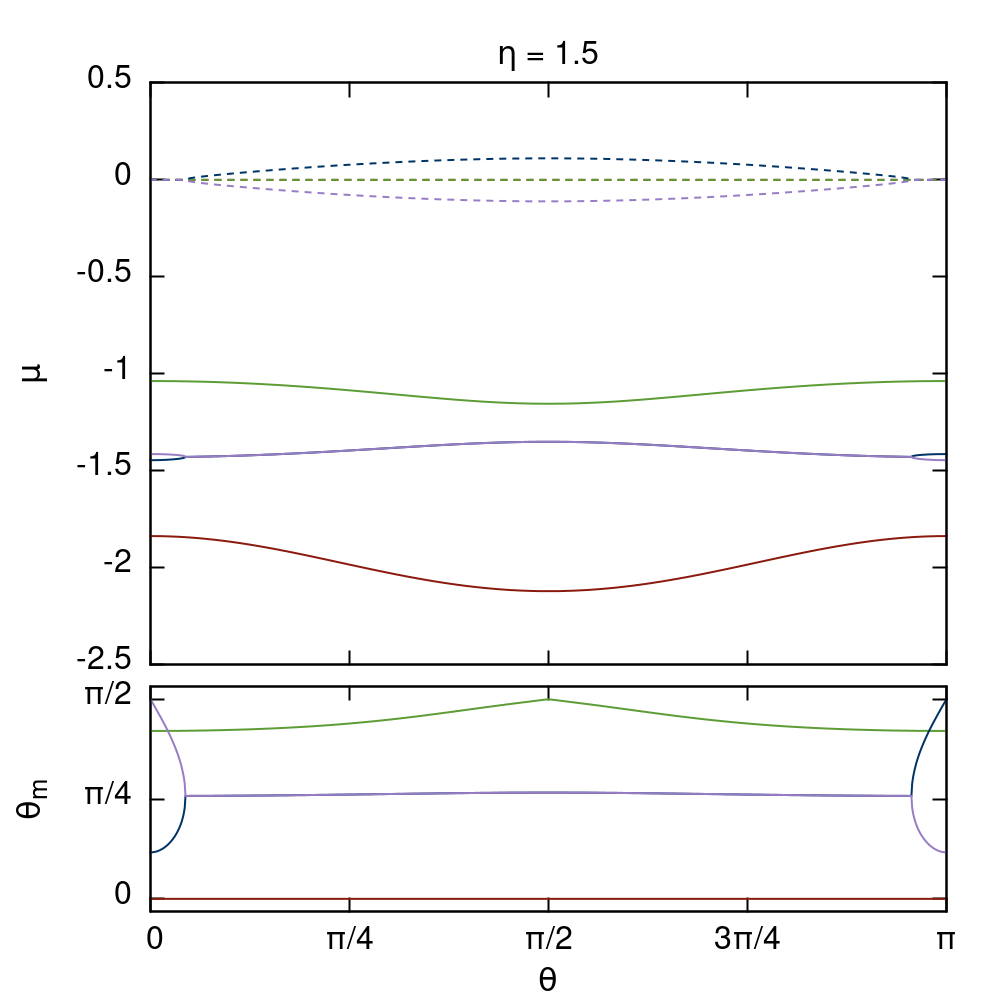}\includegraphics[width=0.5\linewidth]{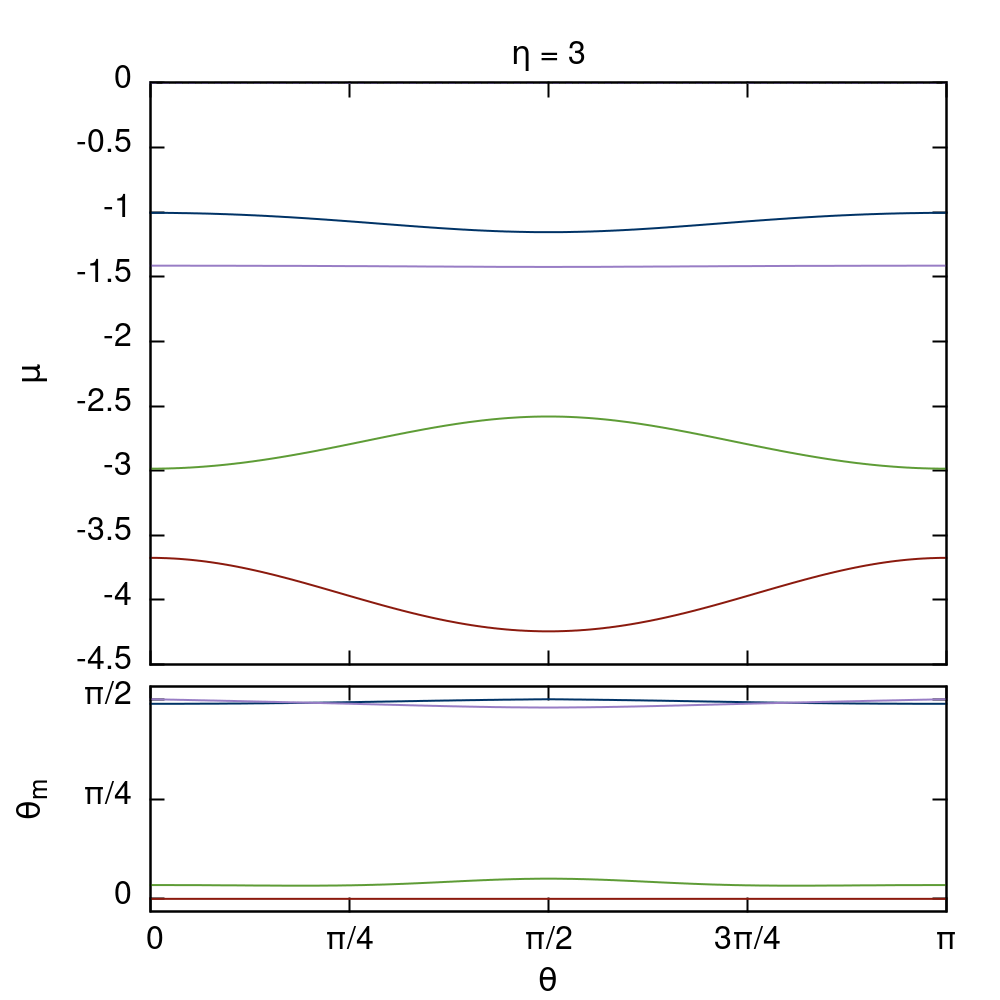}
\end{subfigure}
\caption{\label{Fig:halfLengthScales}
Plot of the linear modes and length scales on a half plane with spatial dependence $\hat{x}_1 = (\cos\theta,\sin\theta,1)/\sqrt{2}$, for successive values of $\eta$ and $\delta_{12} = 0$. Note that rotations $\theta$ are equivalent to rotating $\delta_{12}$ by $2\theta$ instead, due to \eqref{eq:conjsym}. The top panels plot the masses (inverse length scales) $\mu_i = \frac{1}{\lambda_i}$. Each $\mu_i$ is a different colour, with the real part plotted as a solid line and the imaginary part as a dashed line of the same colour. The bottom panel plots the mixing angle $\theta^i_m$ of each mode, where the colours of the modes match the colours of the corresponding mass above.}
\end{figure*}

We can see a plot of this quantity for the basal plane in \figref{Fig:baselLengthScales} and away from the basal plane in \figref{Fig:halfLengthScales}. The lower panels of these plots show the mixing angle of each mode. Note that the modes are particularly mixed in the complex regions of the plot. This is due to decaying length scales appearing in complex conjugate pairs with equivalent modes. This means two modes must have equivalent decay coefficients in the complex region, bringing together two otherwise mostly matter and mostly magnetic modes.

\subsection{Leading modes}

The long-range behaviour of each field, in the $x_1$-direction, is governed by the eigenvector $\vec{v}_1$ whose eigenvalue $\mu_1$ has smallest positive real part. If this eigenvalue is complex we would expect the fields to oscillate as they decay. However, the period of the oscillations is long in comparison to the decay. In addition, the oscillations are dependent on the choice of $\delta_{12}$. These two complications ultimately lead to oscillations being unobservable in practise for the parameters we consider in the full non-linear model.

\begin{figure*}
\includegraphics[width=1.0\linewidth]{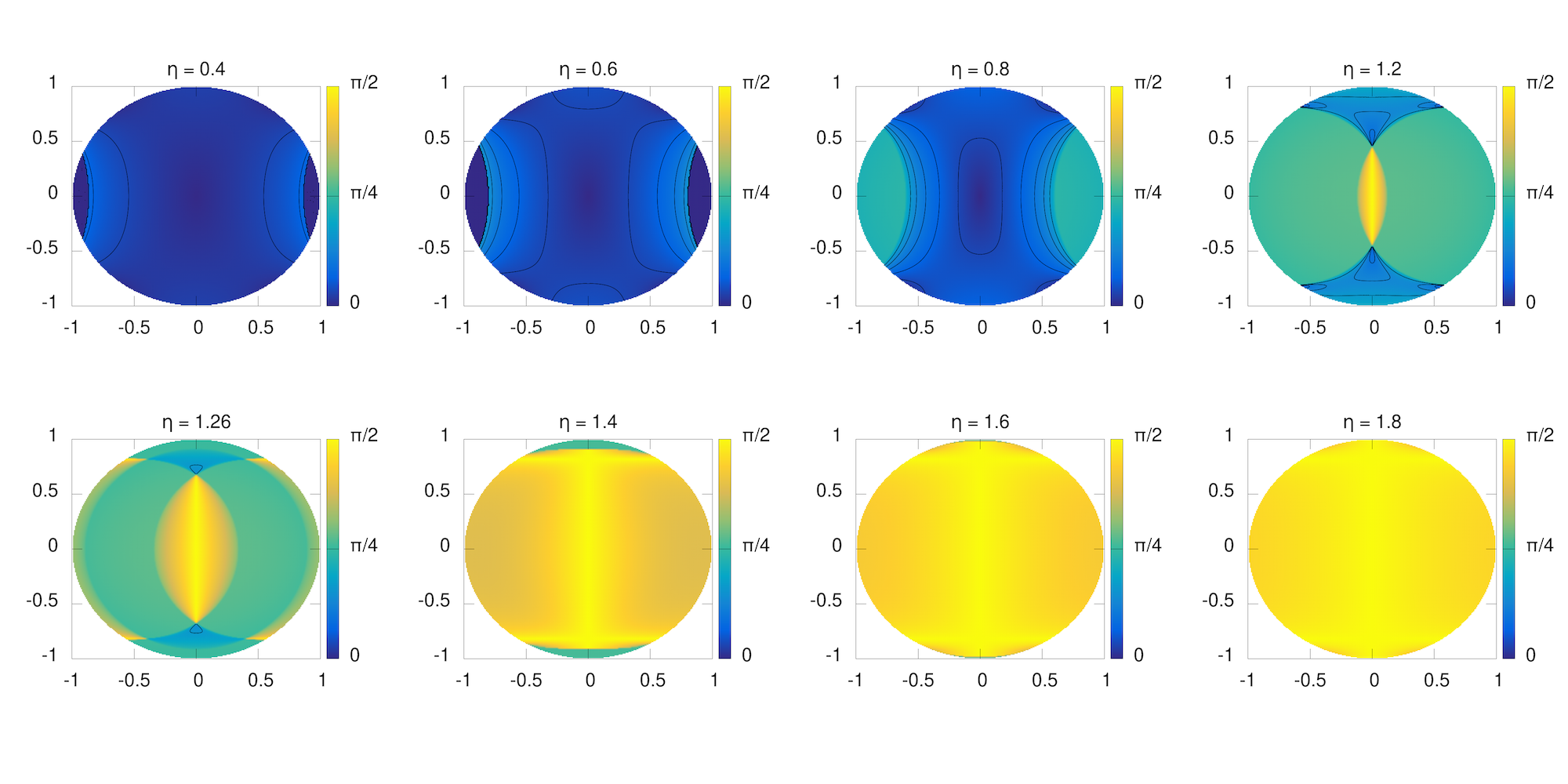}
\caption{\label{Fig:mixing}
Contour plots of the mixing angle $\theta^1_m$ given in \eqref{Eq:mixing}, corresponding to the leading mode, for $\delta_{12} = 0$. The position corresponds to the $(x,y)$ components of the normal unit 3-vector $\hat{x}_1 \in S^2$, where we have projected the northern hemisphere of $S^2$ to a flat disc. Due to reflection symmetry the southern hemisphere will match. Note that changing $\delta_{12}$ is equivalent to rotating the plot around the origin by $\delta_{12}/2$.}
\end{figure*}

For nematic superconductors it is more interesting to consider the mixing angle of the leading mode. The mixing angle $\theta_m^1$ of the leading length scale is plotted in \figref{Fig:mixing}. 
For the nematic model we can see in \figref{Fig:mixing} that for small $\eta$ matter modes dominate and for large $\eta$ magnetic modes dominate. However for $\eta \approx 1$ we observe that the leading mode is highly coupled for certain orientations. Hence, we would expect extreme orientation dependence for the interactions for these parameters.

Finally, the dominant eigenvalue will determine the direction of the magnetic field at long-range. If there is a disparity between the field direction of the non-linear part of the defect (for example the direction of external field for a Meissner state), then the magnetic field will exhibit twisting as it decays from $x_1=0$ (non-linear dominated) to $x_1 \rightarrow \infty$ (linear dominated). Note that this assumes that the leading mode is excited by the defect. If we again return to the basal plane, the purple mode in each plot is a purely magnetic mode, corresponding to in-plane magnetic field in the direction $\hat{x}_2 = (-\sin\theta, \cos\theta, 0)$. Hence if we apply an external magnetic field in the orthogonal direction to this (in the $\hat{x}_3$-direction), this mode will never be excited, meaning a different linear mode must dominate at long range.

\section{Meissner state}
We turn now to the effect of applying a weak external field $H$ to a nematic superconductor, requiring us to solve the full nonlinear equations of motion in \eqref{eq:eom}. We will model a superconductor/insulator boundary as a semi-infinite superconductor $\Omega$ occupying the half space $x_1 \geq 0$, where $\hat{x}_1$ is the inward pointing normal. An external magnetic field $H = H_0 \hat{x}_3$, orthogonal to the boundary normal ($\hat{x}_1 \cdot \hat{x}_3 = 0$) is applied. This excites the superconducting fields, that decay orthogonally from the boundary into the bulk of the system, dimensionally reducing the problem to a 1-dimensional variational problem on $x_1\in [0,\infty)$.

We first perform a transformation of coordinates from the crystalline basis $(x,y,z)$ to the excitation basis $(x_1,x_2,x_3)$. This coordinate transformation is performed by transforming the anisotropy matrices in \eqref{eq:F} according to \eqref{eq:trid}. We assume our fields are only spatially dependent on $x_1$ and then seek minimizers of the Gibbs free energy $G$ within $\Omega$,
\begin{equation}
G = \int_\Omega \mathcal{F} - H \int_\Omega B. 
\label{eq:G}
\end{equation}
Note that $\int_\Omega B$ is quantised up to a surface term as it is equivalent to the first Chern number (winding number), multiplied by a constant. Hence, the equations of motion are independent of $H$ in the bulk of the superconductor, which only has an effect on the boundary of the system. These boundary conditions are a result of $x_1<0$ being an insulator and are given in the Appendix. As we focus on the long range behaviour of the fields, we will ignore additional boundary terms that  result from modification of the pairing near the surface \cite{samoilenka2020boundary,samoilenka2020microscopic}.

While we have assumed that the fields are invariant in $(x_2,x_3)$ directions, we cannot make the standard assumption that $A_3 = 0$. While this would hold in an isotropic model, in an anisotropic model this would imply $B_2 = 0$, which due to non-trivial coupling of modes would not in general be a solution of the equations of motion. This requirement suggests that the magnetic field will not always be in a fixed direction, but may twist as it penetrates the material.

The set of parameters that defines a given solution comprises the normal vector $\hat{x}_1$, the external magnetic field $H = H_0 \hat{x}_3$, the chosen parameter value for $\eta$ and finally the chosen phase difference $\delta_{12}$ on the right boundary in the bulk ($x_1>>0$). Note that the choice of phase difference on the boundary is not physical, as due to $\theta_{12}$ being massless, there is no restriction on the value it takes. Hence, one should minimise over the parameter $\delta_{12}$. Note that if $\Omega$ were truely infinite, $\delta_{12}$ would have no effect on the energy, as the massless decay mode for $\theta_{12}$ would allow it to decay linearly at no energy penalty over the infinite distance.

Finally, we must ensure that the external field strength $H_0 < H_{c_1},H_{c_2}$. Namely, the external field should be weak enough such that vortices or the normal state are not energetically preferred. This is the case for all our simulations; we will discuss how to explicitly calculate $H_{c_1}$ and $H_{c_2}$ in a later section.

\subsection{Meissner state solutions}
All Meissner state solutions were found by choosing the parameters $(\hat{x}_1, \hat{x}_3, H_0, \eta)$, then transforming the anisotropy matrices as described in \eqref{eq:trid} according to our chosen orthonormal basis $\{\hat{x}_1, \hat{x}_2, \hat{x}_3\}$. We then solved the resulting 1-dimensional boundary problem (where $H$ purely fixes the boundary conditions), using a Newton flow algorithm \cite{speight2020skyrmions} which we describe briefly in appendix A.

\begin{figure*}
\includegraphics[width = 1.0\linewidth]{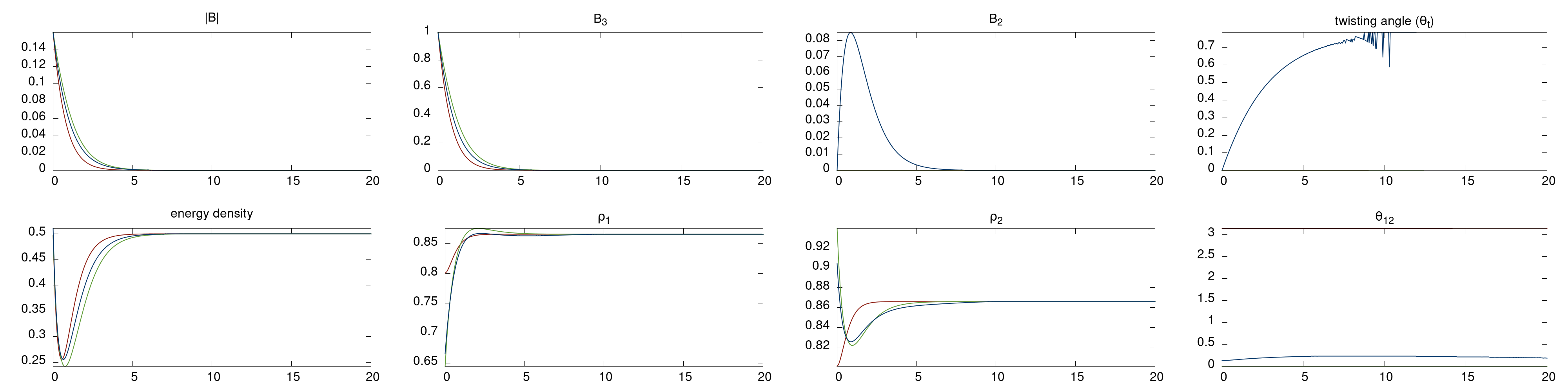}
 \caption{\label{Fig:MeisnerBasal1p6}
Plot of the Meissner state, with an external field of strength $H_0 = 1$ and $\eta = 1.6$. The applied field is applied in the three directions $\hat{x}_3 = (0,1,0)$ (red), $\hat{x}_3 = (0,0,1)$ (green) and $\hat{x}_3 = (0,1,1)/\sqrt{2}$ (blue). We have plotted the total magnetic field $|B|$, magnetic field parallel to the applied magnetic field $B_3$ and orthogonal $B_2$, as well as the angle between the total magnetic field and the applied field (twisting angle). Finally we have plotted the normalised free energy density, condensate magnitudes $\rho_1$ and $\rho_2$, and the phase difference $\theta_{12}$.}
\end{figure*}

We first present Meissner state solutions for the basal plane for $\eta = 1.6$ in \figref{Fig:MeisnerBasal1p6} with normal $\hat{x}_1 = (1,0,0)$ and $H_0 = 1.0$, for various directions $\hat{x}_3$ of applied magnetic field. For $\hat{x}_3 = (0,1,0)$, plotted in red, we find the minimal value for the phase difference on the boundary to be $\delta_{12} = \pi$. If we consider the corresponding length scales, plotted in \figref{Fig:baselLengthScales}, we would expect the leading length scale to be complex. However, the mixed eigenvalue (linear mode) couples $B_2$, $\rho_1$ and $\rho_2$ and is not excited. This can be seen in the solution as there are only two decoupled modes remaining, one purely magnetic for $B_3$ and one purely matter for $\rho_1 + \rho_2$. Hence the solution acts similarly to the isotropic GL model, where $\psi_1$ and $\psi_2$ have equivalent coherence length, which is exactly what is observed in the numerical solution.

If we then consider $\hat{x}_3 = (0,0,1)$, plotted in green, the minimal value for phase difference is $\delta_{12} = 0$, giving a coupled real mode that is excited. This coupled mode is mixed between $B_3$ and $\rho_1 - \rho_2$, so that a strong magnetic field implies a large disparity between the condensate magnitudes. We observe exactly this in the numerical solution, where $\rho_1$ is lower than its ground state value and $\rho_2$ is higher than its ground state value. In addition, another feature that often occurs in tandem with mode mixing is field inversion, which we observe in the condensates and magnetic field.

Finally for $\hat{x}_3 = (0,1,1)/\sqrt{2}$, plotted in blue, we also get $\delta_{12} = 0$, but with all modes excited. As the modes with magnetic component have different length scales, and hence decay at different rates, we observe magnetic field twisting. This can be observed locally as $|B_2|>0$, representing the magnetic field orthogonal to the direction of the applied magnetic field $\hat{B}_3$.  We can represent this local magnetic twisting by an angle,
\begin{equation}
\tan\theta_t = B_2/B_3,
\end{equation}
which we call the twisting angle.  Note that $\theta_t = 0$ means the magnetic field is aligned with $H$, while this choice is arbitrary this choice makes the most sense due to the boundary condition $B(0)=H$ or $\theta_t(0) = 0$. We have plotted this for the Meissner state in blue and see that it changes as the fields decay into the bulk of the material. The linearization predicts for $\delta_{12} = 0$ that the leading length scale corresponds to the mixed mode, with magnetic component in the direction of $B_3$. This suggests that the twisting angle will decay towards $\pi/4$, which matches the numerical result.

\begin{figure*}
\includegraphics[width = 1.0\linewidth]{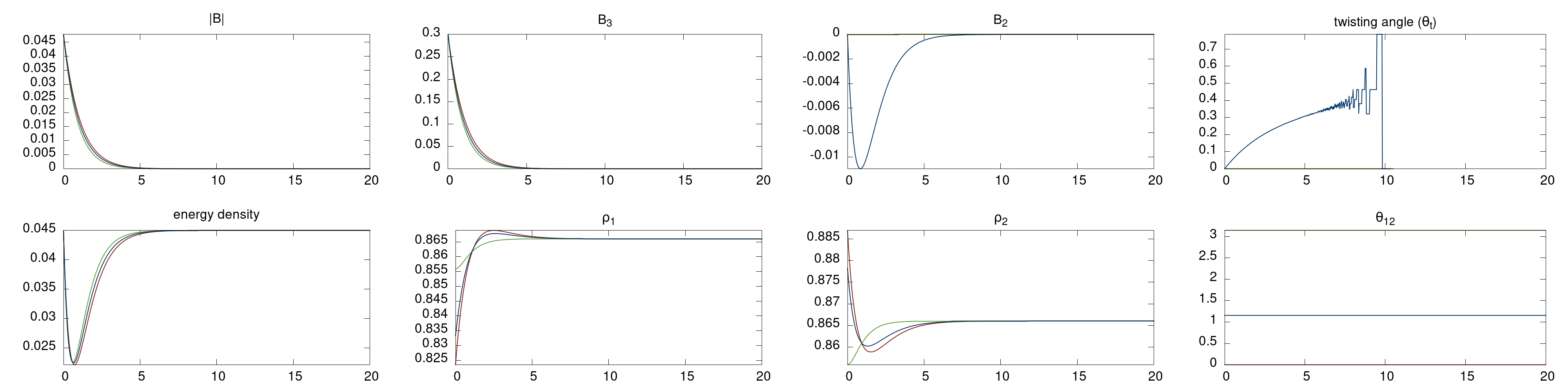}
 \caption{\label{Fig:MeissnerHalf1p0}
Plot of the Meissner state with boundary normal $\hat{x}_1 = (1,0,1)/\sqrt{2}$ and external field strength $H_0 = 0.3$, $\eta = 1.0$. The applied field is applied in the three directions $\hat{x}_3 = (-1,0,1)/\sqrt{2}$ (red), $\hat{x}_3 = (0,1,0)$ (green) and $\hat{x}_3 = (-\frac{1}{2},\frac{1}{\sqrt{2}},\frac{1}{2})$ (blue). We have plotted the total magnetic field $|B|$, magnetic field parallel to the applied magnetic field $B_3$ and orthogonal $B_2$, as well as the angle between the total magnetic field and the applied field (twisting angle). Finally we have plotted the normalised free energy density, condensate magnitudes $\rho_1$ and $\rho_2$, and the phase difference $\theta_{12}$.}
\end{figure*}

Other results in the basal plane match the above analysis, so we now consider solutions away from the basal plane. In figure \figref{Fig:MeissnerHalf1p0} we can see solutions for $\eta = 1$ with normal $\hat{x}_3 = (1,0,1)/\sqrt{2}$. The applied magnetic field is $H = H_0 \hat{x}_3$ with $H_0 = 0.3$. For $\hat{x}_3 = (-1,0,1)/\sqrt{2}$, plotted in red with $\delta_{12} = 0$, we see that the length scales have a leading mode that is mixed in the direction of the applied field. This is excited, causing disparity between the two condensates. For $\hat{x}_3 = (0,1,0)$, plotted in green with $\delta_{12} = \pi$, only a purely magnetic and purely matter mode are excited, causing no twisting and the condensates have equivalent behaviour. Finally $\hat{x}_3 = (\frac{-1}{2}, \frac{1}{\sqrt{2}}, \frac{1}{2})$, plotted in blue with $\delta_{12} = 1.156$. The blue line is very different from the basal plane case, with leading length scale having a combination of $\hat{x}_2$ and $\hat{x}_3$ and as predicted we see the twisting angle change as the fields decay.

The Meissner state solutions exhibit clear magnetic field twisting that is predicted by the linearization. However, we have seen that to understand how to apply the linearisation requires minimising over $\delta_{12}$ first. This limits its usefulness and also mutes some of the interesting behaviour like oscillations. Nonetheless, the predictable disparity between condensate magnitudes and the existence of magnetic field twisting are key experimental signatures of nematic systems. Our analysis also suggests that giving the phase difference mode a mass, by adding a term dependent on $\theta_{12}$ to $F_p$ would simplify the process of predicting the Meissner state and will likely add additional behaviour predicted by the linearization.

\section{Vortices and Skyrmions}
We turn now to topologically nontrivial energy minimizers. We will focus on minimizers that exhibit winding $n = 1-4$ in the form of single vortices and vortex bound states. By this we mean that we will model the cross-sections of vortex strings in the 3-dimensional model defined in the free energy \eqref{eq:F}. We will assume that after transforming our coordinate system, according to \eqref{eq:trid}, that the vortex string is in the $x_3$-direction. Hence, we assume that our fields $\psi_\alpha(x_1,x_2)$ and $A_i(x_1,x_2)$ vary only in the cross-section of the string and are translation invariant in the $x_3$ or string direction.

In an isotropic Gingzburg-Landau model, all choices of $\hat{x}_3$ are equivalent due to rotational symmetry. In contrast, the anisotropy of a nematic system (defined by \eqref{eq:nematic}), breaks rotational symmetry, meaning $\hat{x}_3$ becomes a parameter of the model. Note that only $\hat{x}_3$ is a parameter, as the choice of $\hat{x}_1$ and $\hat{x}_2$ just serves to pick a coordinate system on the cross-section.

The standard approach is then to set $A_3 = 0$ which fixes the magnetic field to be in the $x_3$-direction ($B_1 = B_2 = 0$). However, we have already shown that in general this is not a solution of the  equations of motion. Hence, we will retain all 3 components of $A$, which will allow the magnetic field to twist direction around the vortex. Note that the magnetic flux has the topological quantization condition,
\begin{equation}
\frac{1}{2\pi}\int_{-\infty}^\infty B_3 \, dx_1 \, dx_2 = n,
\end{equation}
where $n$ is the winding number of the system, representing the number of vortices. The other components in contrast, must topologically integrate to zero $\int B_1 = \int B_2 = 0$.

The model's parameters exist within a 4-dimensional family: the potential parameter $\eta > 0$, the boundary phase difference $\delta_{12} \in [0,2\pi]$, and the unit normal to the plane $\hat{x}_3$. The $n=1$ and $2$ vortex solutions were previously found for $\eta = 3$, $\hat{x}_3 = \hat{\zvec}$, $\delta_{12} = 0$ \cite{zyuzin2017nematic}, however they assumed that $A_3 = 0$. While in general this assumption is false, it does hold for the specific choice $\hat{x}_3 = \hat{\zvec}$ (the basal plane) as we will show, hence their results hold despite the assumption. In \cite{zyuzin2017nematic} it was  demonstrated that vortex solutions take the form of a pair of spatially separated fractional flux vortices called Skyrmions (for a detailed discussion of flux quantization in multicomponent system see e.g. \cite{babaev2002vortices} ). We will study these solutions in more detail, for various parameters, as well as solutions for higher winding number. Note that the connection between Skyrmion solutions and models that exhibit coupled length scales has been previously considered in \cite{winyard2019skyrmion}.

\subsection{Skyrmions}
Skyrmions form when integer flux vortices split into spatially separated fractional vortices in each component $\psi_\alpha$, such that their zeros (points at which $\psi_\alpha = 0$) are not coincident. Since any such field configuration never attains the value $(\psi_1,\psi_2)=(0,0)\in\C^2$, we may construct from it a gauge invariant field $\Phi:\R^2\to S^2$,
\begin{equation}
    \Phi(x_1,x_2)=\frac{(\bar\psi_1\psi_2+\psi_1\bar\psi_2,i(\bar\psi_1\psi_2-\psi_1\bar\psi_2),|\psi_2|^2-|\psi_1|^2)}{|\psi_1|^2+|\psi_2|^2}.
\end{equation}
We may then describe the field using the gauge invariant quantities
$\Phi$, $\rho=\sqrt{|\psi_1|^2+|\psi_2|^2}$ and $J$, the supercurrent, which is conveniently regarded as a one-form on $\R^2$.
In order for a 2-dimensional solution on the physical space $\mathbb{R}^2$ to have finite energy, as $|x|\rightarrow \infty$, $\Phi$ and $\rho$ must tend to constants $\Phi_0 \in \mathbb{C}P^1$, $\rho_0 \in \mathbb{R}$ and $J$ should tend to $0$.
Since $\Phi$ tends to a constant on the circle at spatial infinity, we may extend it continuously to a maps, still denoted $\Phi$, from teh one-point compactification $\R^2\cup\{\infty\}\cong S^2$ to $S^2$. Any such map has an integer valued winding number $Q(\Phi)$, the number of times the map wraps the domain two-sphere around the target two-sphere, and this quantity is a topological invariant of the map.

Rewriting the magnetic field using the gauge invariant quantities we get,
\begin{equation}
B = \left( \frac{1}{2} \Phi^\star \omega  - d\left( \frac{J}{\rho^2} \right) \right),
\end{equation}
where $\omega$ is the usual area form on $S^2$, and $\Phi^*\omega$ is its pullback to $\R^2$ by the map $\Phi$. It then follows, by Stokes's Theorem, that the total magnetic flux through the plane is
\begin{equation}
\int_\Omega B = \frac{1}{2} \int_\Omega \Phi^\star \omega =: 2\pi \mathcal{Q}(\Phi),
\end{equation}
where we have observed that the winding number of $\Phi$ equals the total signed area of its image divided by $4\pi$ (the area of $S^2$). For numerical purposes, it is convenient to use the integral formula
\begin{equation}
\mathcal{Q}(\Phi) = \int_{\mathbb{R}^2} \frac{i \epsilon_{ji}}{2\pi|\Psi|^4}\left( |\Psi|^2 \partial_i \Psi^\dagger \partial_j \Psi + \Psi^\dagger \partial_i \Psi \partial_j \Psi^\dagger \Psi\right) d^2 x,
\end{equation}
where $\Psi=(\psi_1,\psi_2)$ (see \cite{garaud2013chiral} for a detailed derivation in the $N$-component case). Note that the expression on the right is invariant under $\Psi\mapsto\lambda\Psi$ for any function $\lambda:\R^2\to\C\backslash\{0\}$, so this really is a function of $\Phi$. 
We call $\mathcal{Q}$ the Skyrme charge or Skyrmion number. Note that this integral is well-defined  only if the cores of the fractional vortices \emph{do not} coincide and that, in this case, it is precisely the number of magnetic flux quanta in the field configuration.

\subsection{Solutions}
\begin{figure*} 
\begin{tabular}{|c|c|}
      \hline
      $n=1$, $\delta_{12} = 0$ & $n=-1$, $\delta_{12} = 0$\\
 \includegraphics[trim=25 60 40 175,clip,width=0.5\linewidth]{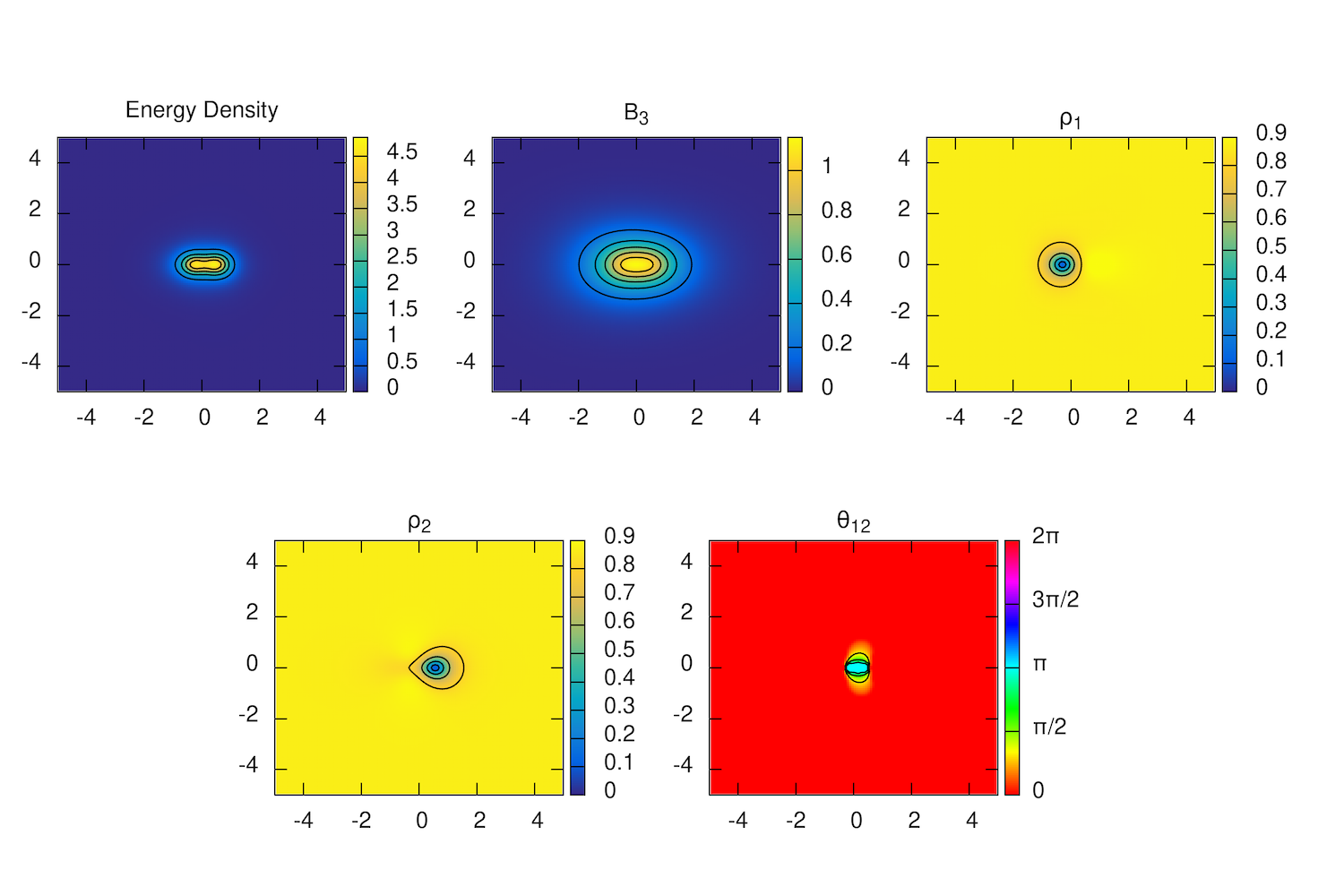} & \includegraphics[trim=25 60 40 175,clip,width=0.5\linewidth]{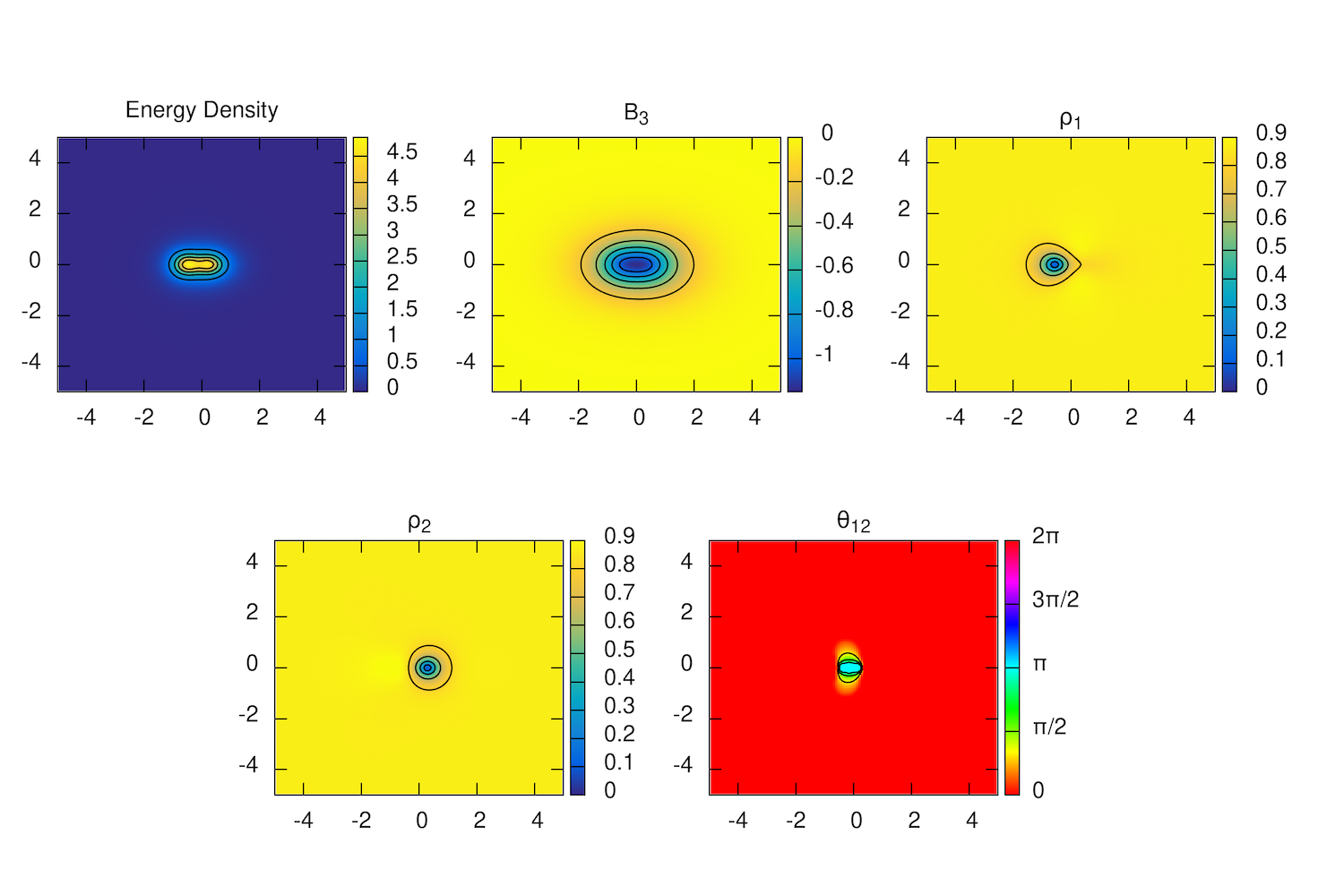} \\
      \hline
       $n=1$, $\delta_{12} = \pi/2$ & $n=-1$, $\delta_{12} = \pi/2$\\
\includegraphics[trim=25 60 40 175,clip,width=0.5\linewidth]{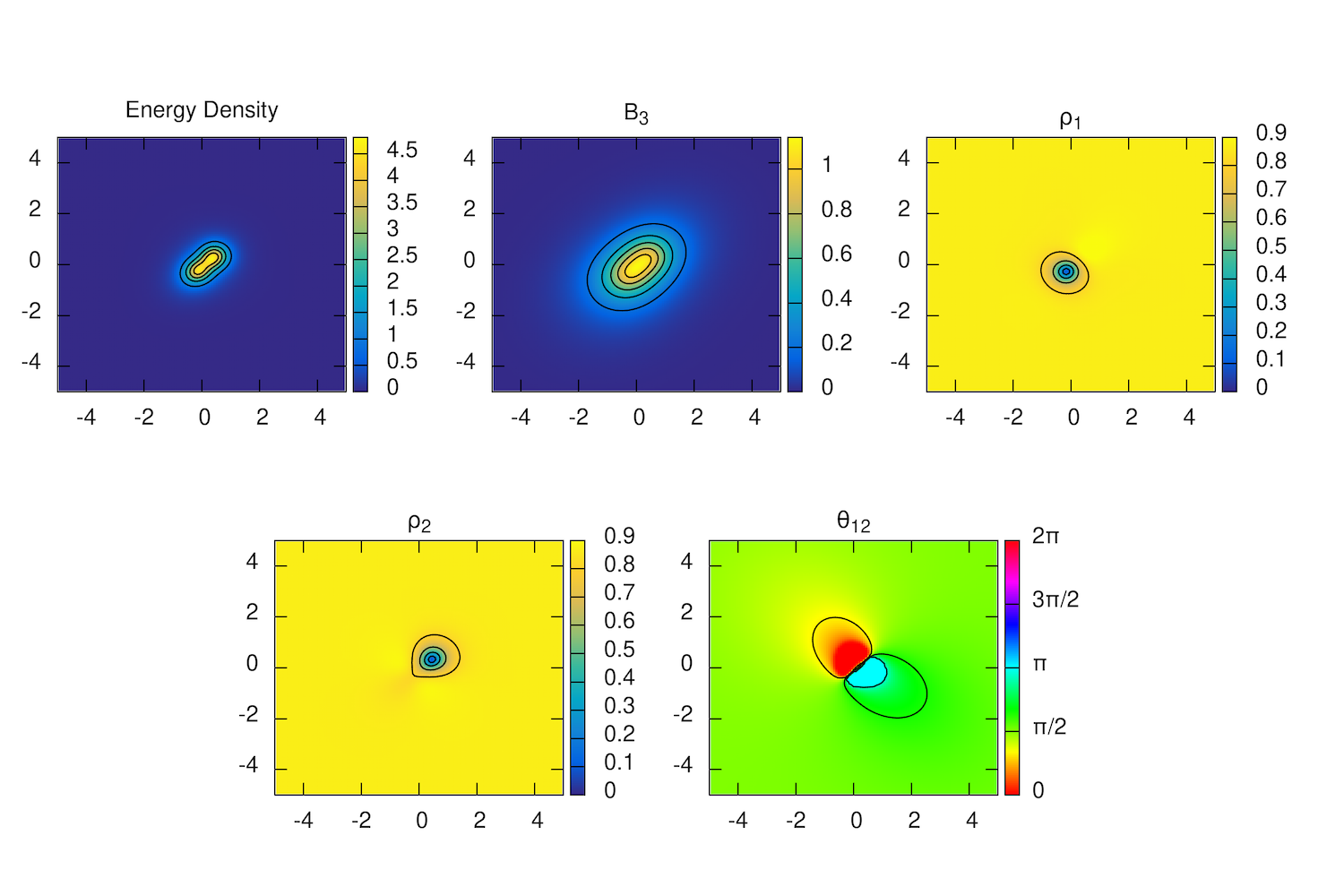} & \includegraphics[trim=25 60 40 175,clip,width=0.5\linewidth]{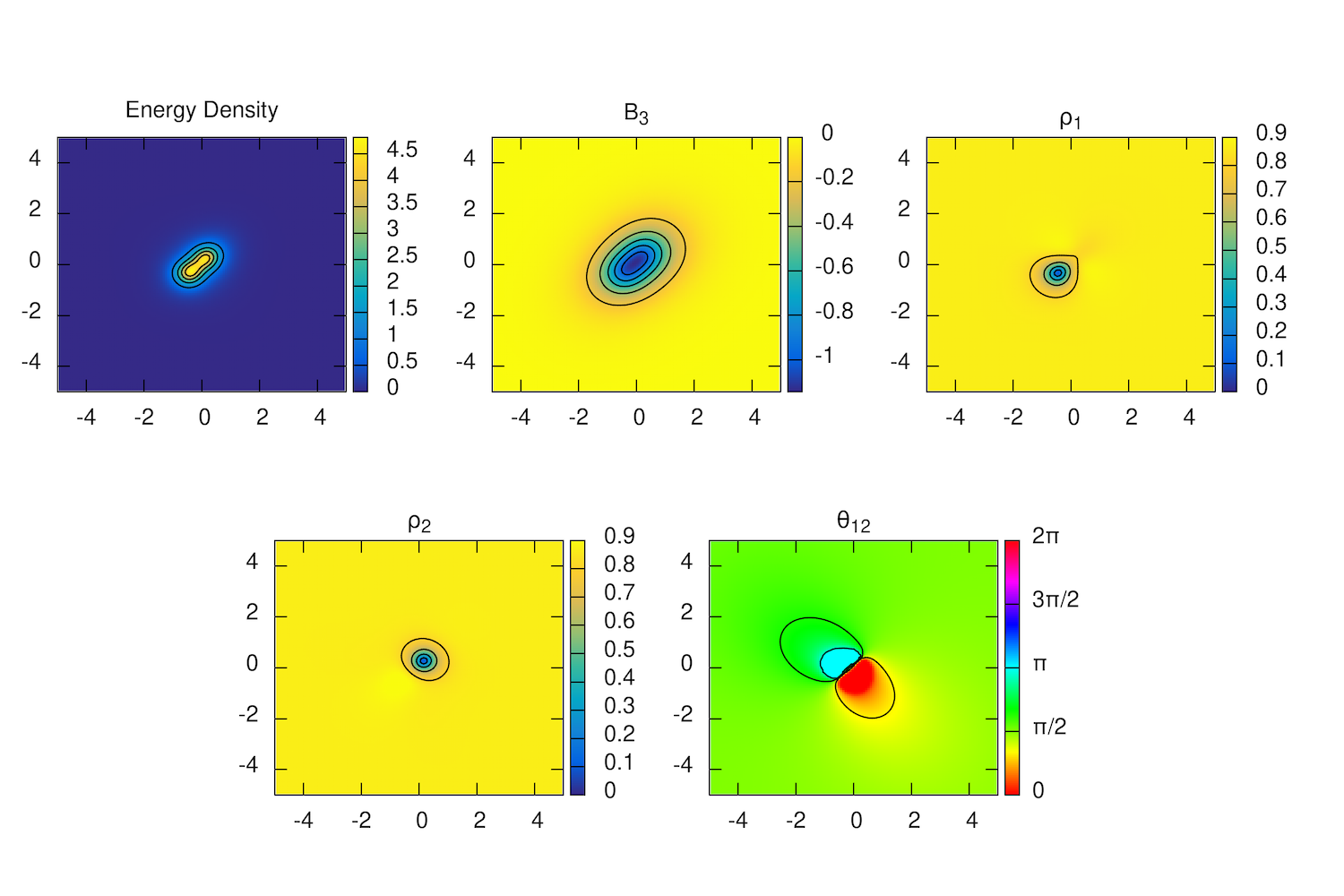} \\
      \hline
\end{tabular}
 \caption{\label{Fig:charge1}
Contour plots of degenerate single vortex/anti-vortex solutions in the basal plane $\hat{x}_3 = (0,0,1)$. The top row fixes the boundary phase difference to be $\delta_{12} = 0$, while the bottom row fixes it to be $\delta_{12} = \frac{\pi}{2}$. For each solution we have plotted the (normalised) free energy density, magnetic flux $B_3$, condensate magnitudes $\rho_1$ and $\rho_2$ and the phase difference $\theta_{12}$. The solutions form Skyrmions such that $Q=1$. Note, there is no in plane magnetic field in the basal plane $B_1 = B_2 = 0$.}
\end{figure*}

To find the vortex solutions we transform the anisotropy matrices according to \eqref{eq:trid} and use the same Newton flow algorithm described in appendix B.  We will first consider vortex solutions in the basal plane, with normal $\hat{x}_3=(0,0,1)$. The $n=1$ solution for $\eta = 3$ with boundary phase difference $\delta_{12}=0$ is plotted in figure \figref{Fig:charge1}. While this solution has been previously considered \cite{zyuzin2017nematic}, the intricate symmetry of the system was not discussed. So we can ask which, if any, of the symmetries in the symmetries section leave this vortex solution unchanged? Such a symmetry must be orientation preserving (else it maps $n=1$ configurations to $n=-1$ configurations), and must leave $\theta_{12}$ unchanged (else it changes the boundary condition). The sole candidate for $\delta_{12}=0$ is the composition of \eqref{eq:refsym} and \eqref{eq:conjsym} (in either order as they commute),
\begin{align}
\wt\psi_\alpha(x_1,x_2)&=\ol{\psi_\alpha(x_1,-x_2)},\\
\wt{A}(x_1,x_2)&=(-A_1(x_1,-x_2),A_2(x_1,-x_2),-A_3(x_1,-x_2)).
\end{align}
The numerical solution in \figref{Fig:charge1} exhibits precisely this reflexion symmetry, about the axis connecting the two fractional vortices. The mixed symmetry in \eqref{eq:conjsym} implies that changing the boundary phase difference $\delta_{12}$ is equivalent to rotating the basal plane. Hence, rotating $\delta_{12}$ in \figref{Fig:charge1} causes the reflexion axis of the solution to rotate by $\delta_{12}/2$, which can be seen in \figref{Fig:charge1}. Also, while the energies of the vortex and anti-vortex are degenerate, the symmetry that produces the anti-vortex from the vortex reflects the free energy density, magnetic flux magnitude $|B_3|$ and phase difference $\theta_{12}$ in the line orthogonal to the symmetry axis of the vortex solution ($y$-axis for $\delta_{12} = 0$) as can be seen in \figref{Fig:charge1}. Finally, we note that there is no spontaneously generated in plane magnetic field in the basal plane. This can be seen by substituting the translationally invariant ansatz for the basal plane into the non-linear equations of motion in \eqref{eq:eom}, leading to $B_1 = B_2 = 0$ being a trivial solution despite $B_3 \neq 0$ and $\psi_\alpha \neq u_\alpha$. In addition, this is predicted by the length scales, as the mixed or coupled magnetic modes in the basal plane couple only $B_3$ and the matter fields, whereas $B_1$ and $B_2$ completely decouple.

\begin{figure*}
\includegraphics[trim=0 150 0 400,clip,width=1.0\linewidth]{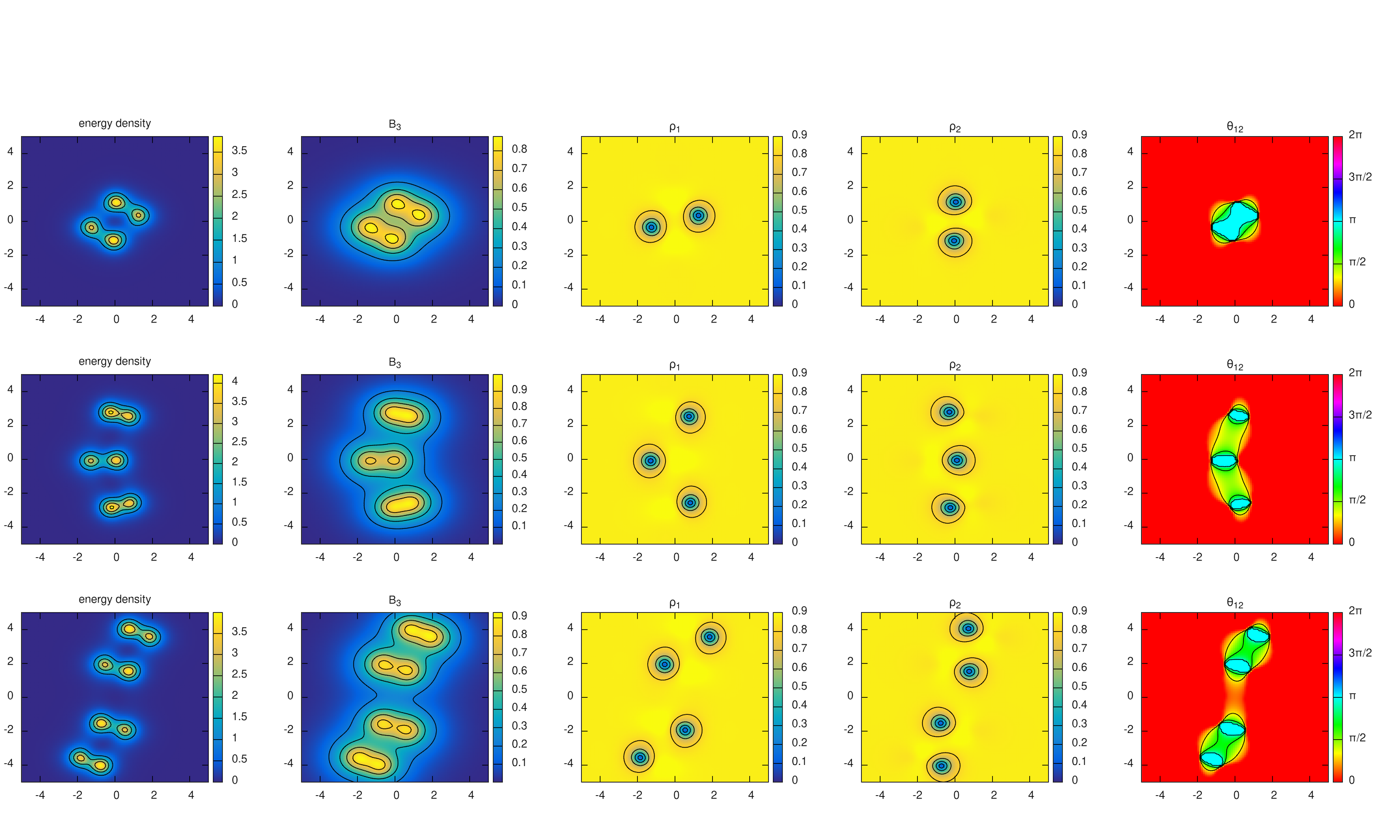}
\caption{\label{Fig:charge2-4}
Contour plot of $n=2-4$ bound states for $\eta = 3$, in the basal plane $\hat{x}_3 = (0,0,1)$ with boundary phase difference value $\delta_{12} = 0$. The normalised energies of the solutions are $\hat{F}_1 = 7.42$, $\hat{F}_2 = 7.30$, $\hat{F}_3 = 7.33$ and $\hat{F}_4 = 7.28$. The solutions exhibit fractional vortex splitting or Skyrmions ($\mathcal{Q} = n$), that form bound states.}
\end{figure*}

We have plotted the $n=2,3,4$ numerical solutions for $\eta = 3$ in the basal plane in \figref{Fig:charge2-4}. Unlike standard Ginzburg-Landau theory, the solutions for $n \geq 2$ form bound states. This can be seen by considering the normalized free energy of each configuration $\hat{F}_n = \frac{F_n - F_0}{n}$, where $F_n$ is the total free energy of the solution with winding number $n$. In the basal plane with $\eta = 3$, we have the energies $\hat{F}_1 = 7.42$, $\hat{F}_2 = 7.30$, $\hat{F}_3 = 7.33$ and $\hat{F}_4 = 7.28$, such that for even winding number $n$, the normalised free energy always decreases. This demonstrates that the bound state has lower energy than infinitely separated vortices, leading to a stable solution. We can see in \figref{Fig:charge2-4} that the $n=1$ Skyrmions form pairs for $n=2$, which subsequently form chains for higher even degree solutions. For odd degree however, one of the vortices cannot form a pair and is either repelled away from the chain or deforms it, causing the normalised energy per degree to increase slightly.

\begin{figure*}
\includegraphics[trim=38 60 0 700,clip,width=1.05\linewidth]{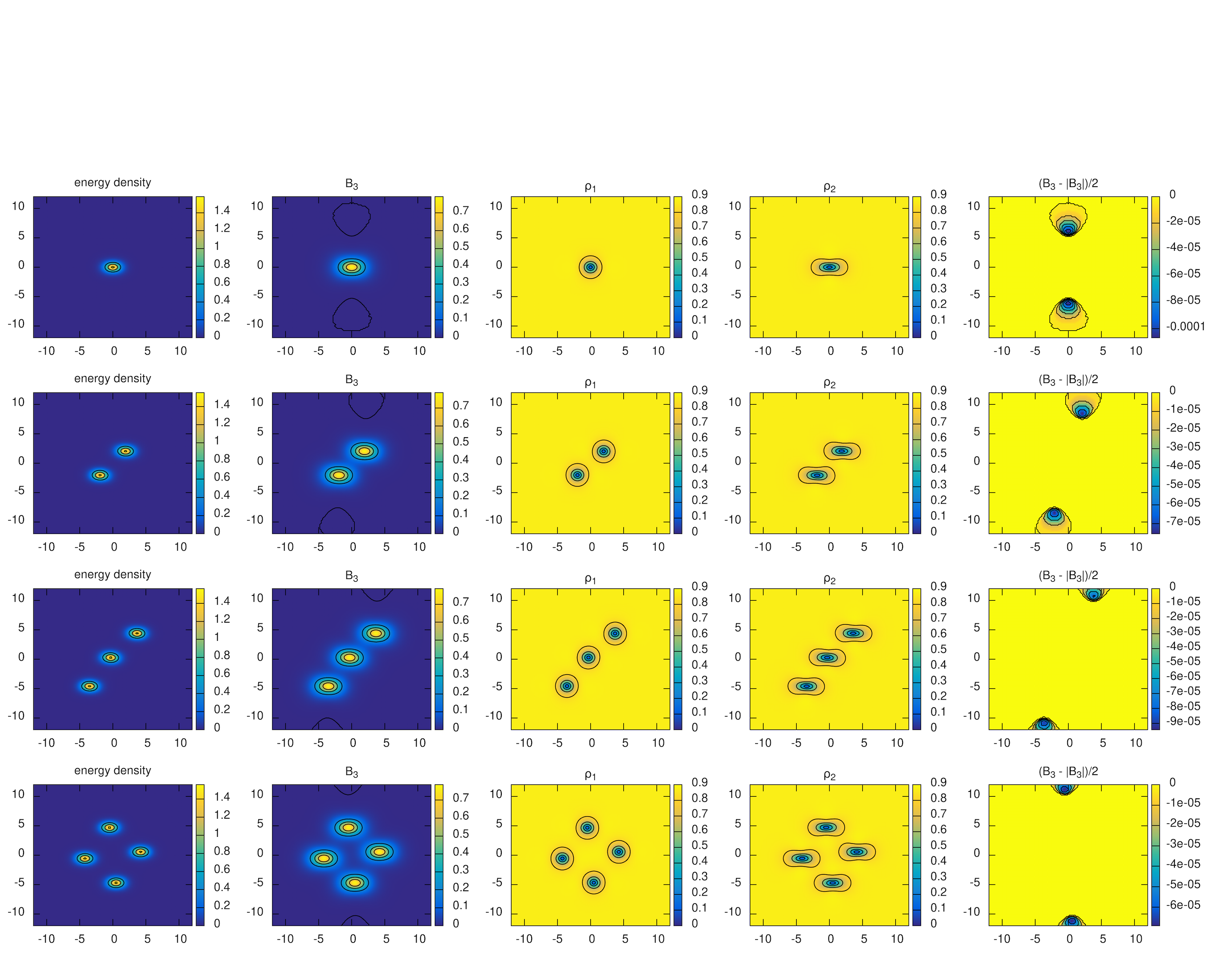}
\caption{\label{Fig:k1_basal}
Contour plots for the $n=1-4$ solutions for $\eta = 1$, in the basal plane $\hat{x}_3 = (0,0,1)$ with boundary phase difference value $\delta_{12} = 0$. The solutions do not form Skyrmions ($Q = 0$), but do exhibit magnetic field inversion.}
\end{figure*}

We have also plotted the solutions for $\eta = 1$ in the basal plane in \figref{Fig:k1_basal}. We see that Skyrmions are not formed ($Q = 0$); instead we get bound states of distorted composite vortices. Due to the highly coupled length scales, seen in \figref{Fig:baselLengthScales}, the vortices do exhibit magnetic field inversion which can cause bound states, as discussed in \cite{silaev2018non,speight2019chiral}.

We now turn to vortex solutions away from the basal plane. If we consider the e.o.m.\ in \eqref{eq:eom} after performing a change of coordinates, we notice that $B_1 = B_2 = 0$ is not trivially a solution if we assume that $B_3,\psi_\alpha \neq 0$ and $\hat{x}_3 \neq \hat{z}$. In addition, if we consider the linearization in a plane away from the basal plane, then the linear modes mix in-plane magnetic fields with the matter fields. This suggests that fluctuating matter fields can induce spontaneous in plane magnetic fields (orthogonal to the vortex line), causing magnetic field twisting.

\begin{figure}
\includegraphics[trim=20 150 100 300,clip,width=1.05\linewidth]{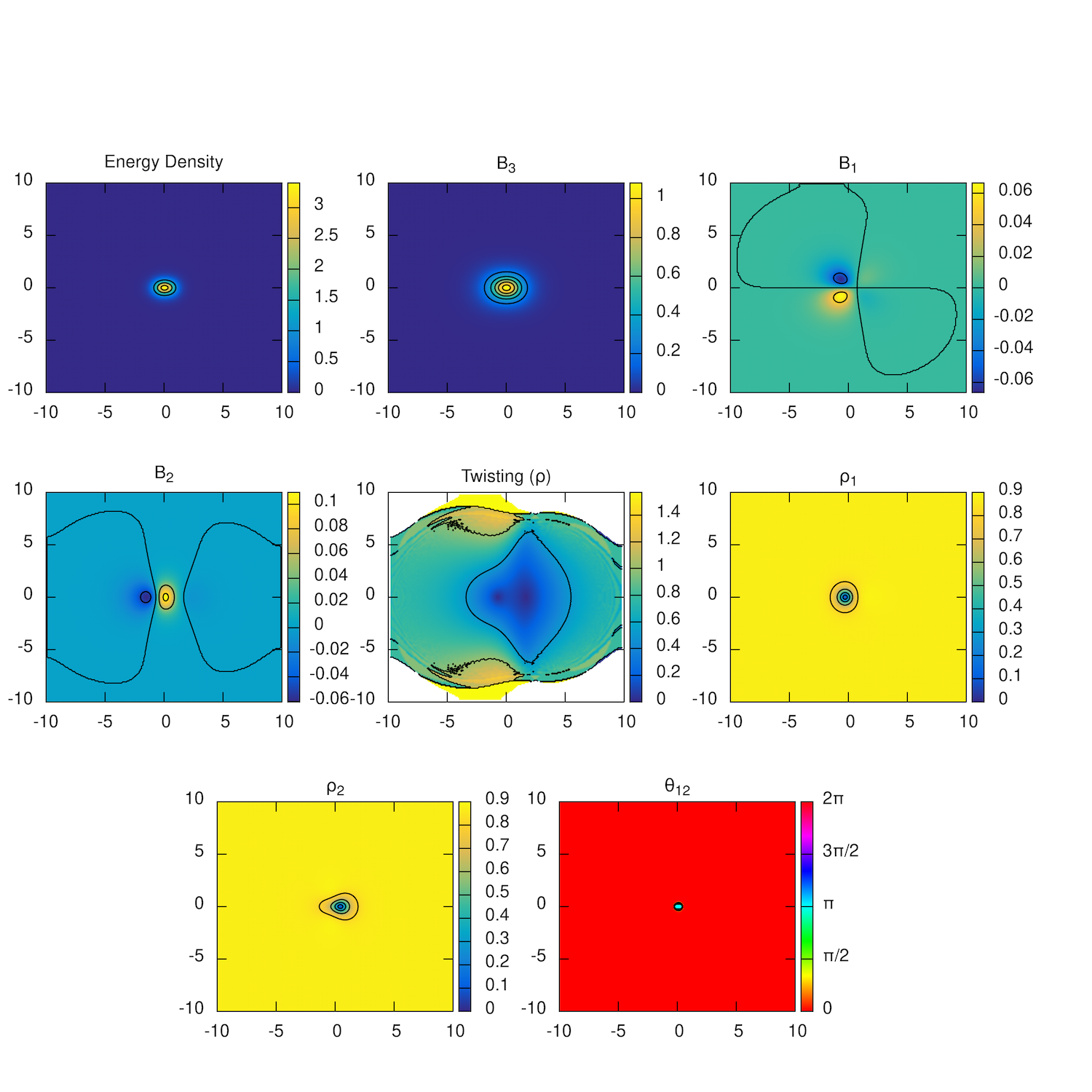}
\caption{\label{Fig:k3_half_N1}
Contour plot for $\eta = 3$, in the half plane $\hat{x}_3 = (0,1,1)/\sqrt{2}$ with boundary phase difference value $\delta_{12} = 0$. The coordinates are $\hat{x}_1 = (1,0,0)$ and $\hat{x}_2 = (0,-1,1)/\sqrt{2}$. Note that spontaneous in-plane magnetic field is generated from the coupled modes, causing the magnetic field to twist significantly.}
\end{figure}

We have plotted the $n=1$ solution for $\eta = 3$ on the half plane $\hat{x}_3 = (0,1,1)/\sqrt{2}$ in \figref{Fig:k3_half_N1}. We can see substantial magnetic field twisting away from the vortex line. Note that the magnetic field is always orthogonal to the plane at the zero of a condensate ($\rho_1 = 0$ or $\rho_2 = 0$). We can represent the twisting as an angle, similar to the Meissner state,
\begin{equation}
\tan\theta_t = \sqrt{B_1^2 + B_2^2}/B_3.
\end{equation}
For $\eta = 3$ we observe spontaneous magnetic fields $B_1$ and $B_2$ greater than $10\%$ of the orthogonal field $B_3$. This effect is easily strong enough to be detected in experiment, and could be used to point towards materials that exhibit nematic superconductivity.

\begin{figure*}
\hspace*{-0.75cm}
\includegraphics[width=1.08\linewidth]{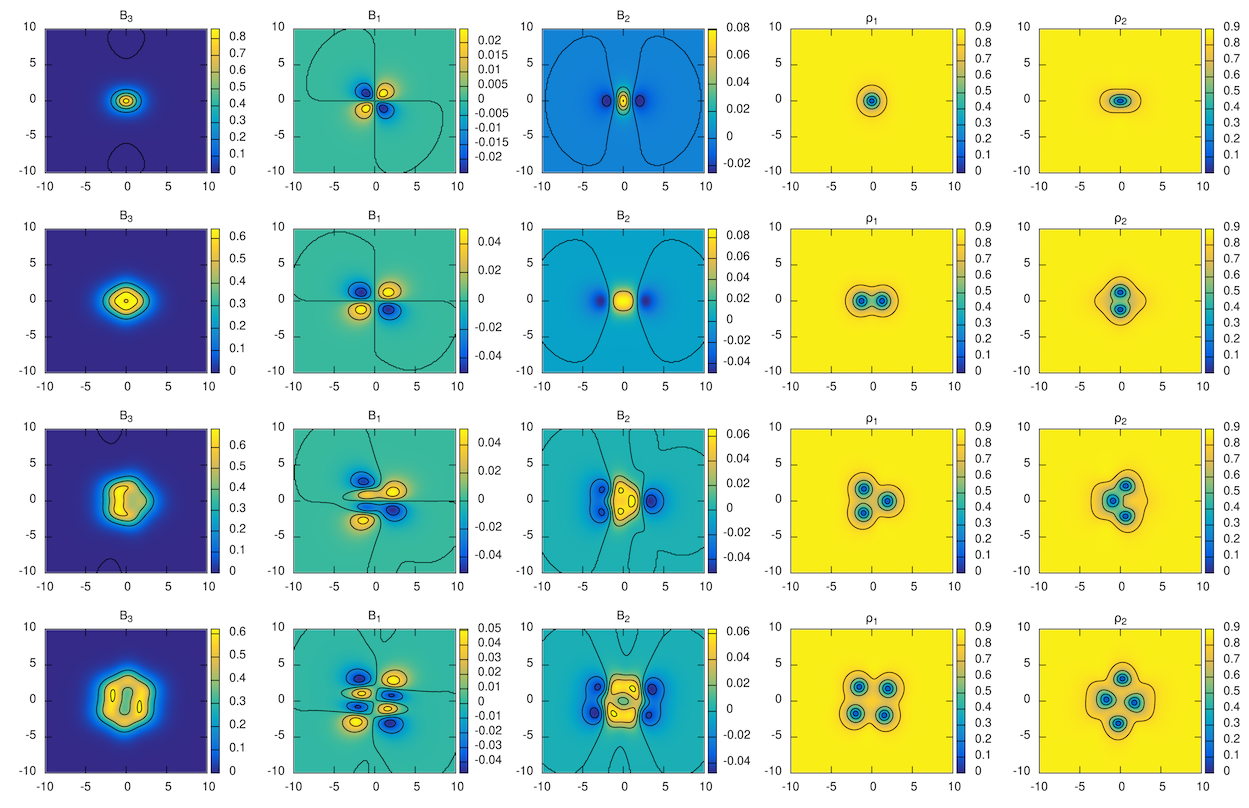}
\caption{\label{Fig:k1_nonbasal}
Contour plots for $n=1-4$ bound state for $\eta = 1$, in the half plane $\hat{x}_3 = (0,1,1)/\sqrt{2}$ with boundary phase difference value $\delta_{12} = 0$. The coordinates are $x_1 = (1,0,0)$ and $x_2 = (0,-1,1)/\sqrt{2}$. Note that spontaneous in pane magnetic field are generated from the coupled modes, causing the magnetic field to twist significantly.}
\end{figure*}

\begin{figure*}
\hspace*{-0.75cm}
\includegraphics[width=1.08\linewidth]{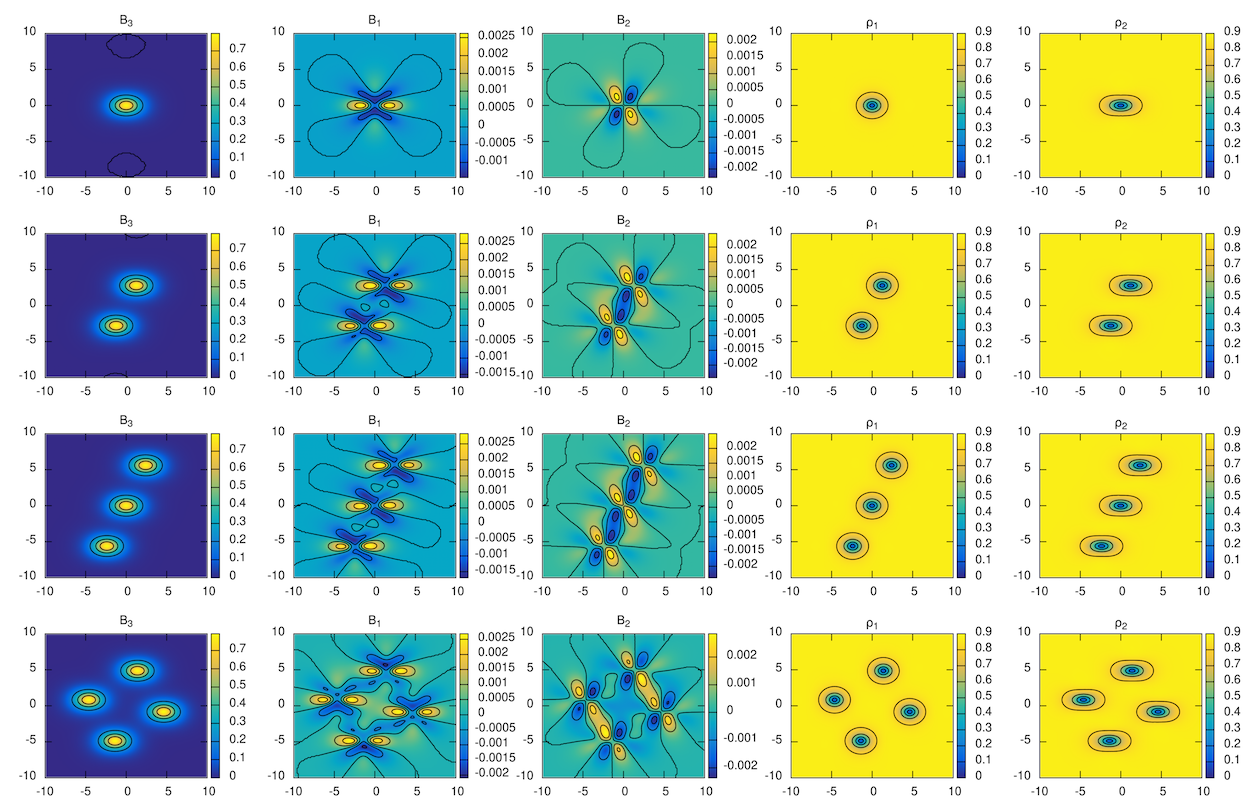}
\caption{\label{Fig:k1_nonbasal2}
Contour plots for $n=1-4$ bound state for $\eta = 1$, in the half plane $\hat{x}_3 = (1,0,1)/\sqrt{2}$ with boundary phase difference value $\delta_{12} = 0$. Note that spontaneous in pane magnetic field are generated from the coupled modes, causing the magnetic field to twist significantly.}
\end{figure*}

The mixed symmetry in \eqref{eq:conjsym}, combines rotations about the $z$-axis, which rotates the normal $\hat{x}_3$, with rotations of the phase difference $\theta_{12}$. This produces a 1-parameter family of energetically equivalent vortex solutions. This suggests that we should try changing orientation or $\delta_{12}$. We know that vortex bound states are mediated by their long-range forces. Thus, if we consider the values of $\eta$ that gave the leading mode to be mixed and heavily orientation dependent, this region will be affected most by changing orientation. Thus by checking \figref{Fig:mixing} we can see that $\eta \approx 1$ is the region we expect the most effect. By comparing the resulting vortex states in \figref{Fig:k1_nonbasal} and \figref{Fig:k1_nonbasal2} we can see a marked difference in the structure. When the normal to the vortex plane is $\hat{x}_1 = (0,1,1)/\sqrt{2}$ in \figref{Fig:k1_nonbasal} we see a tight Skyrmion structure for the bound state. However, when the normal is $\hat{x}_1 = (1,0,1)/\sqrt{2}$ we observe a loosely bound composite vortex structure. Note that these simulations were run with a fixed phase difference $\delta_{12} = 0$ on the boundary. Due to the mixed symmetry, rotating this would be equivalent to rotating the normal.

\begin{figure}
\includegraphics[trim=20 150 100 300,clip,width=1.05\linewidth]{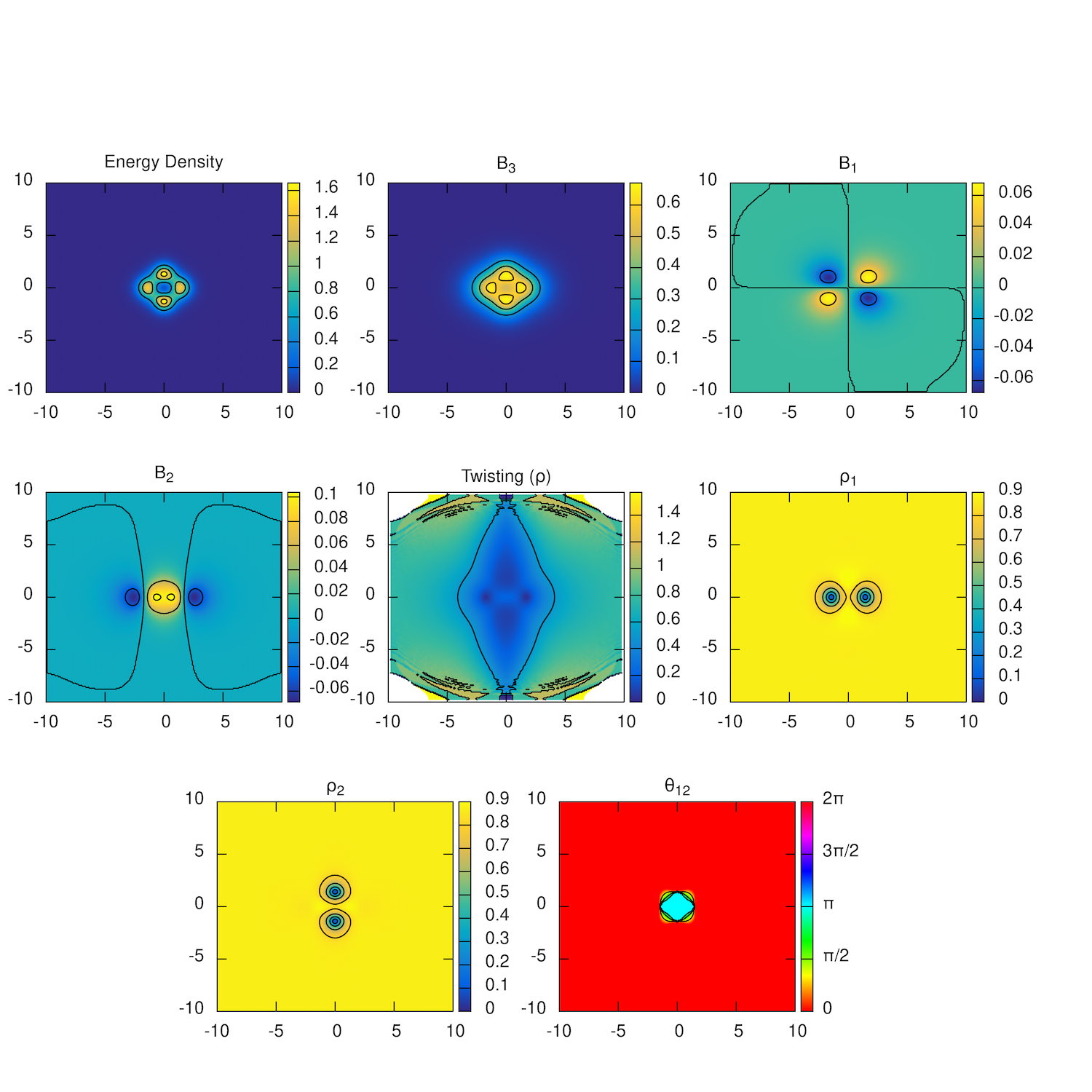}
\caption{\label{Fig:k3_half_N2}
Contour plot of an $n=2$ bound state for $\eta = 3$, in the half plane $\hat{x}_3 = (0,1,1)/\sqrt{2}$ with boundary phase difference value $\delta_{12} = 0$. The coordinates are $\hat{x}_1 = (1,0,0)$ and $\hat{x}_2 = (0,-1,1)/\sqrt{2}$. Note that spontaneous in-plane magnetic field is generated from the coupled modes, causing the magnetic field to twist.}
\end{figure}

\begin{figure}
\includegraphics[trim=20 150 100 300,clip,width=1.05\linewidth]{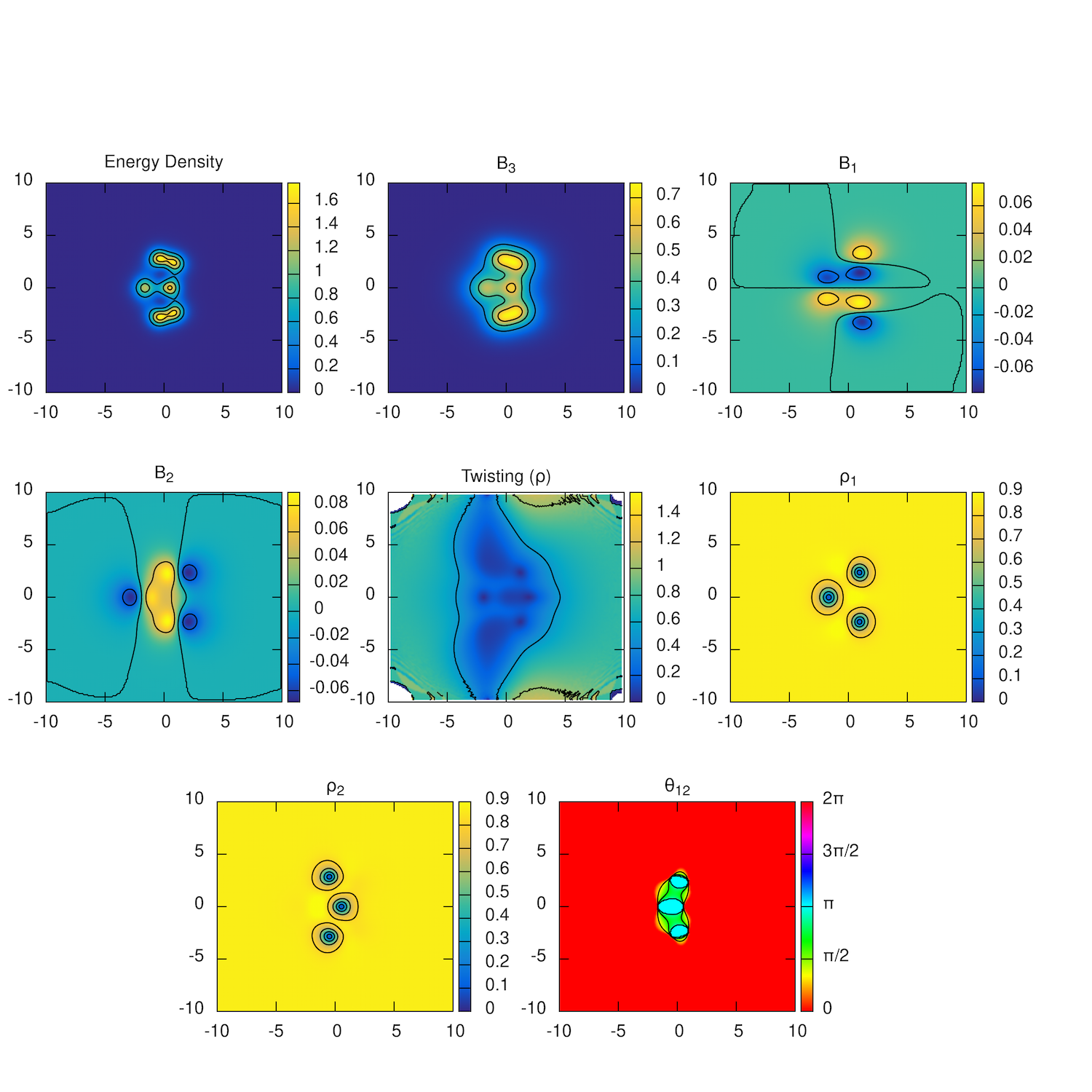}
\caption{\label{Fig:k3_half_N3}
Contour plot of an $n=3$ bound state for $\eta = 3$, in the half plane $\hat{x}_3 = (0,1,1)/\sqrt{2}$ with boundary phase difference value $\delta_{12} = 0$. The coordinates are $\hat{x}_1 = (1,0,0)$ and $\hat{x}_2 = (0,-1,1)/\sqrt{2}$. Note that spontaneous in-plane magnetic field is generated from the coupled modes, causing the magnetic field to twist.}
\end{figure}

\begin{figure}
\includegraphics[trim=20 150 100 300,clip,width=1.05\linewidth]{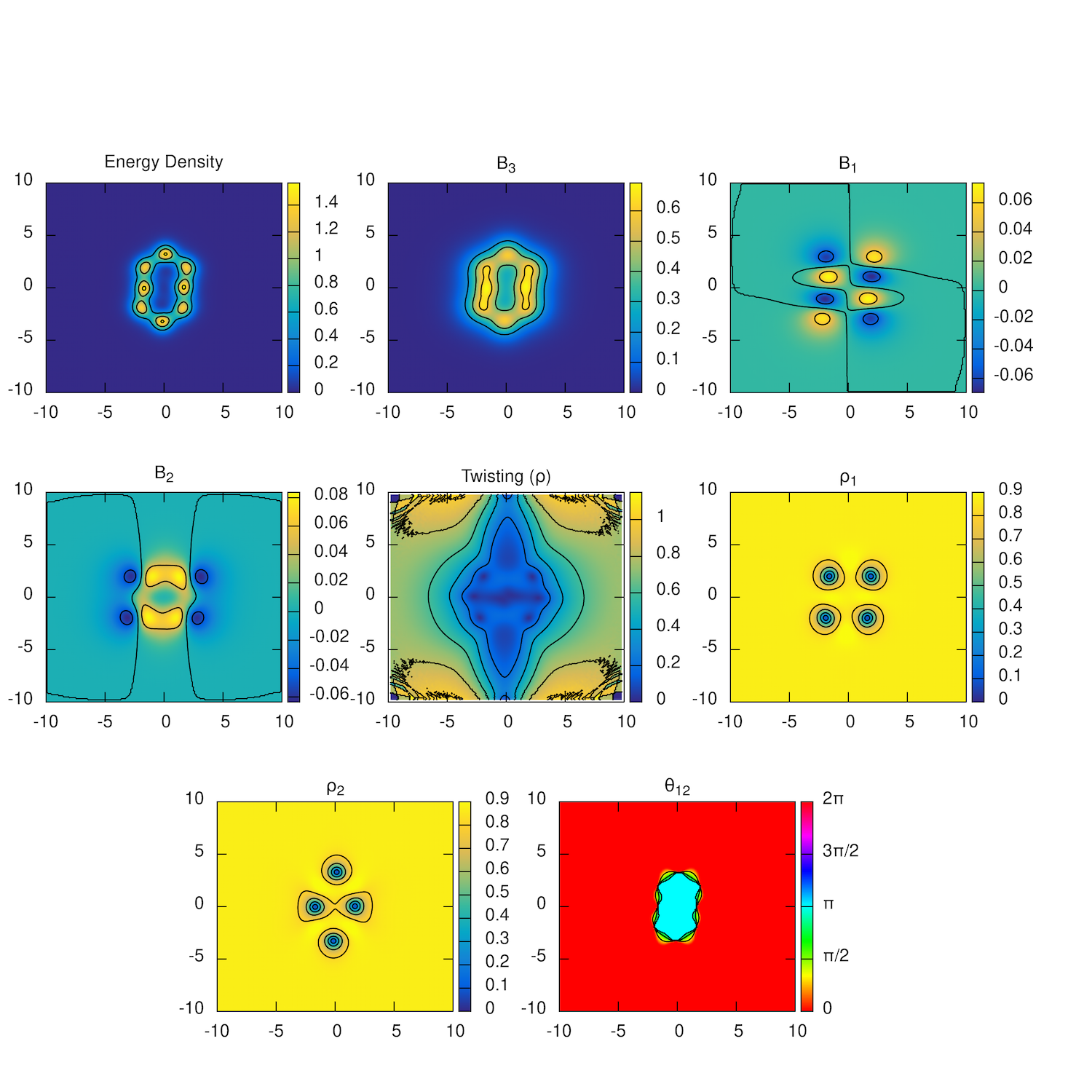}
\caption{\label{Fig:k3_half_N4}
Contour plot of an $n=4$ bound state for $\eta = 3$, in the half plane $\hat{x}_3 = (0,1,1)/\sqrt{2}$ with boundary phase difference value $\delta_{12} = 0$. The coordinates are $\hat{x}_1 = (1,0,0)$ and $\hat{x}_2 = (0,-1,1)/\sqrt{2}$. Note that spontaneous in-plane magnetic field is generated from the coupled modes, causing the magnetic field to twist.}
\end{figure}

If we consider the higher winding solutions for $\eta = 3$ in \figref{Fig:k3_half_N2}-\figref{Fig:k3_half_N4} we see that the bound states are still Skyrmion chains, but of a different form. For $\eta = 1$, plotted in \figref{Fig:k1_nonbasal} the bound states are now Skyrmion clusters, unlike in the basal plane. Both of these parameters also exhibit significant magnetic field twisting.

Finally, we note that if the vortex plane is rotated such that the normal lies in the basal plane, we find that the interesting physics disappears and the solutions act very similar to an isotropic model. This is unsurprising, as the length scale calculation demonstrated mixed modes with magnetic component purely in the $z$-direction and hence are never excited. The magnetic mode in $y$ and $x$-directions decouple, allowing the condensate to act according to its decoupled length scales.

\section{Magnetisation and Critical Fields}

We turn now to the effect of applying an increasing external field $H \in \mathbb{R}^3$ to a nematic superconductor. The material has three possible phases separated by critical values of external magnetic field strength $H_0 = |H|$. For $H_0 < H_{c_1}$ (lower critical field) the material exhibits the Meissner state, discussed above, where the magnetic response is limited to the boundary of the system. For $ H_{c_1} < H_0 < H_{c_2}$ vortices or Skyrmions enter the system.
Finally for $H_0 > H_{c_2}$, the material becomes a normal conductor $\psi_\alpha = 0$, with $B = H$. The only caveat to this, is if the critical field for vortex state formation is larger than the thermodynamical critical magnetic
field, then the system acts as a type I superconductor, exhibiting a single phase transition from the Meissner state directly to the normal state. In this section we will approximate the values of $H_{c_1}$ and $H_{c_2}$.

\subsection{Lower critical field $H_{c_1}$}
The lower critical field $H_{c_1}$ determines the point at which it becomes optimal, in the bulk, to have a vortex structure, as opposed to the familiar constant superconducting ground state $\psi_\alpha = u_\alpha$. Specifically, we seek the minimum value of $H_0$ such that we can find a vortex solution whose bulk Gibbs free energy per unit area is equivalent to that of the homogeneous state $\psi_\alpha = u_\alpha$, $A_i = 0$.

Given an external field direction $\hat{H}$, we can calculate the lower critical field $H_{c_1}$ by considering the normalised Gibbs free energy $\hat{G}_n = G_n - G_0$ of the possible vortex states in the plane with normal $\hat{H}$. Note that $G_n$ is the minimal Gibbs free energy for the winding number $n$. Hence, we seek the value of external field strength $H_0$ such that $G_n = 0$ for some $n$. By rearranging \eqref{eq:G} we can see that this occurs at 
\begin{equation}
H_{c_1} = \min_{n}\left\{ \hat{F}_n / (2\pi) \right\}.
\end{equation}
where $\min_{n}$ is the minimum with respect to the winding number $n$ and $\hat{F}_n$ is the normalised free energy. 

\begin{figure}
\includegraphics[width=1.0\linewidth]{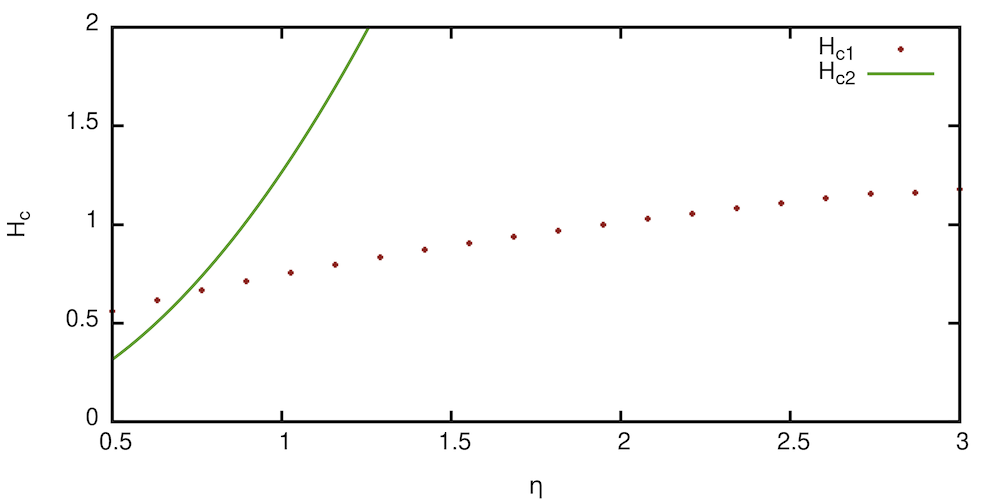}
\caption{\label{Fig:Hc}
A plot comparing $H_{c_1}$ (lower critical field) and $H_{c_2}$ (upper critical field) for the basal plane.}
\end{figure}

The above method requires finding the global $F$ minima for all winding numbers $n$, which is impractical in practice. Hence, we will assume that $\hat{F}_1$ is sufficiently close to the minimum to give a good approximation for $H_{c_1}$. We have plotted the values for $H_{c_1}$ calculated using the above approximation for the basal plane in \figref{Fig:Hc}.

\subsection{Upper critical field $H_{c_2}$}
To find the upper critical field $H_{c_2}$, we must consider the standard conductor solution, $\psi_\alpha = 0$ and $B = H$, which is always a solution of the equations of motion in \eqref{eq:eom}. If $H_0 \geq H_{c_2}$ this solution becomes linearly stable, but it is unstable below this critical external field strength.

We return to the linearization, however unlike the work presented in the linearization section, we must linearize \eqref{eq:eom} about $\psi = 0$, with $A$ such that $\nabla \times A = H$,
\begin{equation}
-Q^{\alpha\beta}_{ij} D_i D_j \psi_\beta - \eta^2 \psi_\alpha = 0.
\label{eq:eomHc2}
\end{equation}
We can then assume that $H_{c_1}$ is the largest value of $H_0$ for which \eqref{eq:eomHc2} has solutions other than $\psi_\alpha = 0$. We will use the same method as in the linearization section to rotate the orthonormal basis, setting the applied magnetic field direction to be $\hat{x}_3$ such that $H = H_0 \hat{x}_3$ and we define the $SO(3)$ matrix $M$ with rows $(\hat{x}_1,\hat{x}_2,\hat{x}_3)$ where $\hat{x}_1$, $\hat{x}_2$ can be any unit vectors that are orthogonal to $\hat{x}_3$. Hence we can write our new coordinate system along with a rescaling $x_i := \sqrt{\frac{H_0}{2}}M x$, such that we can fix our gauge and write the gauge field,
\begin{equation}
A = \sqrt{\frac{H_0}{2}}(-x_2, x_1, 0).
\end{equation}
This leads to \eqref{eq:eomHc2} becoming,
\begin{equation}
-\frac{H_0}{2}\mathcal{Q}^{\alpha\beta}_{ij}\mathcal{D}_i \mathcal{D}_j \psi_\beta - \eta^2 \psi_\alpha = 0.
\end{equation}
This gives the transition $H_{c_2}$ to occur at 
\begin{equation}
H_{c_2} = 2\eta^2/\lambda
\label{eq:Hc2}
\end{equation}
where $\lambda$ is the lowest eigenvalue of the operator $\mathcal{O}$,
\begin{equation}
\mathcal{O}\left( \begin{array}{c} \psi_1 \\ \psi_2 \end{array} \right) = - \left( \begin{array}{cc} \mathcal{Q}^{11}_{ij} \mathcal{D}_i \mathcal{D}_j & \mathcal{Q}^{12}_{ij} \mathcal{D}_i \mathcal{D}_j \\ \mathcal{Q}^{21}_{ij}\mathcal{D}_i \mathcal{D}_j & \mathcal{Q}^{22}_{ij} \mathcal{D}_i \mathcal{D}_j \end{array}\right)\left( \begin{array}{c} \psi_1 \\ \psi_2 \end{array}\right).
\end{equation}
We seek the lowest eigenvalue of the differential operator $\mathcal{O}$, which can be found analytically for simple $\mathcal{Q}$ or particular choices of $\hat{x}_3$, but in general must be found numerically. We first note that $\left[\mathcal{O},\mathbb{I}_2 \otimes i \mathcal{D}_3\right] = 0$, so we can seek simultaneous eigenstates of $\mathcal{O}$ and $\mathbb{I}_2 \otimes i \mathcal{D}_3$. Hence we can assume,
\begin{equation}
\psi = \psi(x_1,x_2)e^{i k x_3},
\end{equation}
for some $k \in \mathbb{R}$. We will assume that $k = 0$, so that $\psi$ is translation invariant in the direction of $\hat{x}_3$.  We can then make use of ladder operators, an approach used for numerous other models \cite{zhitomirskii1989magnetic, agterberg1998vortex} to find $H_{c_2}$, which we define as,
\begin{equation}
a := \frac{i}{2}(\mathcal{D}_1 + i\mathcal{D}_2), \quad a^\dagger = \frac{i}{2}(\mathcal{D}_1 - i \mathcal{D}_2),
\end{equation}
and the number operator $N := a^\dagger a$. These satisfy the relations,
\begin{equation}
[N,a^\dagger] = a^\dagger, \; [N,a] = -a, \; [a,a^\dagger] = 1.
\end{equation}
Hence, writing the operator in terms of the ladder operators,
\begin{equation}
\mathcal{O} = L_{\alpha\beta}(a^\dagger)^2 + P_{\alpha\beta} a^2 + R_{\alpha\beta} (a^\dagger a + a a^\dagger) + S_{\alpha\beta}
\end{equation}
where we have defined,
\begin{align}
L_{\alpha\beta} &= \mathcal{Q}^{\alpha\beta}_{11} - \mathcal{Q}^{\alpha\beta}_{22} + i\left( \mathcal{Q}^{\alpha\beta}_{12} + \mathcal{Q}^{\alpha\beta}_{21}\right)\\
P_{\alpha\beta} &= \mathcal{Q}^{\alpha\beta}_{11} - \mathcal{Q}^{\alpha\beta}_{22} - i \left(\mathcal{Q}^{\alpha\beta}_{12} + \mathcal{Q}^{\alpha\beta}_{21}\right)\\
R_{\alpha\beta} &= \mathcal{Q}^{\alpha\beta}_{11} + \mathcal{Q}^{\alpha\beta}_{22}\\
S_{\alpha\beta} &= i\left( \mathcal{Q}^{\alpha\beta}_{12} - \mathcal{Q}^{\alpha\beta}_{21}\right).
\end{align}
We now define the function $\left|0\right> := e^{-(x_1^2 + x_2^2)/2}$ noting that as required $a\left|0\right> = 0$. We then seek eigenfunctions of $\mathcal{O}$ of the form,
\begin{equation}
\psi_\alpha = \sum_{n=0}^{\infty}c_n^\alpha \left|n\right>
\end{equation}
where we have defined,
\begin{equation}
\left|n\right> := \frac{1}{\sqrt{n!}}(a^\dagger)^n\left|0\right>.
\end{equation}
With respect to this basis the ladder operators take the form,
\begin{align}
a &= \left( \begin{array}{ccccc} 0 & \sqrt{1} & 0 & 0 & \dots \\
0 & 0 & \sqrt{2} & 0 & \dots \\
0 & 0 & 0 & \sqrt{3} & \dots \\
0 & 0 & 0 & 0 & \dots \\
\vdots & \vdots & \vdots & \vdots & \end{array}\right)\\
a^\dagger &= \left( \begin{array}{ccccc} 0 & 0 & 0 & 0 & \dots \\
\sqrt{1} & 0 & 0 & 0 & \dots \\
0 & \sqrt{2} & 0 & 0 & \dots \\
0 & 0 & 0\sqrt{3} & 0 & \dots \\
\vdots & \vdots & \vdots & \vdots & \end{array}\right)
\end{align}
We then truncate the infinite matrices to size $(n+1)\times(n+1)$, approximating the ladder operators to produce a $2(n+1)\times 2(n+1)$ matrix operator $\mathcal{O}_n$ which approximates $\mathcal{O}$. We can then numerically calculate an approximation of $\lambda$ by finding the smallest eigenvalue $\lambda_n$ of $\mathcal{O}_n$ for increasing $n$ stopping when $|\lambda_{n} - \lambda_{n-1}| < \eps$, for a chosen tolerance $\eps$, which we took to be $\eps = 10^{-6}$.

While we have presented a general method for any $Q$, we now restrict to the nematic model presented in \eqref{eq:nematic}. In particular if we assume $H$ is orthogonal to  the basal plane $\hat{x}_3 = (0,0,\pm 1)$, the sequence $\lambda_n$ is constant for all $n \geq 2$ as $\mathcal{O}$ has an exact ground state of the form,
\begin{equation}
\psi = \left( \begin{array}{c} c_0 \left|0\right> \\ c_2 \left|2\right>\end{array}\right).
\label{eq:subspace}
\end{equation}
This is because, for $\hat{x}_3 = (0,0,\pm 1)$,
\begin{equation}
\mathcal{O} = \left( \begin{array}{cc} 2(a^\dagger a + a a^\dagger) & 4\beta_\perp a^2\\ 4 \beta_\perp(a^\dagger)^2 & 2(a^\dagger a + a a^\dagger)\end{array}\right),
\end{equation}
which acts on the two-dimensional subspace \eqref{eq:subspace} as
\begin{equation}
\mathcal{O} : \left( \begin{array}{c} c_0 \\ c_2 \end{array}\right) \mapsto \left( \begin{array}{cc} 2 & 4\beta_\perp \sqrt{2}\\ 4\beta_\perp \sqrt{2} & 10 \end{array}\right) \left( \begin{array}{c} c_0 \\ c_2 \end{array}\right).
\end{equation}
The smallest eigenvalue of this $2\times 2$ matrix is,
\begin{equation}
\lambda = 6 - 4\sqrt{1 + 2 \beta_\perp^2} = 6 - \frac{4}{3}\sqrt{11},
\end{equation}
which, when substituted into \eqref{eq:Hc2}, gives $H_{c_2}( (0,0,1) ) = \frac{3\eta^2}{9 - 2\sqrt{11}} \approx 1.2676 \eta^2$. We have plotted this against the values of $H_{c_1}$ in \figref{Fig:Hc} For other choices of $\hat{x}_3$ the ground state is an infinite series in $\left|n\right>$.

\begin{figure}
\includegraphics[width=1.0\linewidth]{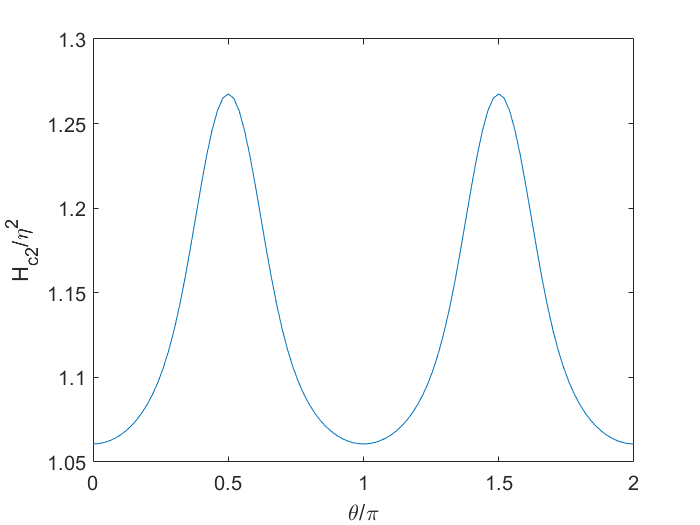}
\caption{\label{Fig:Hc2}
The upper critical field $H_{c2}$ as a function of orientation for applied fields directed along ${\hat{x}_3}=(\cos\theta\cos\phi,\cos\theta\sin\phi,\sin\theta)$. A symmetry of the model implies that $H_{c2}$ is independent of $\phi$.}
\end{figure}

We have plotted the results, which demonstrate the anisotropy of $H_{c_2}$ in \figref{Fig:Hc2}. Note that $H_{c2}$ is only dependent on the $z$ component of $\hat{x}_3$ due to the symmetry in \eqref{eq:ariel}. The numerical results suggest it is only weakly anisotropic: its maximum value (attained when $\hat{x}_3=(0,0,\pm1)$) is $1.195$ times its minimum value (attained when $\hat{x}_3$ lies in the $xy$ plane).

\section{Lattice Solutions}
In this section we consider vortex lattice solutions in the presence of an external field $H$. Namely, we find the unit cell of the periodic vortex solution in the bulk, with external field strength $H_{c_1} < H_0 < H_{c_2}$. The standard approach for a Ginzburg-Landau model is to consider a unit cell of degree $n=1$ with either triangular ($\alpha = \pi/3$) or square ($\alpha = \pi/2$) symmetry. However, as we are considering an anisotropic model, there is no reason to expect that a lattice with such high symmetry will be the global minimizer. Hence, the correct approach is to minimize energy, not just w.r.t.\ the periodic fields, but also w.r.t.\ the geometry of the unit cell itself. We present here a new general method of finding the optimal unit cell, without assuming the symmetry of the underlying lattice.

We note that vortex lattices in the basal plane of nematic superconductors have recently been considered using a finite dimensional field ansatz \cite{how2020half}, optimised over a general cell. The ansatz, a superposition of isolated vortices, was motivated by the GL equations linearised around the superconducting state, so the results are expected to be valid only for $H$ close to $H_{c_2}$. This contrasts with  our numerical approach, which works for the full range of applied fields. The paper \cite{how2020half} proposes some very surprising vortex lattice structures, in which one of the condensates acquires extra zeros compensated by zeros of opposite winding. These solutions are found in a very distant parameter regime from the one we consider here, however.  We will investigate the claims made in \cite{how2020half} using our more general method in a later paper.

\begin{figure}
\includegraphics[width=0.5\linewidth]{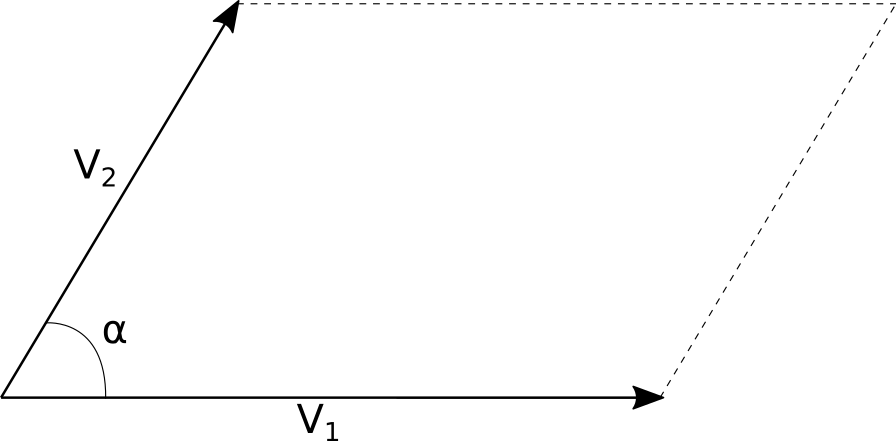}
\caption{\label{fig:par}
Diagram of the geometry of a general unit cell of a vortex lattice. The cell is defined by the two vectors $v_1$ and $v_2$.}
\end{figure}

We first assume that far from the boundary of the system (deep in the bulk), the gauge invariant quantities $(\rho_\alpha,\theta_{12},B)$ are periodic in the plane orthogonal to the applied field, and translation invariant in the direction of the applied field. As in previous sections, we begin by rotating our coordinate system such that $H=(0,0,H_0)$ and all spatial dependence is with respect to $(x_1,x_2)$. This implies that the constituent fields $(\psi_\alpha, A_i)$ are doubly periodic functions of $(x_1,x_2)$ up to gauge. We also note that we can represent a general periodic structure as a tessellation of the $x_1x_2$ plane by a general parallelogram (unit cell), as seen in figure \ref{fig:par}, formed by two vectors $v_1$ and $v_2$ with angle $\alpha$.

In the bulk of a superconductor (excluding boundary effects), the optimal lattice is the one that minimizes the total Gibbs free energy of the system $G_{sys}$. If such a solution is periodic $G_{sys}$ can be calculated from the unit cell as $G_{sys} = \area_{sys}G/\area$, where $\area_{sys}$ is the area of the system, $G$ is the Gibbs free energy of the field configuration in a single unit cell and $\area$ is the area of a unit cell. Hence, we seek local minimizers of the Gibbs free energy per unit area $G/\mathcal{A}$ w.r.t, the fields $(\psi_\alpha, A_i)$ defined on a general flat periodic 2-torus $T_\Lambda^2 = \mathbb{R}^2/\Lambda$ with the geometry of the torus $T_\Lambda^2$ is represented by $v_i$ in \figref{fig:par},
\begin{equation}
\Lambda = \left\{ n_1 v_1 + n_2 v_2 \,|\, n_1, n_2 \in \mathbb{Z}, v_1, v_2 \in \mathbb{R}^2\right\}.
\end{equation}
The fields defined on the unit cell have periodic boundary conditions, up to winding, such that they have boundary conditions,
\begin{align}
\psi_\alpha( x + v_i ) &= \psi_\alpha (x)e^{i f_i},\nonumber\\
A_j ( x + v_i ) &=  A_j(x) + \partial_j f_i.
\end{align}
Note that while the fields are not periodic, the above boundary conditions do leave all physical quantities periodic,
\begin{align}
B(x + v_i) = B(x), \quad \,J(x + v_i) = J(x),\nonumber\\ \rho(x + v_i) = \rho(x). \quad\quad\quad\quad
\end{align}
Hence, we just need any fixed real function $f_i(x)$, which leads to the correct winding for the unit cell.

We can encode the geometry of the unit cell by the matrix $L \in GL_+(2,\mathbb{R})$ with columns $v_1$, $v_2$ or, equivalently, by the pair $(M,\area)\in SL(2,\R)\times (0,\infty)$ where
$M:=\sqrt{\area}L^{-1}$  and $\area=\det L$ is the area of the unit cell. 

To simulate the fields on the unit cell, we will simplify the above formulation by transforming to a more convenient coordinate system in the $x_1x_2$ plane. Let us define $X_1,X_2$ so that $(x_1,x_2)=X_1v_1+X_2v_2$. The unit cell spanned by $v_1,v_2$ is now covered by $(X_1,X_2)\in[0,1]\times[0,1]$ where $X_i = \frac{M_{ij}}{\sqrt{\area}} x_j$. It is convenient to also rescale the spatial coordinate orthogonal to the $x_1x_2$ plane,
defining $X_3=x_3/\sqrt\area$. Then
\begin{widetext}
\begin{align*}
F(M,\area,\psi_\alpha,A) &= \int_{[0,1]^2} \left\{\frac{1}{2} (M Q^{\alpha\beta} M^T)_{ij}\overline{D_i \psi_\alpha} D_j \psi_\beta + \frac{1}{2\area} \left((\partial_1 A_2 - \partial_2 A_1)^2 + (M_{2j} \partial_j A_3)^2 + (M_{1j} \partial_j A_3)^2\right)
+\area F_p(\psi_\alpha)\right\} dX_1 \, dX_2,
\label{eq:Esq} 
\end{align*}
\end{widetext}
leading in turn to the Gibbs free energy,
\begin{align}
\nonumber G(M,\area,\psi_\alpha,A) &= F - \int_{[0,1]^2} H_0(\partial_1 A_2 - \partial_2 A_1)\, dX_1 \, dX_2 \\
&= F - 2n\pi H_0.
\end{align}
where $n$ is the winding number of the field configuration.

We seek minimizers of $G/\area$ with respect to the fields and the
shape $M\in SL(2,\R)$ and area $\area\in(0,\infty)$ of the unit cell. Note that the degree or winding number of a given unit cell $n$ is fixed. This means that strictly speaking we must find the global minimizer of all $G_n/\area$, where $G_n$ is the Gibbs free energy of a unit cell of degree $n$ and then minimize over $n$.
In practice, we find minima of $G_n/\area$ until we get a repeated minimizer, that is, until we find a minimizer of $G_n/\area$ whose cell and field configuration is two cells of the $G_{n/2}/\area$ minimizer joined together, and hence $G_n/\area = G_{n/2}/\area$.

\subsection{Numerical method}

To numerically find minimizers of $G(M,\area,\psi_\alpha,A)$ we discretize the standard square unit cell as described in the vortex section, but with periodic boundary conditions,
\begin{align}
\psi_\alpha( X + (1,0) ) &= \psi_\alpha (X)e^{i n2\pi n X_2},\nonumber\\
\psi_\alpha( X + (0,1) ) &= \psi_\alpha (X),\nonumber\\
A_2 ( X + (1,0) ) &= A_1(X) + 2\pi n \nonumber \\
A_i ( X + (1,0) ) &= A_2(X)\quad i \neq 2 \nonumber \\
A_i ( X + (0,1) ) &= A_i(X)
\label{eq:latticeBCs}
\end{align}
where $n$ is the winding number of the unit cell. We set the fields $\phi = (\psi, A)$ on the unit square torus and the geometry of the unit cell $(M,\area)$ to be some initial condition, avoiding anything too symmetric so as not to bias the results. Then, fixing the unit cell $(M,\area)$, we find a local minimum w.r.t. the collected fields $\phi$, using arrested Newton flow for a particle subject to the potential $G_{dis}(M,\area,\phi)$. This is continued for a small fixed number of steps. We then fix the field configuration $\phi$ and area $\area$, and minimize $G_{dis}(M,\area,\phi)/\area$ with respect to $M \in SL(2,\mathbb{R})$ to a very small tolerance. We will discuss this step in more detail in the next subsection. Finally, we fix the fields $\phi$ and the shape $M$ and minimize $G_{dis}/\area$ w.r.t. $\area$. This last step can be performed exactly using elementary calculus.

The above process is repeated, switching between minimizing $G/\area$ w.r.t.\ the collected fields $\phi$, the shape $M$ and the area $\area$. Once a given tolerance is reached for all 3, we stop the minimization process.

\subsection{Finding the minimal shape}
To find the minimal shape of a unit cell with a given configuration $(\psi_\alpha, A)$, we must solve an optimization problem. We first note that the only terms of $G$ that are dependent on the shape $M$ are the gradient term and the in-plane magnetic terms of the free energy. In fact,
\begin{equation}
G=\frac12 M_{ac}P_{ac,bd}M_{bd}+C_{ab}M_{ab}+D
\end{equation}
where $D$ contains the terms independent of $M$ and $P = P^1 + P^2$ and $C$ are given by,
\begin{align}
P^1_{ac,bd}&:= Re\int_{[0,1]^2}Q^{\alpha\beta}_{cd} \overline{D_a \psi_\alpha}D_b \psi_\beta\\
P^2_{ac,bd} &:= \int_{[0,1]^2} \frac{1}{\area}\delta_{ab}\partial_c A_3 \partial_d A_3\\
C_{ab} &:= \int_{[0,1]^2} A_3 Im\left( Q^{\alpha\beta}_{3a} \overline{\psi}D_b \psi_\beta \right).
\end{align}
It is convenient to identify $M$ with the vector,
\begin{equation}
m := \left( M_{11}, M_{12}, M_{21}, M_{22} \right)\in\R^4,
\end{equation}
by thinking of the pair $(a,b)$ as a single index ranging over $\{(1,1),(1,2),(2,1),(2,2)\}$. Then
\begin{equation}
G(m) = \frac{1}{2} m^T P m + C^T m +D,
\end{equation}
where $P = P^1 + P^2$ has been reinterpreted as a real symmetric $4\times 4$ matrix, $C\in\R^4$, using the same re-indexing trick, and $D$ contains the energy terms that are independent of $m$. Hence we must minimize $G(m)$ over $\R^4$ subject to the constraint $\det M = 1$ (or $M \in SL(2,\mathbb{R})$), that is, 
\begin{equation}
\frac{1}{2} m^T J m = 1, \quad \quad J := \left( \begin{array}{cccc} 0 & 0 & 0 & 1\\ 0 & 0 & -1 & 0 \\ 0 & -1 & 0 & 0 \\ 1 & 0 & 0 & 0 \end{array}\right).
\label{eq:constraint}
\end{equation}
Note that $J^2 = I_4$. To consider minimisers subject to the above constraint, we add a Lagrange multiplier term to the energy $G(m)$,
\begin{equation}
\lambda\left( 1 - \frac{1}{2} m^T J m \right),
\end{equation}
Hence, we seek $m\in\R^4$ such that
\begin{equation}
P m + C -\lambda J m,\qquad m^T J m=2
\label{eq:linearEq}
\end{equation}
for some $\lambda \in \mathbb{R}$.

Finding solutions to \eqref{eq:linearEq} is a challenging problem in general. However, it is simplified in the special case of the basal plane. If we assume that the vortex plane normal is $\hat{z}$, then $\partial_i A_3 = 0$ and hence $C = 0$. Then \eqref{eq:linearEq} is an eigenvalue problem (where we have used the fact that $J^2 = I_4$),
\begin{equation}
J P m = \lambda m.
\end{equation}
In other words $m$ is an eigenvector of $J P$ and $\lambda$ is given by the corresponding eigenvalue. This allows us to minimize $G(m)$ w.r.t. $m$ explicitly by: 
\begin{itemize}
\item constructing $J P$,
\item finding its 4 eigenvectors,
\item selecting the eigenvector with smallest positive real eigenvalue $\lambda$,
\item normalizing the eigenvector s.t.\ $m^T Jm = 2$.
\end{itemize}
If we aren't in the basal plane however, we cannot assume $C = 0$ and hence we will find the minimizer of $G(m)$ using a gradient flow algorithm. We first find the eigenvector that corresponds to the smallest eigenvalue of $JP$ and use this as an initial condition. We then numerically evolve the vector $m$, calculating the time derivative at each step as,
\begin{equation}
\frac{dm}{dt} = - \left( Pm + C - \frac{Jm}{2}( m^T P m + m^T C) \right).
\end{equation}
Once the gradient reaches a small tolerance we stop the algorithm.

Finally, whether through explicit calculation or gradient flow, we have found the vector $m$, and hence the unit cell shape
$M\in SL(2,\R)$ that minimizes $G$. Hence we can read off the new period lattice as being the span of the columns of
\begin{equation}
L = \sqrt{\area}M^{-1}, \quad \quad M = \left( \begin{array}{cc} m_1 & m_2 \\ m_3 & m_4 \end{array}\right).
\end{equation}

Having minimized $G$ with respect to the fields and then the shape $M$ of the unit cell, the last step in each iteration of our algorithm is to minimize $G/\area$ with respect to $\area$. Since
\begin{widetext}
\begin{align*}
\frac{G}{\area}=\frac{1}{2\area^2}\int_{[0,1]^2} \left(
(\partial_1A_2-\partial_2 A_1)^2 + m^T P^2 m \right)dX_1dX_2
+\frac{1}{\area}\left(\frac12m^TP^1 m+C^Tm-2n\pi H_0\right)+
\int_{[0,1]^2}F_p(\psi)dX_1dX_2,
\end{align*}
\end{widetext}
this step has a unique solution provided $H_0$ is sufficiently large.

\subsection{Numerical solutions}
We are interested in understanding the structure of  bulk solutions when $H_{c_1} < H_0 < H_{c_2}$, namely vortex lattices. In addition, we want to understand how the lattice changes as the strength of the external field $H_0$ changes. As the critical fields are parameter and orientation dependent, we first choose our $\eta$ parameter and the external field direction $\hat{H}$, performing a change of basis to $(x_1,x_2,x_3)$, s.t. $\hat{H} = \hat{x}_3$. We then approximate $H_{c_1}$ and $H_{c_2}$ using the methods described above. Hence, given our parameters, we want to understand the 1-dim family of solutions that minimize $G/\area$ and are parametrized by $H_0$. Hence, we find the minimal lattice for $H_0 = H_{c_1} + \frac{1}{2}(H_{c_2} - H_{c_1})$ using the method described above. Then using this solution as an initial condition we vary $H_0$ up to $H_{c_2}$ to find half the family and then down to $H_{c_1}$ to find the other half. We then have a set of solutions that represent how the fields and geometry of the periodic vortex lattice change as the external field is increased. Note that for $H \leq H_{c_1}$ it is optimal for $\area$ to diverge, making simulations challenging when very close to $H_{c_1}$.

By tracking $G/\area$ for the family of solutions, we can numerically find $H_{c_1}$ and $H_{c_2}$ allowing us to check the accuracy of our approximations above. In particular, $H_{c_1}$ occurs when the normalised Gibbs free energy per unit area $\hat{G}_n/\area = (G_n - G_0)/\area$ = 0, where $G_n$ is the global minimiser of degree $n$. Note that $G_0$ is the energy corresponding to the homogeneous superconducting state with $B = 0$ and $\psi_\alpha = u_\alpha$.

To check $H_{c_2}$, we consider the normal state, where $\psi_\alpha = 0$ and $B = H$. This leads to a normal state Gibbs free energy of $\hat{G}_{norm} = -F_P(u_\alpha) - H^2$. $H_{c_2}$ is then the value of $H_0$ such that $\hat{G}_{norm} = \hat{G}(H_0)$.

Using the above critical field values, we can predict the magnetic response of a material, or the magnetic flux per unit area that penetrates a superconducting material as $H_0$ changes. For the homogeneous superconducting ground state $\frac{1}{\area_{\Omega}}\int_{\Omega} B = 0$, for the vortex lattice state $\frac{1}{\area_{\Omega}}\int_{\Omega} B = \frac{2 \pi n}{\area}$ and finally the normal sate $\frac{1}{\area_{\Omega}}\int_{\Omega} B = H_0$, where $\Omega$ is the system and $n$ is the degree or winding number of the unit cell that is the global minimiser of $G/\area$.

\begin{figure*}
\hspace*{-1cm} 
\includegraphics[width=1.1\linewidth]{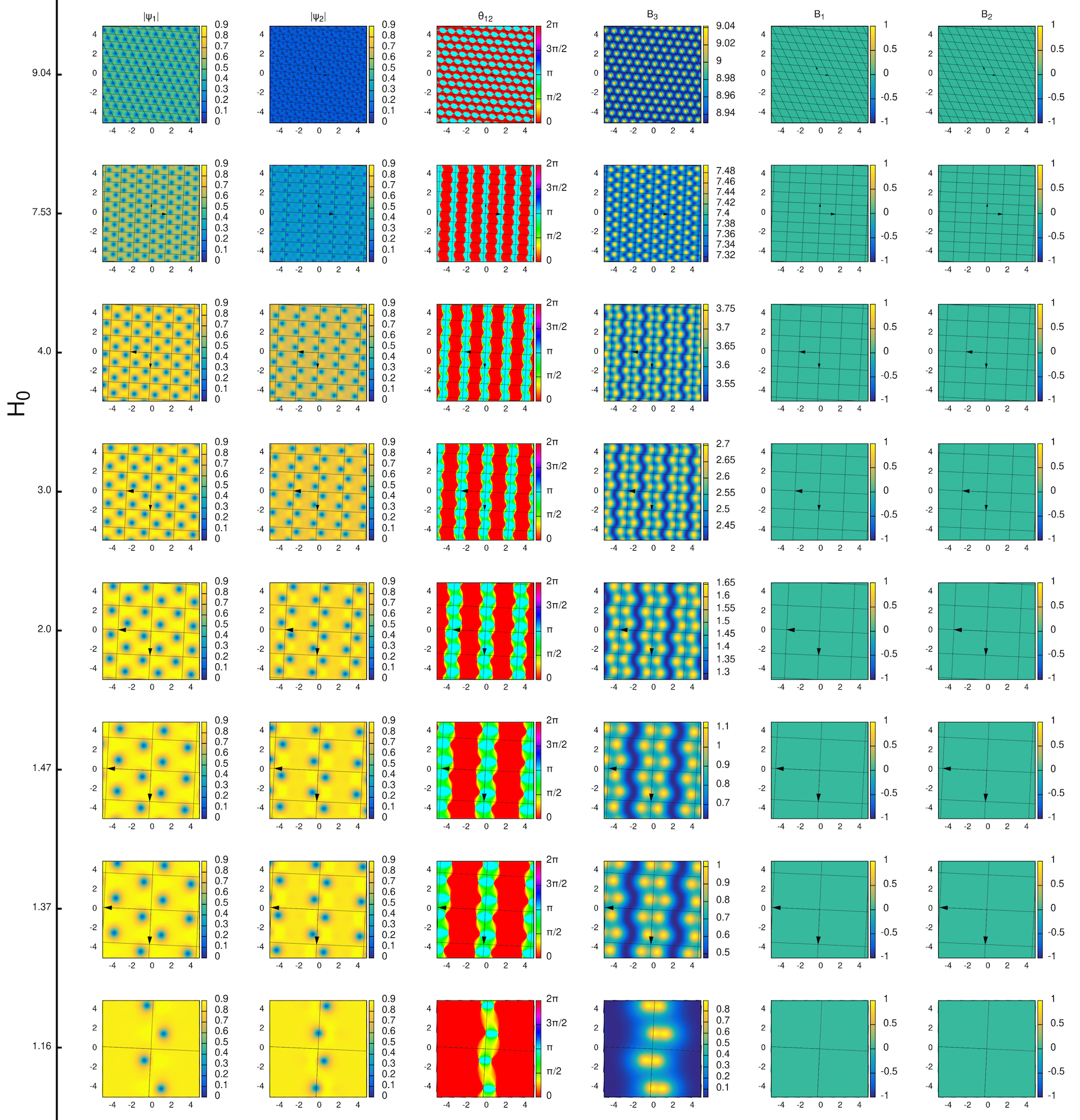}
\caption{\label{Fig:k3_basal_lat}
Contour plots of the vortex lattice for $\eta = 3$ in the basal plane $\hat{x}_3 = (0,0,1)$ with increasing external field strength $H = H_0 \hat{x}_3$. Increasing $H_0$ initially causes the Skyrmion chains to squash together. Then the length of the chains is squashed. Finally there is a phase transition to a standard triangular lattice. The unit cells are marked by black lines and tessellated in the plane. Note that the deviation in the order parameter and magnetic field is decreasing for high external field.}
\end{figure*}

\begin{figure}
\includegraphics[trim=0 0 0 0,clip,width=1.0\linewidth]{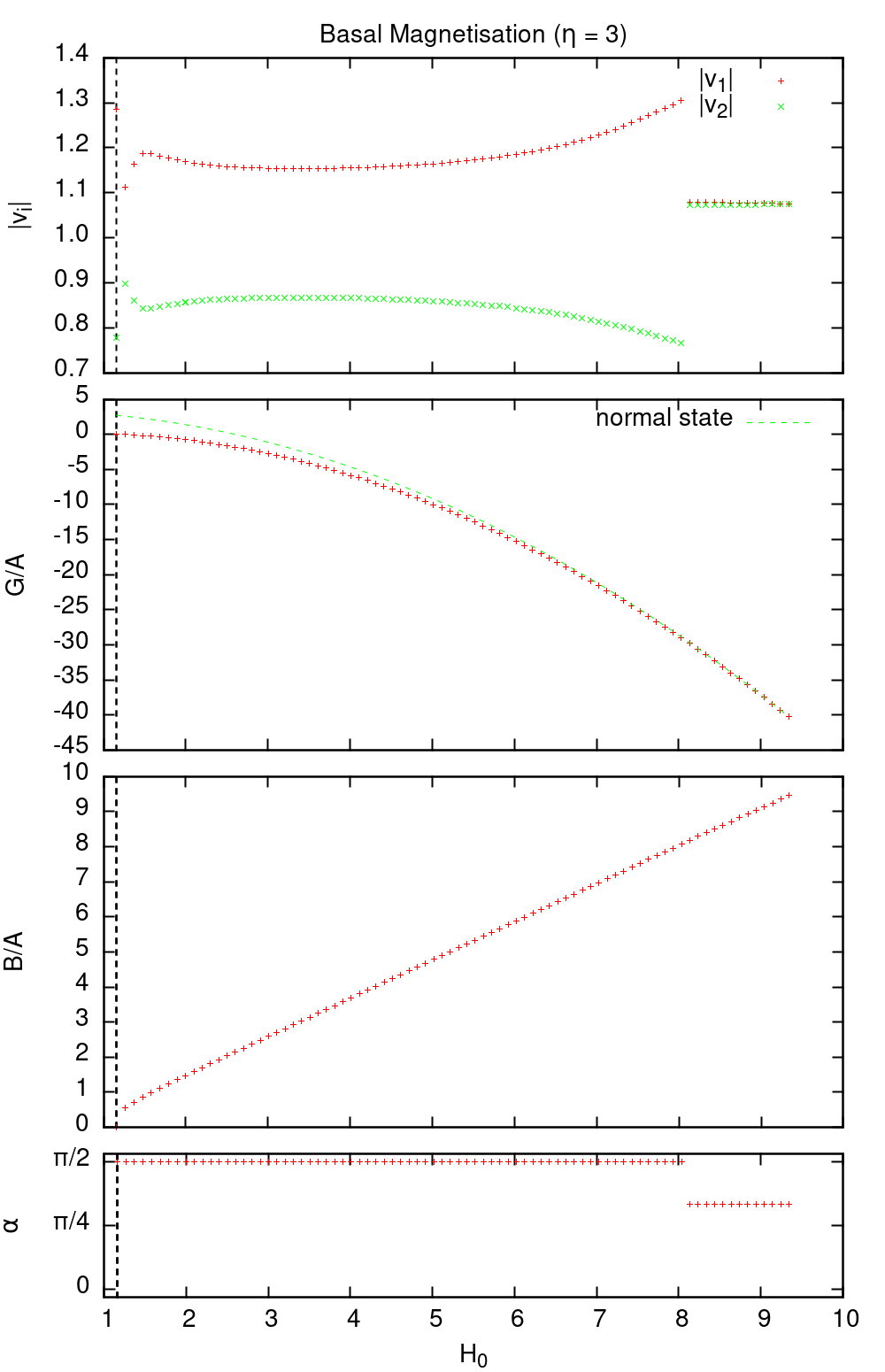}
\caption{\label{Fig:k3_basal_mag}
Plot of the geometry of the minimal unit cell for $\eta = 3$ in the basal plane $\hat{x}_3 = (0,0,1)$ for increasing external field  strength $H=H0x3$. The top plot shows the relative lengths of the vectors $v_1$, $v_2$ that form the unit cell. $G/\mathcal{A}$ is the normalised Gibbs free energy per area which is approximately 0 at $H0=Hc1$ (dashed line). The third plot gives the magnetic response which is proportional to the inverse of the area $B/\mathcal{A} = 2\pi n/\mathcal{A}$ where the winding is initially $n=2$, but for the top row is $n=1$. Finally $\alpha$ is the angle of the unit cell.}
\end{figure}

The result of this procedure for $\eta = 3$, in the basal plane $\hat{x}_3 = (0,0,1)$, where the external field is orthogonal to the plane $H = H_0 x_3$, is plotted in \figref{Fig:k3_basal_lat}. The corresponding magnetic response, unit cell geometry and normalised Gibbs free energy are plotted in \figref{Fig:k3_basal_mag}. The unit cell for the global minimiser is initially rectangular ($\alpha= \pi/2$) with degree $n=2$. The field configuration takes the form of chains of Skyrmions, such that the Skyrme charge is $Q = 2$ for each unit cell. For $H_0$ near $H_{c_1}$ we observe well separated chains, but as $H_0$ increases the chains get closer and then squash together. This can be seen in the resulting field configuration plots in \figref{Fig:k3_basal_lat} and also in the top plot of \figref{Fig:k3_basal_mag}. We see $|v_1|$ the separation of the chains shrinking comparatively to $|v_2|$ the length of a link in the chain. Once the chains are particularly close we then see a slight squashing of the chain link length comparative to the separation. Finally, at $H_0 \approx 8.0$ there is a phase transition as the unit cell becomes triangular ($\alpha = \pi/3$) with degree $B=1$ and $|v_1| \approx |v_2|$. This means we go from two qualitative length scales: the chain separation and the chain link length, to one length scale: the vortex separation. Finally, note that in the basal plane, as with the vortex bound states, there is no  generation of spontaneous in-plane magnetic field, as predicted by the linearisation.

It is worth noting that the normalised Gibbs free energy per unit area plotted in \figref{Fig:k3_basal_mag} goes to zero at $H=H_{c_1}$ marked by the dashed line. This confirms our approximation of $H_{c_1}$ from the previous section. In addition, as expected the order parameter becomes suppressed as $H_0$ approaches $H_{c_2}$ and the orthogonal magnetic field $B_3$ approaches $H_0$ as $H_0$ approaches $H_{c_2}$ with reduced deviation. 

\begin{figure*}
\hspace*{-1cm}
\includegraphics[width=1.1\linewidth]{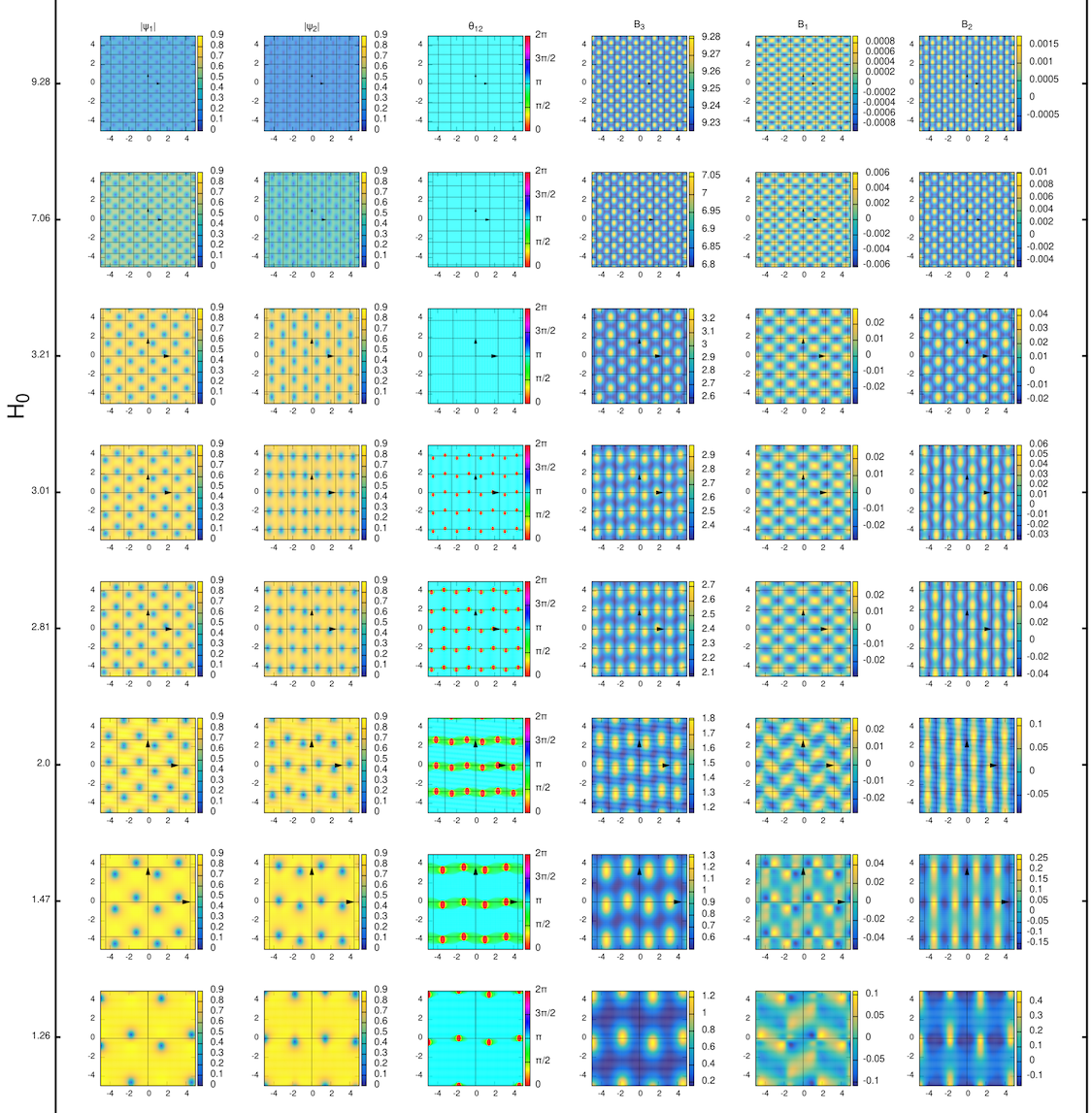}
\caption{\label{Fig:k3_half_lat}
Contour plots of the vortex lattice for $\eta = 3$ on the half plane $\hat{x}_3 = (0,1,1)/\sqrt{2}$ for increasing external field strength $H = H_0 x_3$. $H_0$ (noted on the left) increases up the page causing the chains to squash together. The unit cells are marked by grey lines and tessellated in the plane. The unit cell has degree $n=2$ and Skyrme charge $Q=2$. At $H_0 = 3.25$ the Skyrmions disappear ($Q=0$), leaving composite vortices. The coordinates are $x_1 = (1,0,0)$, $x_2 = (0,1,-1)/\sqrt{2}$.}
\end{figure*}

We also consider the results of performing this process for $\eta = 3$, on the tilted plane $\hat{x}_3 = (0,1,1)/\sqrt{2}$, giving rise to the field configurations plotted in \figref{Fig:k3_half_lat}. We again observe the formation of Skyrmion chains, with a rectangular unit cell ($\alpha = \pi/2$) of degree $n=2$. The biggest difference is the spontaneous generation of in plane magnetic field $B_1$, $B_2 \neq 0$. We again see the chain separation decrease as the external field is increased. Then the chain lengths are squashed and eventually the vortices are forced together, forming composite vortices in a triangular lattice rather than Skyrmions. This transition occurs at a smaller value of $H_0$ than for the basal plane.

\section{Conclusion}
We have discussed   properties of nematic superconductors with a focus on their response to an external magnetic field $H$ and the nature of topological excitations. We first established that the fundamental length scales are nontrivial. In standard superconductors the fundamental length scales are the coherence lengths, associated with the modulus of the order parameter,
and magnetic field penetration depth, describing monotonic decay of the magnetic field. In nematic superconductors the modes are mixed, leading to each physical quantity being described in general by multiple (complex) length scales. This leads to  magnetic fields spontaneously twisting in space. This was demonstrated by considering a boundary problem for the Meissner state, which exhibited both spontaneous fields and magnetic field twisting.

We then considered topological excitations in the bulk, showing that the form of the solutions depend on the parameters of the model and the orientation of the vortex string. There is a large parameter region that admits Skyrmions as solutions and this region depends on the orientation of the solution relative to the crystal axes. We also showed that spontaneous orthogonal fields are excited that cause the magnetic field to twist away from the vortex line. This magnetic field twisting occurred whenever the vortex line was not orthogonal to the basal plane and was up to $10\%$ of the maximum strength of the magnetic field parallel to the vortex line. This should easily be detectable in muon spin rotation experiment and give a clear hallmark of nematic superconductivity.

We then used these results to calculate $H_{c_1}$, which we showed was anisotropic. We then confirmed previous results concerning the weak anisotropy of $H_{c_2}$.

Finally, we introduced a new method to find vortex lattices, by finding the periodic unit cell of the lattice with no assumption of the symmetry of that lattice. This new method can be used to find the vortex lattice solutions in any anisotropic model, without biasing the symmetry of the result. We found for nematic systems that the unit cell tended to contain two flux quanta (Skyrmions) that formed chains for low applied fields $H$. Then as the external field strength was increased the chains squashed together tightly. Finally, the system underwent a phase transition and the vortices became composite and were forced into a traditional triangular lattice. Note that the unit cells with two quanta consist of well separated half-quantum vortices and are markedly different from the double-quanta vortex lattices found, for example, in chiral p-wave superconductors \cite{garaud2016lattices}.

The vortex states discussed above could be detected in scanning squid probes, scanning Hall probes and small-angle neutron scattering. However, the most distinct signal may come from muon spin relaxation experiments. We will directly consider the signal resulting from such an experiment in a follow up paper shortly \cite{companion}.
\appendix

\section{Newton flow numerical method}
Throughout the paper we make use of a newton flow method to approximate local minimisers of several energy functionals w.r.t.  the fields $(\psi_1, \psi_2, A)$. These different energy functionals describe the same system but with different assumptions, conditions or boundary conditions applied.  In particular we have three different cases:
\begin{itemize}
\item Meissner state - a 1-dimensional boundary problem with natural boundary conditions (see section V).  We choose our parameters $(\hat{x}_1, \hat{x}_3, H_0, \eta)$, transforming the anisotropy matrices as described in \eqref{eq:eom}, according to our chosen orthonormal basis $\{\hat{x}_1, \hat{x}_2, \hat{x}_3\}$.  This results in a 1-dimensional energy in terms of three dynamic fields $(\psi_1,\psi_2,A)$.
\item Vortex clusters - a 2-dimensional problem on a regular grid with fixed boundary conditions (see section VI).  We first choose our parameters, transforming the anisotropy matrices according to \eqref{eq:trid}. This gives a free energy dependent on three gauge dependent dynamic fields $(\psi_1,\psi_2,A)$.
\item Vortex lattices - a 2-dimensional problem on a non-trivial unit cell with periodic (up to winding) boundary conditions  (see section VIII). We choose our parameters and fix the geometry of the unit cell (see section VIII) such that our resulting fields are on a regular unit square with the boundary conditions given in \eqref{eq:latticeBCs}.
\end{itemize}
It is important to note that for all simulations, due to magnetic field twisting, we must retain all three components of the gauge field $A$, as we cannot assume the magnetic field direction is fixed.

Having performed the above transformation on the fields and/or space for the given problem, we seek local minimisers of the transformed energy functional with respect to the fixed parameters. Hence,  we discretize the resulting fields on a regular grid of $N_1^d$ lattice sites with spacing $h > 0$, where $d$ is the dimension of the particular problem.  We approximate the 1st and 2nd order spatial derivatives using central 4th order finite difference operators, yielding a discrete approximation $F_{dis}$ to the functional $F(\phi)$ in \eqref{eq:F} (after bring transformed as described above), where $\phi = (\psi_\alpha, A_i)$ are the collected fields. If we consider the function $F_{dis} : \mathcal{C}\rightarrow \mathbb{R}$, where the discretized configuration space is the manifold $\mathcal{C} = (\mathbb{C}^2\times\R^3)^{N_1^d} \approx \mathbb{R}^{7N_1^d}$. We then seek local minima of $F_{dis}$ subject to the boundary conditions of the chosen problem:
\begin{itemize}
\item Meissner state - $\rho_\alpha = u_\alpha$, $\varphi_1 = 0$, $\varphi_2 = - \delta_{12}$ and $A_i = 0$ on the right boundary of the computational grid and the left boundary conditions are given by the natural conditions described in appendix B.
\item Vortex bound states - $\rho_\alpha = u_\alpha$, $\varphi_1 = n \, \theta$, $\varphi_2 = n \, \theta - \delta_{12}$ and $A_i = 0$ on the boundary of the computational grid, where $n$ is the degree or winding number and $\theta$ is the polar angle in the plane. 
\item Vortex lattices - the boundary conditions are described in \eqref{eq:latticeBCs}.
\end{itemize}
We then evolve the system in \eqref{eq:F}, using a gradient decent method, in particular an arrested Newton flow algorithm (described in detail in \cite{speight2020skyrmions}), solving for the motion of a particle in $\mathcal{C}$ under the potential $F_{dis}$,
\begin{equation}
\ddot{\phi} = - \mbox{grad} F_{dis}(\phi),
\end{equation}
starting at an initial configuration $\phi(0)$ and $\dot{\phi} = 0$. Evolving the algorithm causes the configuration to relax towards a local minimum. At each time step $t \mapsto t + \delta t$, we check to see if the direction of the force on the particle opposes its velocity. If $\ddot{\phi}_{i}(t+dt) \cdot \ddot{\phi}_{i}(t) < 0$, then the we set $\dot{\phi}_{i}(t) = 0$ and continue relaxing the configuration. The flow was terminated once the discrete approximate was sufficiently close to a local minimum, namely when every component of $\mbox{grad} F_{dis}(\phi)$ was zero within a given tolerance.

\section{Natural Boundary Conditions}
In order to numerically compute the Meissner state on the half-line $\Omega$ we must minimise the Gibbs free energy,
\begin{equation}
G = \int_\Omega (\mathcal{F} - HB) + \int_{\partial \Omega} \mathcal{F}_{\text{surf}} =: \int_\Omega \mathcal{G}.
\end{equation}
 We will denote the dynamical fields as $\phi_a$,$a= 1, ... , 7$ (consisting of the real and imaginary parts of $\psi_\alpha$ and the components of $A = (A_1, A_2, A_3)$. The variation of $G$ is then,
\begin{align}
\delta G = \int_\Omega \left( \frac{\partial \mathcal{G}}{\partial \phi_a} - \partial_i \left( \frac{\partial \mathcal{G}}{\partial( \partial_i \phi_a)}\right) \right) \delta \phi_a \\
+ \int_{\partial \Omega} \left( \frac{\partial \mathcal{F}_{\text{surf}}}{\partial \phi_a} - n_i \frac{\partial \mathcal{G}}{\partial(\partial_i \phi_a)} \right) \delta \phi_a,
\end{align}
where we have used the divergence theorem, and recalled that $\nvec$ is an \emph{inward} pointing normal to $\partial \Omega$. Demanding that $\delta G = 0$ for all variations requires both of these integrals vanish identically, and hence that $\phi_a$ satisfy the usual Euler-Lagrange equations in $\Omega$ together with the boundary conditions,
\begin{equation}
\frac{\partial \mathcal{F}_{\text{surf}}}{\partial \phi_a} - n_i \frac{\partial \mathcal{G}}{\partial (\partial_i \phi_a)} = 0
\end{equation}
on $\partial \Omega$. 

In general boundary conditions should be calculated microscopically and they are strongly affected by the Friedel oscillations of the density of states near the surface \cite{samoilenka2020microscopic}. For our model we ignore the surface terms, $\mathcal{F}_{\text{surf}} = 0$, as  we are interested in the functional form of the long range of asymptotic field behaviour away from the boundary, which is determined by bulk normal modes. Thus we reduce the boundary conditions to the following,
\begin{align}
n_i Q^{1\beta}_{ij}D_j\psi_\beta = 0,\quad & \quad n_i Q^{2\beta}_{ij}D_j \psi_\beta = 0\\
\partial_i A_i = 0 \quad, & \quad B = H.
\end{align}
Imposing translational invariance $\psi_\alpha = \psi_\alpha (X)$, $A_i = a(X) n_i^\perp + b(X) n_i^\star + c(X) n_i$ and $B(X) = (0,-b',a')$ where $X = n_i x_i$ and assuming that the external magnetic field $H = (0,0,|H|)$ is parallel to $n^\perp$, hence always orthogonal to $X$. This gives the boundary conditions at $X=0$ to be,
\begin{align}
\nvec\cdot Q^{\alpha\beta}\nvec\left( \psi'_{\beta}(0) + ic(0) \psi_\beta(0)\right)\\ + i\nvec \cdot Q^{\alpha\beta}( \nvec^\perp a(0) + \nvec^\star b(0) )\psi_\beta(0) = 0\\
b'(0) = c'(0) = 0,\\
a'(0) = H.
\end{align}
For the other boundary at $X=L$, we assume $L$ so large that the fields decay to their bulk ground state values. Hence we fix the field values, demanding that $a'=b'=c'=0$, $\psi_1 = u_1$ and $\psi_2 = u_2$.

\section{Rescaling of Fields and Parameters}
Our starting point is the model presented in \cite{zyuzin2017nematic} with the following free energy density,
\begin{align}
\mathcal{F} &= \sum_{s = \pm} \left\{ - |\Delta_s|^2 + |D_x \Delta_s|^2 + |D_y \Delta_s|^2 + \beta_z |D_z \Delta_s|^2 \right. \nonumber\\ &
\left. + \beta_\perp \ol{D_{-s} \Delta_s} D_s \Delta_{-s} + \frac{1}{2} |\Delta_s|^4 + \frac{\gamma}{2} |\Delta_s|^2 |\Delta_{-s}|^2 \right\},
\label{eq:oldF}
\end{align}
where $\Delta_\pm = |\Delta_\pm|e^{i\varphi_\pm}$ is the complex order parameter and we have covariant derivatives $D_\pm = D_x \pm i D_y$ where the standard covariant derivative is $D_i = -(i/\eta) \partial_i + a_i$, where $a$ is the gauge field.

We can rescale the theory presented in \eqref{eq:oldF} using the following rescaled fields,
\begin{align}
\mathfrak{F} &= \frac{1}{2}\eta^2 \mathcal{F},\\
A &= -\frac{1}{\eta} a, \\
\psi_1 &= \Delta_+, \\
\psi_2 &= \Delta_-,
\end{align}
giving the following the 3-dimensional free energy $F = \int_{\mathbb{R}^3} \mathfrak{F}$ given in \eqref{eq:F} with anisotropy given in \eqref{eq:nematic} and potential given in \eqref{eq:Fp}.

We are interested in modelling a layered material such as $Bi_2Se_3$. The assumption that the Fermi velocity in the plane of the layers $v$ is equivalent to the orthogonal Fermi velocity $v_z$ leads to the following parameter values \cite{zyuzin2017nematic},
\begin{equation}
\beta_\perp = \frac{1}{3}, \quad \beta_z = \frac{4}{3}, \quad \gamma = \frac{1}{3},
\end{equation}
which are the parameters we will use throughout the paper.

\section{Acknowledgements}
We thank Julien Garaud and Alexander Zyuzin for useful discussions. The work of MS and TW is supported by the UK Engineering and Physical Sciences Research Council through grant EP\/ P024688\/ 1. TW is also supported by an academic development fellowship, awarded by the University of Leeds. EB is supported by the Swedish Research Council Grants No. 2016-06122, 2018-03659 and Olle Engkvists Stiftelse. The numerical work of this paper was performed using the code library Soliton Solver, developed by one of the authors, and was undertaken on ARC4, part of the High Performance Computing facilities at the University of Leeds.

\bibliography{bibliography}{}
\bibliographystyle{unsrt}

\end{document}